\documentclass[11pt]{article}
\usepackage{amssymb,latexsym,amsmath,graphicx}
\usepackage{mathtools}
\usepackage{latexsym}
\usepackage{mathrsfs}
\usepackage{amsfonts}
\usepackage{float}
\usepackage{booktabs}
\usepackage{bbm}
\usepackage{dsfont}
\usepackage[colorlinks=true, urlcolor=blue, linkcolor=blue]{hyperref}

\usepackage{lscape}

\newcommand{\argmin}{\operatornamewithlimits{argmin}}

\newtheorem{thm}{Theorem}
\newtheorem{lem}{Lemma}

\usepackage{epstopdf}
\usepackage{xcolor}

\begin{document}

\begin{center}
	\Large{{\bf Nonparametric functional data classification and bandwidth selection in the presence of MNAR class variables}} 
\end{center}

\begin{center}
	{\bf Majid Mojirsheibani}\footnote{
		 Corresponding author.
 Email:  majid.mojirsheibani@csun.edu~\\
		This work was supported by the NSF under Grant DMS-2310504 of Majid Mojirsheibani}\\
	{\bf Department of Mathematics, California State University, Northridge, CA\\ 91330, USA\\
	{\rm \large The final version will be published in Statistics}}
\end{center}

\vspace{5mm}
\begin{abstract}
The problem of nonparametric functional data classification and bandwidth selection is considered when the response variable, also called the class label, might be missing but not at random (MNAR). This setup is broadly acknowledged to be more challenging than the simpler case of missing at random  setup.  With the focus on kernel methods, we develop nonparametric classification rules and also propose bandwidth selection procedures for the proposed estimators. To study the performance of the resulting classifiers, we look into the rates of convergence of the misclassification error of the proposed method to that of the theoretically optimal classifier. The 
final-sample performance of our classifiers is assessed via numerical studies.
\end{abstract}

\allowdisplaybreaks

\section{Introduction}\label{intro}
Consider the following two-group classification problem. Let $(\mbox{\large $\boldsymbol{\chi}$}, Y)\in  \mathbb{X} \times \{0,\,1\}$ be a random pair where $\mbox{\large $\boldsymbol{\chi}$}$ is a functional covariate taking values in some abstract space $\mathbb{X}$, and $Y$ is the class variable (label)  that must be predicted based on $\mbox{\large $\boldsymbol{\chi}$}$.  More precisely, the aim of classification is to find a map $\mathcal{T}:\mathbb{X}\to \{0,1\}$  for which the probability of misclassification, i.e., 
$ L(\mathcal{T}) := P\{\mathcal{T}(\mbox{\large $\boldsymbol{\chi}$})\neq Y\}$,
is as small as possible. The classifier with the lowest misclassification error, also referred to as the Bayes classifier, is given by
\begin{equation}\label{Bayes}
\mathcal{T}^{\mbox{\tiny opt}}(\mbox{\scriptsize $\boldsymbol{\chi}$})=\left\{
\begin{array}{ll}
1 ~ & ~ \hbox{if} ~~ \mathcal{R}(\mbox{\scriptsize $\boldsymbol{\chi}$}):= E\big[Y\big|\,\mbox{\large $\boldsymbol{\chi}$}=\mbox{\scriptsize $\boldsymbol{\chi}$}\big] ~ > ~ \frac12 \\
0 ~ & ~ \hbox{otherwise;}
\end{array}
\right. 
\end{equation}
see, for example,  C\'erou and Guyader (2006), Abraham et al. (2006), and Devroye, et al.\,(1996; Ch. 2). 
In practice, finding $\mathcal{T}^{\mbox{\tiny opt}}$ is virtually impossible since the above conditional expectation is always unknown and must be estimated. However, suppose that we have access to $n$ independent and identically distributed (iid) data points, 
$
\{(\mbox{\large $\boldsymbol{\chi}$}_1,Y_1), \cdots, (\mbox{\large $\boldsymbol{\chi}$}_n,Y_n)\},$ where $(\mbox{\large $\boldsymbol{\chi}$}_i,Y_i)\stackrel{\mbox{\tiny iid}}{=}(\mbox{\large $\boldsymbol{\chi}$},Y), ~ i=1,\cdots, n
$
and let $\widehat{\mathcal{R}}(\mbox{\scriptsize $\boldsymbol{\chi}$})$ be any estimator of the regression function $E\big[Y\big|\,\mbox{\large $\boldsymbol{\chi}$}=\mbox{\scriptsize $\boldsymbol{\chi}$}\big]$. We can then consider the plug-in classifier
\begin{equation}\label{plugin}
\widehat{\mathcal{T}}_n(\mbox{\scriptsize $\boldsymbol{\chi}$})=\left\{
\begin{array}{ll}
1 ~ & ~ \hbox{if} ~~ \widehat{\mathcal{R}}_n(\mbox{\scriptsize $\boldsymbol{\chi}$}) ~ > ~ \frac12 \\
0 ~ & ~ \hbox{otherwise.}
\end{array}
\right.  
\end{equation}
The performance of the classifier (\ref{plugin}) depends on the estimator of the conditional expectation. 
In this paper we consider {\it local-averaging} nonparametric estimators of the regression function $\mathcal{R}(\mbox{\scriptsize $\boldsymbol{\chi}$})=E[Y|\,\mbox{\large $\boldsymbol{\chi}$}=\mbox{\scriptsize $\boldsymbol{\chi}$}]$ when some of the $Y_i$'s may be missing according to a {\it Missing-Not-At-Random} (MNAR) setup where the probability that $Y_i$ is missing can depend on both $\mbox{\large $\boldsymbol{\chi}$}_i$ and $Y_i$. This setup is generally acknowledged to be a challenging problem in missing data literature and is very different from the simpler case of data missing at random where the absence of $Y_i$  only depends on $\mbox{\large $\boldsymbol{\chi}$}_i$ (but not $Y_i$).

In what follows, the focus will be mainly on kernel-based classifiers. Before going any further, we note that when there are no missing values, the classical Nadaraya-Watson kernel estimator of $\mathcal{R}(\mbox{\scriptsize $\boldsymbol{\chi}$})$ is of the form
\begin{equation}\label{D1}
\widehat{\mathcal{R}}_n(\mbox{\scriptsize $\boldsymbol{\chi}$}; h):= \widehat{E}\big[Y\big|\,\mbox{\large $\boldsymbol{\chi}$}=\mbox{\scriptsize $\boldsymbol{\chi}$}\big] = \sum_{i=1}^n w_{n,i}(\mbox{\scriptsize $\boldsymbol{\chi}$};h)\, Y_i,~~~\mbox{where}~~ 
w_{n,i}(\mbox{\scriptsize $\boldsymbol{\chi}$};h)=\frac{\mathcal{K}\big(h^{-1}d(\mbox{\scriptsize $\boldsymbol{\chi}$}, \mbox{\large $\boldsymbol{\chi}$}_i)\big)}{\sum_{j=1}^n \mathcal{K}\big(h^{-1}d(\mbox{\scriptsize $\boldsymbol{\chi}$}, \mbox{\large $\boldsymbol{\chi}$}_j)\big)}
\end{equation}
where $\mathcal{K}:\mathbb{R}_+\to\mathbb{R}_+$ is the kernel used with the bandwidth  $h\equiv h_n>0$. Classical results on the convergence properties (point-wise and uniform) of the estimator (\ref{D1}) are discussed in Ferraty and Vieu (2006) and Ferraty et al (2010). Similarly, when there are no missing data, notable results on functional data classification include the work of Biau et al. (2005), Abraham et al. (2006), C\'erou and Guyader (2006), Ferraty  and Vieu  (2006), Biau et al. (2010),  Delaigle and Hall (2012), Carroll et al. (2013), and Meister (2016).


Now suppose that some of the $Y_i$'s can be missing according to a MNAR mechanism. Then the estimator $\widehat{\mathcal{R}}_n(\mbox{\small $\boldsymbol{\chi}$}; h)$ in (\ref{D1}) is not available any more and, furthermore, the estimator based on the complete cases alone, i.e., 
\begin{equation}\label{mcc1}
\mathcal{R}^{\mbox{\tiny cc}}_n(\mbox{\small $\boldsymbol{\chi}$}; h) \,:= \,\sum_{i=1}^n \Delta_i Y_i \,\mathcal{K}\big(h^{-1}d(\mbox{\scriptsize $\boldsymbol{\chi}$}, \mbox{\large $\boldsymbol{\chi}$}_i)\big)\div\sum_{i=1}^n \Delta_i \,\mathcal{K}\big(h^{-1}d(\mbox{\scriptsize $\boldsymbol{\chi}$}, \mbox{\large $\boldsymbol{\chi}$}_i)\big)
\end{equation} 
where $\Delta_i$\,=\,0 when $Y_i$ is missing (and $\Delta_i$\,=\,1 otherwise), turns out to be the wrong estimator in the sense that it estimates the quantity $E(\Delta Y|\mbox{\large $\boldsymbol{\chi}$}=\mbox{\scriptsize $\boldsymbol{\chi}$})\big/E(\Delta|\mbox{\large $\boldsymbol{\chi}$}=\mbox{\scriptsize $\boldsymbol{\chi}$})$ which is not the same as $E(Y|\mbox{\large $\boldsymbol{\chi}$}= \mbox{\scriptsize $\boldsymbol{\chi}$})$, in general, under the MNAR setup. Furthermore, methods based on imputation  (such as regression imputation) to reconstruct the missing values can also produce poor results in nonparametric classification.

To present our methodology, we also need to address the choice of the {\it selection probability} (also called the {\it nonresponse propensity}), i.e., the quantity $\pi(\mbox{\scriptsize $\boldsymbol{\chi}$}, y) =  P\big\{\Delta=1 \big| \mbox{\large $\boldsymbol{\chi}$} =\mbox{\small $\boldsymbol{\chi}$}, Y=y\big\}$, where the random variable  $\Delta$\,=\,1 if $Y$ is observable (and $\Delta$\,=\,0 otherwise).  Classical methods such as those of Greenlees et al (1982) assumed a fully parametric model for both the underlying distribution and the selection probability. Such model assumptions have been relaxed in recent years and semiparametric methods that assume a parametric model for either the outcome or the selection probability (but not both) have been proposed. For example,  when $\mbox{\small $\boldsymbol{\chi}$}\in \mathbb{R}^d$, $d$\,$\geq$\,2,  Shao and Wang (2016) use instrument variables to estimate the selection probability without specifying the outcome model for the distribution of $y|\mbox{\small $\boldsymbol{\chi}$}$. In the same vein (still for $\mbox{\small $\boldsymbol{\chi}$}\in \mathbb{R}^d$), Morikawa et al (2017) use kernel regression estimators to avoid parametric outcome model assumptions, while Tang et al (2003)  and Zhao and Ma (2022) estimate the outcome model without specifying the selection probability. Miao et al (2024) employ follow-up subsamples to deal with identifiability without using instrumental variables.  Uehara et al  (2023) 	consider a semiparametric response model and use instrument variables to estimate the selection probability. Chen et al (2020) study a semiparametric model with unspecified  missingness mechanism  and  develop maximum pseudo likelihood estimation procedures when the response conditional density is an exponential family.

For the important case of predictive models (such as classification) we  consider a versatile logistic-type selection probability  model that works as follows.  Let  $\varphi>0$ be a real-valued function on $\mathbb{R}$  and define
\[
\pi_{\varphi}(\mbox{\scriptsize $\boldsymbol{\chi}$},y)\,=\,\frac{1}{1+\exp\big\{g(\mbox{\scriptsize $\boldsymbol{\chi}$})\big\}\cdot \varphi(y)}.
\]
Then we consider the following selection probability which is in the spirit of Kim and Yu (2011)
\begin{equation}\label{NonIgnore}
P\left\{\Delta=1\,\big| \mbox{\large $\boldsymbol{\chi}$}=\mbox{\scriptsize $\boldsymbol{\chi}$}, Y=y\right\}\,= \,E\left[\Delta\,\big| \mbox{\large $\boldsymbol{\chi}$}=\mbox{\scriptsize $\boldsymbol{\chi}$}, Y=y\right]\,=\,
\frac{1}{1+\exp\big\{g(\mbox{\scriptsize $\boldsymbol{\chi}$})\big\}\cdot \varphi^*(y)} \,~:=~ \pi_{\varphi^*}(\mbox{\scriptsize $\boldsymbol{\chi}$}, y),~~
\end{equation}
where $\varphi^*$ represents the true function  $\varphi$ that can depend on unknown parameters 
and $g$ on $\mathbb{X}$ is a completely unspecified real-valued function.  A popular choice of $\varphi$ is $\varphi(y)=\exp(\gamma y)$ for some unknown parameter $\gamma$. It is also well known that estimating the unknown quantities in (\ref{NonIgnore}) can be challenging due to identifiability issues.  

To deal with these issues, here we follow Miao et al (2024) and consider the situation where one has access to  a small follow-up subsample of response values selected from the set of non-respondents.  From an applied point of view,  an attractive feature of our approach is that the follow-up subsample size can be negligibly small; this means that the  seemingly unpleasant need for a follow-up subsample can in practice be a non-issue.
 This fact is further asserted in our numerical studies where sometimes a subsample of size as small  as 2, 1, or even 0 will do!

Regression and classification under missing response setups is challenging, yet progress has already been made in the literature. In the case of MNAR data,  one can refer to the results of Bindele and Zhao (2018) for the estimation of $\beta$ in the model
$E(Y |\mbox{\large $\boldsymbol{\chi}$} = \mbox{\scriptsize $\boldsymbol{\chi}$}) = g(\mbox{\scriptsize $\boldsymbol{\chi}$}, \beta)$, where $g$ is known, those of 
Niu et al (2014) and Guo et al (2019) for linear regression, and the work of Mojirsheibani (2022) on confidence bands. 
Other closely related results include the work of Ling et al (2015), Bouzebda et al (2024), and Ferraty et al (2013).  

Our proposed methods in this paper have potential applications to the field of machine learning and statistical classification in semi-supervised learning where one must deal with large amounts of missing  labels in the data.  Researchers in machine learning have made efforts to develop procedures for utilizing the unlabeled cases (i.e., the missing  $Y_i$'s) in order to construct better classification rules; see Wang and Shen (2007). However, most such results assume that the response variable is missing completely at random; see, for example, Azizyan et al (2013). Our proposed estimators and classifiers can also be used in {\it ensemble} classification methods such as those of Biau et al (2016).

Our contributions in this paper are as follows. 
(a) We develop kernel-based classification rules in the presence of MNAR response variables (labels). The proposed methodology is a two-step procedure, where  the first step involves the construction of a family of kernel-based classifiers where the members of the family are indexed by the kernel bandwidth $h$ as well as the nonignorability parameter of the missing probability mechanism.  In the second step, a search is performed to find the member of a {\it cover} of this family  that has the smallest empirical misclassification error.    (b) We also propose bandwidth estimators that are suitable for classification with missing values. To assess the theoretical performance of the resulting classification rules, we study strong convergence properties of the proposed classifiers and look into their rates of convergence. We also carry out some numerical studies on the performance of the proposed classifiers. These studies further confirm the theoretical findings of this paper.

The rest of this paper is organized as follows.  In section \ref{Main-1} we present our initial  classifiers and study their theoretical performance when the bandwidth is a free parameter. Section \ref{bandw} proposes a bandwidth estimator in the presence of missing data. This estimator is then plugged into the initial classifier to produce our
main classification rule. Numerical studies are presented in Section \ref{examples}. All proofs are deferred to section \ref{PRF}.

\section{Main results} \label{Main-1}
\subsection{The proposed approach} \label{Main-1-1}
To construct our proposed classifiers, 
we start by using the following representation of the regression function 
$\mathcal{R}(\mbox{\scriptsize $\boldsymbol{\chi}$})=E[Y|\mbox{\large $\boldsymbol{\chi}$}=\mbox{\scriptsize $\boldsymbol{\chi}$}]$  (a proof of this appears in section\,\ref{PRF} under Lemma \ref{LEM-00})
\begin{equation}\label{repr2}
\mathcal{R}(\mbox{\scriptsize $\boldsymbol{\chi}$}) \,\equiv\,\mathcal{R}(\mbox{\scriptsize $\boldsymbol{\chi}$};\varphi^*)\,:=\,\eta_1(\mbox{\scriptsize $\boldsymbol{\chi}$}) + \frac{\psi_1(\mbox{\scriptsize $\boldsymbol{\chi}$};\varphi^*)}{\psi_2(\mbox{\scriptsize $\boldsymbol{\chi}$};\varphi^*)}\cdot \left(1- \eta_2(\mbox{\scriptsize $\boldsymbol{\chi}$})\right),
\end{equation}
where $\varphi^*$ is as in (\ref{NonIgnore}) and the functions $\psi_k$ and $\eta_k$, $k=1,2,$ are conditional expectations given by 
\begin{equation} \label{psieta}
\psi_k(\mbox{\scriptsize $\boldsymbol{\chi}$}; \varphi^*) \,:= E\big[\Delta Y^{2-k} \varphi^*(Y)\big|\mbox{\large $\boldsymbol{\chi}$}=\mbox{\scriptsize $\boldsymbol{\chi}$}\big]~~~\mbox{and}~~~\eta_k(\mbox{\scriptsize $\boldsymbol{\chi}$}) := E\big[\Delta Y^{2-k} \big|\mbox{\large $\boldsymbol{\chi}$}=\mbox{\scriptsize $\boldsymbol{\chi}$}\big],~~~\mbox{for}\,~k=1,2.
\end{equation}
Therefore, in view of (\ref{repr2}), the optimal classifier in (\ref{Bayes}) can also be expressed as 
\begin{equation}\label{NEW_Bayes}
\mathcal{T}^{\mbox{\tiny opt}}(\mbox{\scriptsize $\boldsymbol{\chi}$}) \,=\, \mathcal{T}_{\varphi^*}(\mbox{\scriptsize $\boldsymbol{\chi}$}) := \left\{
\begin{array}{ll}
1 ~ & ~ \hbox{if} ~~ \mathcal{R}(\mbox{\scriptsize $\boldsymbol{\chi}$};\varphi^*) ~ > ~ \frac12 \\
0 ~ & ~ \hbox{otherwise,}
\end{array}
~~~~~\mbox{where ~$\mathcal{R}(\mbox{\scriptsize $\boldsymbol{\chi}$}; \varphi^*)$ is as in (\ref{repr2}).}
\right. 
\end{equation}
Now let 
$$
\mathbb{D}_n=\{(\mbox{\large $\boldsymbol{\chi}$}_1,Y_1,\Delta_1),\dots,  (\mbox{\large $\boldsymbol{\chi}$}_n,Y_n,\Delta_n)\},~~~\mbox{where $\Delta_i=0$ if $Y_i$ is missing (and $\Delta_i$\,=\,1 otherwise)}
$$ 
represent an iid sample of size $n$ (data). Then in the hypothetical situation where $\varphi^*$ is completely known in (\ref{repr2}), one can consider the following estimator of $\mathcal{R}(\mbox{\scriptsize $\boldsymbol{\chi}$};\varphi^*)$
\begin{eqnarray}\label{mhat43}
\widehat{\mathcal{R}}(\mbox{\scriptsize $\boldsymbol{\chi}$}; \varphi^*, h) &=& 
\widehat{\eta}_{1}(\mbox{\scriptsize $\boldsymbol{\chi}$};h) +
\frac{\widehat{\psi}_{1}(\mbox{\scriptsize $\boldsymbol{\chi}$}; \varphi^*,h)}{\widehat{\psi}_{2}(\mbox{\scriptsize $\boldsymbol{\chi}$}; \varphi^*,h)}\, 
\big(1-\widehat{\eta}_{2}(\mbox{\scriptsize $\boldsymbol{\chi}$};h)\big),
\end{eqnarray}
where $\widehat{\psi}_{k}(\mbox{\scriptsize $\boldsymbol{\chi}$}; \varphi^*,h)$ and $\widehat{\eta}_{k}(\mbox{\scriptsize $\boldsymbol{\chi}$};h)$, $k=1,2$, are the kernel estimators given by
\begin{eqnarray*} 
\widehat{\psi}_{k}(\mbox{\scriptsize $\boldsymbol{\chi}$};\varphi^*,h)&=&
\frac{\sum_{i=1}^n \Delta_i Y^{2-k}_i \varphi^*(Y_i)\,
\mathcal{K}(h^{-1}d(\mbox{\scriptsize $\boldsymbol{\chi}$}, \mbox{\large $\boldsymbol{\chi}$}_j))}{\sum_{i=1}^n \mathcal{K}(h^{-1}d(\mbox{\scriptsize $\boldsymbol{\chi}$}, \mbox{\large $\boldsymbol{\chi}$}_j))},~~~~k=1, 2,\label{PSI1222.hat}\\
\widehat{\eta}_{k}(\mbox{\scriptsize $\boldsymbol{\chi}$};h) &=&
\frac{\sum_{i=1}^n \Delta_i Y^{2-k}_i \mathcal{K}(h^{-1}d(\mbox{\scriptsize $\boldsymbol{\chi}$}, \mbox{\large $\boldsymbol{\chi}$}_j))}{\sum_{i=1}^n\mathcal{K}(h^{-1}d(\mbox{\scriptsize $\boldsymbol{\chi}$}, \mbox{\large $\boldsymbol{\chi}$}_j))}, ~~~k=1, 2,\label{ETA1222m.hat}
\end{eqnarray*}
where  the function $\mathcal{K}:\mathbb{R}_+\to\mathbb{R}_+$ is the kernel used with the bandwidth  $h>0$. 
Of course the estimator in (\ref{mhat43}) is not available because $\varphi^*$ is unknown and must be estimated. To that end, we employ a data splitting approach that works as follows.  
Start by randomly splitting the data  into a training sample $\mathbb{D}_m$ of size $m$ and a validation set $\mathbb{D}_{\ell}$ of size $\ell=n-m$. It is assumed that as $n\to \infty$, both $\ell\to\infty$ and $m\to \infty$; the choices of $m$ and $\ell$ will be discussed later. 
In what follows, we define the index sets corresponding to $\mathbb{D}_m$ and $\mathbb{D}_\ell$ by
\[
\boldsymbol{{\cal I}}_m=\Big\{i\in\{1,\cdots,n\}\,\Big|\,(\mbox{\large $\boldsymbol{\chi}$}_i,Y_i,\Delta_i)\in \mathbb{D}_m\Big\}~~\mbox{and}~~~
\boldsymbol{{\cal I}}_\ell=\Big\{i\in\{1,\cdots,n\}\,\Big|\,(\mbox{\large $\boldsymbol{\chi}$}_i,Y_i,\Delta_i)\in \mathbb{D}_\ell\Big\}.
\]
Next, let $\mathcal{F}$ be the class of functions to which the unknown function $\varphi^*$ in (\ref{NonIgnore}) belongs and for each $\varphi\in \mathcal{F}$ consider  the following counterpart of (\ref{mhat43}) constructed based on the training set $\mathbb{D}_m$ only
\begin{eqnarray}\label{mhat3}
\widehat{\mathcal{R}}_{m}(\mbox{\scriptsize $\boldsymbol{\chi}$};\varphi,h) &=& 
\widehat{\eta}_{m,1}(\mbox{\scriptsize $\boldsymbol{\chi}$};h) +\frac{\widehat{\psi}_{m,1}(\mbox{\scriptsize $\boldsymbol{\chi}$}; \varphi,h)}{\widehat{\psi}_{m,2}(\mbox{\scriptsize $\boldsymbol{\chi}$}; \varphi,h)}\, 
\big(1-\widehat{\eta}_{m,2}(\mbox{\scriptsize $\boldsymbol{\chi}$};h)\big)
\end{eqnarray}
with the convention $0/0 = 0$, where 
\begin{eqnarray} 
\widehat{\psi}_{m,k}(\mbox{\scriptsize $\boldsymbol{\chi}$};\varphi,h)&=&
\frac{\sum_{i\in \boldsymbol{{\cal I}}_m}\Delta_i Y^{2-k}_i  \,\varphi(Y_i)\,\mathcal{K}(h^{-1}d(\mbox{\scriptsize $\boldsymbol{\chi}$}, \mbox{\large $\boldsymbol{\chi}$}_j))}{\sum_{i\in \boldsymbol{{\cal I}}_m} \mathcal{K}(h^{-1}d(\mbox{\scriptsize $\boldsymbol{\chi}$}, \mbox{\large $\boldsymbol{\chi}$}_j))},~~~k=1, 2,~~~ \varphi\in\mathcal{F},\label{PSI12.hat}\\[4pt]
\widehat{\eta}_{m,k}(\mbox{\scriptsize $\boldsymbol{\chi}$};h) &=&
\frac{\sum_{i\in \boldsymbol{{\cal I}}_m}\Delta_i Y^{2-k}_i \mathcal{K}(h^{-1}d(\mbox{\scriptsize $\boldsymbol{\chi}$}, \mbox{\large $\boldsymbol{\chi}$}_j))}{\sum_{i\in \boldsymbol{{\cal I}}_m} \mathcal{K}(h^{-1}d(\mbox{\scriptsize $\boldsymbol{\chi}$}, \mbox{\large $\boldsymbol{\chi}$}_j))}
,~~~k=1, 2, \label{ETA12m.hat}  
\end{eqnarray}
and for each $\varphi$ in $\mathcal{F}$ define the corresponding classifier
\begin{equation}\label{gFin}
\widehat{\mathcal{T}}_{m,\varphi,h}(\mbox{\scriptsize $\boldsymbol{\chi}$})=\left\{
\begin{array}{ll}
1 ~ & ~ \hbox{if} ~~ \widehat{\mathcal{R}}_{m}(\mbox{\scriptsize $\boldsymbol{\chi}$};\varphi,h)  ~ > ~ \frac12 \\
0 ~ & ~ \hbox{otherwise.}
\end{array}
\right. 
\end{equation}

\vspace{3mm}\noindent
To estimate $\varphi^*$, we will use the approach based on the approximation theory of totally bounded function spaces. This turns out to be a suitable approach when studying the performance of our proposed regression estimators via their  $L_p$ norms. More precisely, 
let   $\mathcal{F}$ be a given class of functions $\varphi$\,:$~[0, 1] \rightarrow (0,B],$ for some $B<\infty$.   Fix $\varepsilon>0$ and let the finite collection of functions $\mathcal{F}_{\varepsilon}=
\{\varphi_1, \dots, \varphi_{\mbox{\tiny $N(\varepsilon)$}}\}$,\,  $\varphi_i$\,:$~[0, 1] \to (0,B],$ be an $\varepsilon$-cover of $\mathcal{F}$, i.e., for each $\varphi\in \mathcal{F}$, there is a $\varphi'\in \mathcal{F}_{\varepsilon}$ such that $\lVert \varphi-\varphi'\rVert_{\infty} < \varepsilon$; here, $\parallel\cdot\parallel_{\infty}$ is the usual supnorm. The cardinality of the smallest $\varepsilon$-cover of $\mathcal{F}$  is called the {\it covering number} of the family $\mathcal{F}$ and will be denoted by $\mathcal{N}_{\varepsilon}(\mathcal{F})$. If $\mathcal{N}_{\varepsilon}(\mathcal{F})<\infty$ holds for every $\varepsilon>0$, then the family $\mathcal{F}$ is said to be {\it totally bounded} (with respect to $\parallel\cdot\parallel_{\infty}$).  
The quantity $\log(\mathcal{N}_{\varepsilon}(\mathcal{F}))$ is called Kolmogorov's $\epsilon$-entropy of the set $\mathcal{F}$. For more on this see, for example, the monograph by van der Vaart and Wellner (1996; p. 83). 

\vspace{3mm}\noindent
Now,  let 0\,$<$\,$\varepsilon_n$\,$\downarrow\,$0 be a  decreasing sequence,  as $n\to\infty$, and let $\mathcal{F}_{\varepsilon_n}=\{\varphi_1,\dots, \varphi_{N(\varepsilon_n)}\}$  $\subset \mathcal{F}$ be any $\varepsilon_n$-cover of $\mathcal{F}$;  the choice of $\varepsilon_n$
will be discussed later. Also, as mentioned in the introduction, here we follow Miao et al (2024) and Kim and Yu (2011) and consider the setup where one has access to response values for a small follow-up subsample selected from the set of non-respondents.  More formally, let $\delta_i$, $i=1,\cdots,\ell$, be iid Bernoulli random variables, independent of the data $\mathbb{D}_n$, with the probability of success 
\begin{equation}\label{PL-prob}
p_n = P\{\delta_i = 1\}, ~~ i=1,\cdots,\ell,~~~\mbox{with}~p_n\to 0, ~\mbox{as $n$  $\to\infty$}. 
\end{equation}
Then we select a non-respondent from the validation set $\mathbb{D}_{\ell}$ to be included in the small follow-up subsample only when $(1-\Delta_i)\,\delta_i$\,=\,1, $i\in \boldsymbol{{\cal I}}_\ell$, where $\Delta_i$\,=\,0 if $Y_i$ is missing. 
Next, for each $\varphi\in \mathcal{F}_{\varepsilon_n}$ consider the empirical misclassification of  the classifier $\widehat{\mathcal{T}}_{m,\varphi,h}(\mbox{\scriptsize $\boldsymbol{\chi}$})$ in (\ref{gFin}), based on the validation set $\mathbb{D}_\ell$, given by
\begin{equation}\label{NEW-Lhat}
\widehat{L}_{m,\ell}(\widehat{\mathcal{T}}_{m,\varphi,h}) ~=~ \frac{1}{\ell}\Bigg[
\sum_{i \in \boldsymbol{{\cal I}}_{\ell}}
 \Delta_i \cdot\mbox{\Large $\mathds{1}$}\big\{\widehat{\mathcal{T}}_{m,\varphi,h}(\mbox{\large $\boldsymbol{\chi}$}_i)\neq Y_i\big\}
\,+ 
\sum_{i \in \boldsymbol{{\cal I}}_{\ell}}
\frac{(1-\Delta_i)\delta_i}{p_n} \cdot\mbox{\Large $\mathds{1}$}\hspace{-0.5mm}\big\{\widehat{ \mathcal{T}}_{m,\varphi,h}(\mbox{\large $\boldsymbol{\chi}$}_i)\neq Y_i\big\}
\Bigg],~~
\end{equation}
where, $\mbox{\Large $\mathds{1}$}\{A\}$ is the indicator function of the set $A$.
Our estimator of $\varphi^*$ is then 
\begin{equation}  \label{SK1}
\widehat{\varphi}~=~\argmin_{\varphi\,\in \mathcal{F}_{\varepsilon_n}}\,\widehat{L}_{m,\ell}(\widehat{\mathcal{T}}_{m,\varphi,h})
\end{equation}
and the final proposed classifier is given by
\begin{equation}\label{gFinal}
\widehat{\mathcal{T}}_{n,\widehat{\varphi},h}(\mbox{\scriptsize $\boldsymbol{\chi}$})=\left\{
\begin{array}{ll}
1 ~ & ~ \hbox{if} ~~ \widehat{\mathcal{R}}_{m}(\mbox{\scriptsize $\boldsymbol{\chi}$};\widehat{\varphi},h)  ~ > ~ \frac12 \\
0 ~ & ~ \hbox{otherwise.}
\end{array}
\right. 
\end{equation}
We discuss the theoretical properties of this classifier in the next section.

\subsection{How good is (\ref{gFinal}) as compared to the optimal classifier in  (\ref{Bayes})? Some theoretical results.}\label{main-ii}
To study the performance of the  classifier given by (\ref{gFinal}) and the convergence of its misclassification error to that of the optimal classifier $\mathcal{T}^{\mbox{\tiny opt}}$ given by  (\ref{Bayes}), we first state a number of assumptions. 
In what follows, $\mathbb{X}$ is the space $L^2([a,b])$, for some $-\infty<a<b<\infty$, and $d$ is the metric induced by the usual $L^2([a,b])$ norm.
Furthermore, $\forall \mbox{\scriptsize $\boldsymbol{\chi}$}\in \mathbb{X}$, define $B(\mbox{\scriptsize $\boldsymbol{\chi}$},h) = \big\{\mbox{\scriptsize $\boldsymbol{\chi}$}' \in \mathbb{X}\,\big|\, d(\mbox{\scriptsize $\boldsymbol{\chi}$}', \mbox{\scriptsize $\boldsymbol{\chi}$})<h \big\}$ to be the open ball of functions centered at  $\mbox{\scriptsize $\boldsymbol{\chi}$}$ with $d$-radius equal to $h$.

\vspace{3.5mm}\noindent
{\bf Assumption (A0)} There is a subset 
${\mathcal{S}}_{\mbox{\tiny $\mathbb{X}$}}\subset \mathbb{X}$ 
satisfying $\mathds{P}\big\{\mbox{\large $\boldsymbol{\chi}$}\in {\mathcal{S}}_{\mbox{\tiny $\mathbb{X}$}} \big\} =1$.

\vspace{3mm}\noindent
{\bf Assumption (A1)} 
There exists a function $\phi$   such that  
$\forall \mbox{\scriptsize $\boldsymbol{\chi}$}\in {\mathcal{S}}_{\mbox{\tiny $\mathbb{X}$}}$ and for all $h>0$,
\[
0<C \phi(h) \leq \mathds{P}\big\{\mbox{\large $\boldsymbol{\chi}$}\in B(\mbox{\scriptsize $\boldsymbol{\chi}$},h)\big\}  \leq C' \phi(h) 
\]
for positive constants $C$ and $C'$. 

\vspace{3mm}\noindent
{\bf Assumption (A2)} {\it [Lipschitz conditions on $\psi_k$.]} Let $\psi_k$ be as in (\ref{psieta}). There are constants $\beta_1, \beta_2 >0$ such that $\forall\,\mbox{\scriptsize $\boldsymbol{\chi}$}_1,\,  \mbox{\scriptsize $\boldsymbol{\chi}$}_2\in {\mathcal{S}}_{\mathbb{X}}$ and  $\forall\,\varphi\in\mathcal{F}$

\vspace{-6mm}
\[
~~~~~\big| \psi_k(\mbox{\scriptsize $\boldsymbol{\chi}$}_1; \varphi) -
\psi_k(\mbox{\scriptsize $\boldsymbol{\chi}$}_2; \varphi)\big|  \leq C_k\,d^{^{\beta_k}}(\mbox{\scriptsize $\boldsymbol{\chi}$}_1, \mbox{\scriptsize $\boldsymbol{\chi}$}_2),~\, k=1,2,
\]

\vspace{-1mm}\noindent
where  
$C_1$ and  $C_2$ are positive constants.

\vspace{3mm}\noindent
{\bf Assumption (A3)} The kernel $\mathcal{K}$ is nonnegative, bounded and Lipschitz on its support $[0,1)$, and with $\mathcal{K}(1)=0$  satisfying  $-\infty <C<\mathcal{K}'(t)<C'<\infty$, for all $t\in [0,1)$, for constants $C$ and $C'$.

\vspace{3mm}\noindent
{\bf Assumption (A4)}  
The function $\phi$ in Assumption (A1) is such that $\exists C>0,\, \exists \eta_0>0$ such that $\forall \eta<\eta_0$, $\phi'_1(\eta)<C$. Furthermore,  with $\mathcal{K}(1)=0$,
$\exists C>0,\, \exists \eta_0>0$ such that $\forall 0<\eta<\eta_0$, the function $\phi$ satisfies $\int_0^{\eta} \phi(t)\, dt > C \eta \phi(\eta)$.

\vspace{3mm}\noindent
{\bf Assumption (A5)} {\it[Assumptions on the covering number of ${\mathcal{S}}_{\mbox{\tiny $\mathbb{X}$}}$]} For any $\tau>0$, let $\mathcal{N}_{\tau}({\mathcal{S}}_{\mbox{\tiny $\mathbb{X}$}})$ be the $\tau$-covering number of ${\mathcal{S}}_{\mbox{\tiny $\mathbb{X}$}}$, i.e., the smallest number of open balls of $d$-radius equal to $\tau$ needed to cover ${\mathcal{S}}_{\mbox{\tiny $\mathbb{X}$}}$.  

\vspace{1mm}
\noindent{\bf (i)} Let $\tau_m := \log m/ m$. For $n$ (and thus $m$) large enough,
$
(\log m)^2/\big(m \phi(h)\big) <  \log [\mathcal{N}_{\tau_m}(\mathcal{S}_{\mbox{\tiny $\mathbb{X}$}})] < m \phi(h)/\log m
$. Furthermore,  $mh \sqrt{\phi(h)}\to 0$ as $m\to\infty$.

\vspace{1mm}
\noindent{\bf (ii)} The Kolmogorov's $\tau_m$-entropy of $\mathcal{S}_{\mbox{\tiny $\mathbb{X}$}}$ and 
$\varepsilon_n$-entropy of $\mathcal{F}$
satisfy the summability condition  $\sum_{m=1}^\infty \exp\{(1-\beta) \log \big[\mathcal{N}_{\varepsilon_n}(\mathcal{F}) \vee \mathcal{N}_{\tau_m}(\mathcal{S}_{\mbox{\tiny $\mathbb{X}$}})\big] < \infty$, for some $\beta>1$.

\vspace{3mm}\noindent
{\bf Assumption (A6)} There is a constant $\pi_{\mbox{\tiny min}}>0$ such that $\pi_{\varphi}(\mbox{\scriptsize $\boldsymbol{\chi}$}, y) > \pi_{\mbox{\tiny min}}$,   for all  $\mbox{\scriptsize $\boldsymbol{\chi}$}\in \mathcal{S}_{\mbox{\tiny $\mathbb{X}$}}$ and all $y$, where $\pi_{\varphi}$ is the selection probability in (\ref{NonIgnore}). 

\vspace{3mm}\noindent
{\bf Assumption (A7)}  $\mathds{E}[\Delta\,\varphi(Y)|\mbox{\large $\boldsymbol{\chi}$}=\mbox{\scriptsize $\boldsymbol{\chi}$}] \geq \varrho_0$, for all $\mbox{\scriptsize $\boldsymbol{\chi}$}\in \mathcal{S}_{\mbox{\tiny $\mathbb{X}$}}$ and each $\varphi\in \mathcal{F},$ for some  constant $\varrho_0 >0$. 

\vspace{3mm}\noindent
{\bf Assumption (A8)} The function $\varphi^*$ belongs to a totally bounded class  $\mathcal{F}$  of functions $\varphi: [0, 1] \to (0, B]$, for some $B<\infty$, where $\varphi^*$ is the true $\varphi$ in (\ref{NonIgnore}).

\vspace{3mm}\noindent
Assumption (A0) is not new and has already been considered in the literature; see, for example, Ferraty et al (2013). 
Assumptions (A1)--(A5) are standard in functional kernel regression; see, for example, Ferraty et al (2010). Assumption (A6) is also common in missing data literature (as in Cheng and Chu (1996) or Ferraty et al (2013)); this assumption essentially states that  $Y$ can be observed (i.e., $\Delta$\,=\,1) with a non-zero probability for all value of $(\mbox{\small $\boldsymbol{\chi}$}, y)$. 
Assumption (A7) is rather mild and can be justified by noticing that $\mathds{E}[\Delta\,\varphi(Y)|\mbox{\large $\boldsymbol{\chi}$}] = \mathds{E}[\varphi(Y) \mathds{E}(\Delta\,| \mbox{\large $\boldsymbol{\chi}$}, Y)\,|\mbox{\large $\boldsymbol{\chi}$}] \geq \pi_{\mbox{\tiny min}} \mathds{E}[\varphi(Y)|\mbox{\large $\boldsymbol{\chi}$}]$ and the fact that $\varphi(y)>0$ for all $y$. Assumption (A8) is technical.

\vspace{3.5mm}\noindent
Our first result below establishes the {\it almost complete (a.\,co.)} of the misclassification error of the proposed classifier in (\ref{gFinal}) to that of the theoretically optimal classifier. In passing, we recall   that a sequence of real-valued random variables $Z_n$ is said to converge {\it a.\,co.} to a constant $c$ if for every $t>0$, $\sum_{n\geq 1}P\{|Z_n-c|>t\}<\infty$; see, for example, Ferraty et al (2010).
To  state our first result, define the quantities 
\begin{eqnarray}\label{LnL*}
L_n(\widehat{\mathcal{T}}_{n, \widehat{\varphi},h})\,=\, P\left\{\widehat{\mathcal{T}}_{n,\widehat{\varphi},h }(\mbox{\large $\boldsymbol{\chi}$}) \neq Y\Big|\, \mathbb{D}_n\right\}~~~\mbox{and}~~~~
L^*  \,=\, P\{\mathcal{T}^{\mbox{\tiny opt}}(\mbox{\large $\boldsymbol{\chi}$}) \neq Y\} \,=\, P\{\mathcal{T}_{\varphi^*}(\mbox{\large $\boldsymbol{\chi}$}) \neq Y\},\,
\end{eqnarray}
where $\mathcal{T}_{\varphi^*}$ is as in (\ref{NEW_Bayes}) and $\widehat{\mathcal{T}}_{n,\widehat{\varphi},h}$ is the classifier in  (\ref{gFinal}).

\begin{thm}\label{THM-BBC} 
Let $\widehat{\mathcal{T}}_{n,\widehat{\varphi},h}(\mbox{\scriptsize $\boldsymbol{\chi}$})$ be the classifier given  in (\ref{gFinal}) and suppose that assumptions (A0)--(A8) hold.  Let 
$\varepsilon_n$\,$\downarrow$\,$0$ be any sequence of positive constants satisfying  $(\ell p_n^2)^{-1} \log[\mathcal{N}_{\varepsilon_n}(\mathcal{F})]\to 0$,  
as $n$ (thus $m$ and $\ell$) $\to\infty$. Then one has
	\begin{eqnarray*}
L_n(\widehat{\mathcal{T}}_{n, \widehat{\varphi},h}) - L^*	&=& 
      \mathcal{O}(h^\alpha) 	+ \mathcal{O}(\varepsilon_n)
		+ \mathcal{O}_{a.\,co.}\left( \sqrt{\frac{\log[\mathcal{N}_{\tau_m}(\mathcal{S}_{\mbox{\tiny $\mathbb{X}$}})]}{m\cdot \phi(h)}} \right) 
		+   \mathcal{O}_{a.\,co.}\left(	\sqrt{\frac{\log[\mathcal{N}_{\varepsilon_n}(\mathcal{F})]}{\ell p_n^2}}	\right)  
	\end{eqnarray*}
where $\alpha>0$ is a constant not depending on $n$, $\tau_m=\log(m)/m$,  and  $\phi(h)$ is as in assumption (A1).
\end{thm}

\section{Bandwidth estimation with missing $Y_i$'s}\label{bandw}
\subsection{The proposed methodology}\label{bandw-1}
The classifier $\widehat{\mathcal{T}}_{n,\widehat{\varphi},h}$ that was developed in the previous section (see  (\ref{gFinal}))  depends on the bandwidth $h$, which is only assumed to satisfy  assumption  (A5)(i) in order for this classifier to perform asymptotically well. In practice one would also like to use a suitable estimator $\widehat{h}$ of $h$ and study the performance of the corresponding classifier, $\widehat{\mathcal{T}}_{n,\widehat{\varphi},\widehat{h}}$. 

It turns out that finding ``good'' estimators of $h$ in classification can be challenging. It is well understood in the literature on kernel classification that the optimal bandwidth that minimizes standard quantities such as the MISE or ISE is not necessarily optimal in classification in the sense of minimizing the misclassification error; see for example  Devroye et al (1996; Sec.\,25.9).  In fact, a counter-example is given in Theorem 25.9 of the cited reference, where it is shown that the optimal bandwidth based on the MISE yields rather poor misclassification errors. The problem can be even more challenging for  MNAR data.
In the framework of our paper, this misclassification error is the quantity $L_n(\widehat{\mathcal{T}}_{n, \widehat{\varphi},h})\,=\, P\big\{\widehat{\mathcal{T}}_{n,\widehat{\varphi},h }(\mbox{\large $\boldsymbol{\chi}$}) \neq Y\big|\, \mathbb{D}_n\big\}$.
As argued in the cited reference above,  the optimal bandwidth, denoted by  $h_{\mbox{\tiny opt}}$, is the one that minimizes the error $L_n(\widehat{\mathcal{T}}_{n, \widehat{\varphi},h})$, which is virtually always unknown; see Devroye et al (1996; Sec.\,25.1).
These difficulties are further compounded by the fact that finding a data-dependent bandwidth  $\widehat{h}_{\mbox{\tiny opt}}$ which is in some sense close to $h_{\mbox{\tiny opt}}$ does not necessarily imply the closeness of the corresponding misclassification errors. Since consistency is often the minimum requirement for any classifier, $\widehat{h}_{\mbox{\tiny opt}}$ must be chosen in such a way that  the resulting classifier will be consistent (Devroye et al\,(1996; p.\,424)). To that end, 
%
%
let $H$ be a given set of possible values for $h$. Also, as in the previous section, let $\mathcal{F}_{\varepsilon_n}$ be an $\varepsilon_n$-cover of $\mathcal{F}$ and for each $\varphi\in\mathcal{F}_{\varepsilon_n}$ denote by $\mathcal{C}_m(\varphi)$ the family of classifiers of the form (\ref{gFin}) indexed by $h$, i.e.,
\begin{eqnarray}\label{CMPHI}
\mathcal{C}_m(\varphi) &:=& \left\{
\widehat{\mathcal{T}}_{m,\varphi,h}:~ \mathbb{R}^d \to \{0, 1\} \Big| \, \widehat{\mathcal{T}}_{m,\varphi,h}~\mbox{as in (\ref{gFin})},~~h\in H
\right\}.
\end{eqnarray}
Now let $\widehat{L}_{m,\ell}(\widehat{\mathcal{T}}_{m,\varphi,h})$ be as  in (\ref{NEW-Lhat}) and consider the estimators of $h$ and $\varphi^*$ given by
\begin{equation} \label{phi.h.hat}
(\widehat{\varphi}, \widehat{h}) \,:= \argmin_{\varphi\in \mathcal{F}_{\varepsilon_n},\, h\in H} \widehat{L}_{m,\ell}(\widehat{\mathcal{T}}_{m,\varphi,h})~= 
\argmin_{\varphi\in \mathcal{F}_{\varepsilon_n},~ \widehat{\mathcal{T}}_{m,\varphi,h}\,\in\, \mathcal{C}_m(\varphi)} \widehat{L}_{m,\ell}(\widehat{\mathcal{T}}_{m,\varphi,h}).
\end{equation}
The corresponding classifier is then
\begin{equation}\label{gFinal2}
\widehat{\mathcal{T}}_{n,\widehat{\varphi},\widehat{h}}(\mbox{\scriptsize $\boldsymbol{\chi}$})=\left\{
\begin{array}{ll}
1 ~ & ~ \hbox{if} ~~ \widehat{\mathcal{R}}_{m}(\mbox{\scriptsize $\boldsymbol{\chi}$};\widehat{\varphi},\widehat{h})  ~ > ~ \frac12 \\
0 ~ & ~ \hbox{otherwise,}
\end{array}
\right. 
\end{equation}
where $\widehat{\mathcal{R}}_{m}(\mbox{\scriptsize $\boldsymbol{\chi}$};\varphi,h) $ is as in (\ref{mhat3}). To study the performance of classifier (\ref{gFinal2}) we 
consider two cases. 

\vspace{4mm}\noindent 
{\it Case (i): $H$ is a finite collection.}\\
As suggested by Devroye et al (1996; Ch.\,25), from a practical point of view it is computationally more attractive to confine the set of possible values of $h$ to a finite collection. More specifically,  suppose that $h\in H  := \big\{ h_1, h_2,\cdots h_{\mbox{\tiny N}}\}$, where one may even allow $N\equiv N(n)$ to grow with $n$. How good is the classifier (\ref{gFinal2}) based on this choice of $H$?  To answer this, let
\begin{eqnarray} \label{Lm-Ln}
L_m(\widehat{\mathcal{T}}_{m,\varphi,h})\,=\,P\big\{  \widehat{\mathcal{T}}_{m,\varphi,h}(\mbox{\large $\boldsymbol{\chi}$}) \neq Y\,\big| \mathbb{D}_m \big\}~~~~\mbox{and}~~~~
L_n(\widehat{\mathcal{T}}_{n, \widehat{\varphi},\widehat{h}})\,=\, P\left\{\widehat{\mathcal{T}}_{n,\widehat{\varphi},\widehat{h} }(\mbox{\large $\boldsymbol{\chi}$}) \neq Y\big|\, \mathbb{D}_n\right\}
\end{eqnarray}
be the misclassification error probabilities  of $\widehat{\mathcal{T}}_{m,\varphi,h}$ and 
$\widehat{\mathcal{T}}_{n, \widehat{\varphi},\widehat{h}}$, respectively, and consider the basic decomposition 
\begin{equation}\label{Basic_Decomp}
L_n(\widehat{\mathcal{T}}_{n, \widehat{\varphi},\widehat{h}}) -L^* ~=~ 
\Big[L_n(\widehat{\mathcal{T}}_{n, \widehat{\varphi},\widehat{h}}) - \inf_{\varphi\in\mathcal{F}_{\varepsilon_n}} \inf_{h\in H} 
L_m(\widehat{\mathcal{T}}_{m,\varphi,h})\Big] + \Big[ \inf_{\varphi\in\mathcal{F}_{\varepsilon_n}} \inf_{h\in H} 
L_m(\widehat{\mathcal{T}}_{m,\varphi,h}) - L^*    \Big],
\end{equation}
where, as before,  $L^*$ is the misclassification error of the theoretically optimal classifier in  (\ref{LnL*}). To deal with the first square-bracketed term above, we 
 first establish the following exponential performance bound on the deviations of $L_n(\widehat{\mathcal{T}}_{n, \widehat{\varphi},\widehat{h}})$ from  $\inf_{\varphi\in\mathcal{F}_{\varepsilon}} \inf_{h\in H} L_m(\widehat{\mathcal{T}}_{m,\varphi,h})$.
\begin{thm}\label{THM-H1}
Consider the classifier in (\ref{gFinal2}) based on a finite set $H$ with cardinality $N$. Then, for any distribution of $(\mbox{\large $\boldsymbol{\chi}$}, Y)\in\mathbb{X}\times\{ 0, 1\}$, every $0< \varepsilon_n\downarrow 0$, as $n\to\infty$, and every $t>0$,
\begin{eqnarray}\label{THM-H1-E}
	P\left\{  L_n(\widehat{\mathcal{T}}_{n, \widehat{\varphi},\widehat{h}}) - \inf_{\varphi\in\mathcal{F}_{\varepsilon_n}} \inf_{h\in H} 
L_m(\widehat{\mathcal{T}}_{m,\varphi,h}) 	\,>\,  t \right\}
	&\leq& 2N\cdot
	\mathcal{N}_{\varepsilon_n}(\mathcal{F})
	\cdot\exp\left\{-c\, p_n^2 \ell\, t^2\right\}
	\end{eqnarray}
	where $c$ is a positive constant not depending on $n$.
\end{thm}
Now, observe that if $\ell p_n^2$\,$\to$\,$\infty$ fast enough (for example, $\ell p_n^2 / \log n\to\infty$, as $n\to\infty$), then Theorem \ref{THM-H1} in conjunction with the Borel-Cantelli lemma yields $L_n(\widehat{\mathcal{T}}_{n, \widehat{\varphi},\widehat{h}}) - \inf_{\varphi\in\mathcal{F}_{\varepsilon}} \inf_{h\in H} 
L_m(\widehat{\mathcal{T}}_{m,\varphi,h})\to^{a.s.}$\,$0$. Unfortunately, this falls short of achieving strong consistency of 
$\mathcal{T}_{n, \widehat{\varphi},\widehat{h}}$ (for the the optimal classifier $\mathcal{T}_{\mbox{\tiny B}}$ in (\ref{LnL*})). This is because although the first square-bracketed term in (\ref{Basic_Decomp}) can converge to zero, there is no guarantee that the second square-bracketed term  can be made small enough unless $H$ is an appropriate infinite collection of  $h$ values; this is considered next.

\vspace{3mm}\noindent 
{\it Case (ii): $H$ is an infinite collection.}\\ 
Once again let $\widehat{\mathcal{T}}_{n,\widehat{\varphi},\widehat{h}}$ be the classifier in (\ref{gFinal2}), where $(\widehat{\varphi}, \widehat{h})$ is as in (\ref{phi.h.hat}), but with $H$ being an infinite class of possible values of $h$ such as $H$\,=\,$(0, \infty)$ or $H$\,=\,$(0, A]$ for some arbitrary  $A>0$. To proceed, 
we need to define the notion of {\it shatter coefficient} of a set. Let $\mathcal{A}$ be a class of sets $A\subset \mathbb{X}$. Then the $n^{\mbox{\tiny th}}$ shatter coefficient of $\mathcal{A}$ is the combinatorial quantity 
\[ \mathcal{S}(\mathcal{A},n)=\max_{\mbox{\small $\boldsymbol{\chi}$}_1,...,\mbox{\small $\boldsymbol{\chi}$}_n\in\, \mathbb{X}} \{ \text{number of different sets in } \{ \{               \mbox{\small $\boldsymbol{\chi}$}_1,...,\mbox{\small $\boldsymbol{\chi}$}_n \} \cap A \vert A\in \mathcal{A} \} \}.
\]

\noindent 
Here,  $\mathcal{S}(\mathcal{A},n)$  measures the richness of the class $\mathcal{A}$. Next let $\mathcal{G}$ be a class of functions of the form $g:\, \mathbb{X} \to \{0,1\}$. Also let 
 $\mathcal{A}_{\mathcal{G}}$ be the class of sets of the form 
\begin{equation}
\label{shatter sets}
A=\{ \{ \mbox{\small $\boldsymbol{\chi}$}\in\, \mathbb{X}\,\big|\, g(\mbox{\small $\boldsymbol{\chi}$})=1\} \times \{  0\} \} \cup \{ \{ \mbox{\small $\boldsymbol{\chi}$}\in\mathbb{X}\,\big|\, g(\mbox{\small $\boldsymbol{\chi}$})=0\} \times \{  1\} \}, \hspace{10mm} g\in\mathcal{G}
\end{equation}
and define the $n^{\mbox{\tiny th}}$ shatter coefficient of the class $\mathcal{G}$ to be 
$\mathcal{S}(\mathcal{G},n)=\mathcal{S}(\mathcal{A}_{\mathcal{G}},n).
$
Clearly the cardinality of $\mathcal{S}(\mathcal{A}_{\mathcal{G}},n)$ depends on the class $\mathcal{G}$. 
%
Now, with $\mathcal{C}_m(\varphi)$ as in (\ref{CMPHI}),
for each $\varphi\in \mathcal{F}_{\varepsilon_n}$ let $\mathcal{A}_{_{\mathcal{C}_m(\varphi)}}$ be the collection 
of all sets of the form
\begin{equation}\label{AMPHI}
A_{m, \varphi, h}=\left\{ \{ \mbox{\small $\boldsymbol{\chi}$}\in\mathbb{X}\,\big|\, \widehat{\mathcal{T}}_{m,\varphi,h}(\mbox{\small $\boldsymbol{\chi}$})=1\} \times \{  0\} \right\} \cup \left\{ \{ \mbox{\small $\boldsymbol{\chi}$}\in\mathbb{X}\,\big|\, \widehat{\mathcal{T}}_{m,\varphi,h}(\mbox{\small $\boldsymbol{\chi}$})=0\} \times \{  1\} \right\}, \hspace{3mm} \widehat{\mathcal{T}}_{m,\varphi,h}\in\mathcal{C}_m(\varphi),
\end{equation}
and observe that 
\begin{equation} \label{SALCm}
\mathcal{S}(\mathcal{C}_m(\varphi),\, \ell)\,=\,\mathcal{S}(\mathcal{A}_{_{\mathcal{C}_m(\varphi)}},\, \ell).
\end{equation}
In passing we also note that in view of (\ref{AMPHI}), for any $\widehat{\mathcal{T}}_{m,\varphi,h}\in\mathcal{C}_m(\varphi),$ we have
\begin{equation} \label{PROB2}
P\left\{  \widehat{\mathcal{T}}_{m,\varphi,h}(\mbox{\large $\boldsymbol{\chi}$}) \neq Y\,\Big|\, \mathbb{D}_m  \right\} \, =\,  P\big\{ (\mbox{\large $\boldsymbol{\chi}$}, Y) \in A_{m,\varphi,h} \,\big| \mathbb{D}_m     \big\}.
\end{equation}
Then we have the following counterpart of Theorem \ref{THM-BBC} for the classifier $\widehat{\mathcal{T}}_{n,\widehat{\varphi},\widehat{h}}$ that uses the bandwidth estimator $\widehat{h}$ instead of $h$.
\begin{thm}\label{THM-BBC-2} 
Let $\widehat{\mathcal{T}}_{n,\widehat{\varphi},\widehat{h}}$ be the classifier in  (\ref{gFinal2}) and suppose that assumptions (A0) -- (A8) hold. If 
$\sum_{n=1}^{\infty} \big\{\mathcal{N}_{\varepsilon_n}(\mathcal{F})\cdot\sup_{\varphi\in \mathcal{F}_{\varepsilon_n}} E\left[
\mathcal{S}\big( 
\mathcal{C}_m(\varphi),\, \ell\big)\right]\big\}^{-b} < \infty
$
holds for some $b>0$ then
\begin{eqnarray}
L_n(\widehat{\mathcal{T}}_{n, \widehat{\varphi},\widehat{h}}) - L^* &=&
\mathcal{O}(h^{\alpha}) 	+ \mathcal{O}(\varepsilon_n)
		+ \mathcal{O}_{a.\,co.}\left( \sqrt{\frac{\log[\mathcal{N}_{\tau_m}(\mathcal{S}_{\mbox{\tiny $\mathbb{X}$}})]}{m\cdot \phi(h)}} \right) \nonumber\\
&&~~~		+ \mathcal{O}_{a.co.}\left(
\sqrt{\frac{\log(  \mathcal{N}_{\varepsilon_n}(\mathcal{F})   ) + 
\log\left( \sup_{\varphi\in \mathcal{F}_{\varepsilon_n}}E\left[\mathcal{S}\big( 
\mathcal{C}_m(\varphi),\, \ell\big)\right]  \right) }{\ell p_n^2}}\right),	\label{BND10-B}	
\end{eqnarray}
		where $L_n(\widehat{\mathcal{T}}_{n, \widehat{\varphi},\widehat{h}})$ is the misclassification error in (\ref{Lm-Ln}).
\end{thm}
Unfortunately the presence of the term $E\big[ \mathcal{S}(\mathcal{C}_m(\varphi),\, \ell) \big]$ in the above bound is  unpleasant because in most situations the term $\mathcal{S}(\mathcal{C}_m(\varphi),\, \ell)$ can be difficult to compute, in which case an upper bound on the shatter coefficient may be convenient. Fortunately, one can bound   $\mathcal{S}(\mathcal{C}_m(\varphi),\, \ell)$ in (\ref{BND10-B}) for particular types of kernels. The following section looks into the popular class of compactly supported kernels.


\subsection{Special case: compact support kernels}\label{bandw-2}
Here we consider kernels of the form 
\begin{equation}\label{comker}
\mathcal{K}(t) = c\,(1-t^2)\cdot\mbox{\large $\mathds{1}$}\{|t|\leq 1\},~~c>0,
\end{equation}
which belongs to the class of compact support kernels of the form 
\[
\mathcal{K}({\bf t}) = \Big(\sum_{i=1}^k a_i \lVert {\bf t} \rVert^{b_i}  \Big)
\cdot \mbox{\large $\mathds{1}$}\{\lVert {\bf t} \rVert\leq 1\},~~~\mbox{with}~~a_i\in\mathbb{R},~ a_k\neq 0, ~~0\leq b_1< b_2<\cdots<b_k<\infty,
\]
see Devroye et al (1996; Sec.\,25).
\begin{thm}\label{THM-RATE2} 
Consider the classifier $\widehat{\mathcal{T}}_{n,\widehat{\varphi},\widehat{h}}$ in  (\ref{gFinal2}) based on the  compact support kernel (\ref{comker}) and suppose that assumptions (A0) -- (A8) hold. Let 
$0<\varepsilon_n$\,$\downarrow$\,$0$ be any sequence of constants satisfying  $(\ell p_n^2)^{-1} \log[\ell\vee m\vee\mathcal{N}_{\varepsilon_n}(\mathcal{F})]\to 0$,  
as $n$ (thus $m$ and $\ell$) $\to\infty$. Then
	\begin{equation*}  
L_n(\widehat{\mathcal{T}}_{n, \widehat{\varphi},\widehat{h}}) - L^* =
\mathcal{O}(h^{\alpha}) 	+ \mathcal{O}(\varepsilon_n)
		+ \mathcal{O}_{a.\,co.}\left( \sqrt{\frac{\log[\mathcal{N}_{\tau_m}(\mathcal{S}_{\mbox{\tiny $\mathbb{X}$}})]}{m\cdot \phi(h)}} \right) 
+	\mathcal{O}_{a.co.}\left(
\sqrt{\frac{\log\big[  \mathcal{N}_{\varepsilon_n}(\mathcal{F})  \vee 
(m^2\ell)\big] }{\ell p_n^2}}\right).
	\end{equation*} 
\end{thm}

\section{Numerical studies}\label{examples}
In this section we present  some numerical studies to assess the performance of the  proposed classifiers under different settings. More specifically, we consider the two-group classification problem where one must predict the class variable, $Y$\,=\,1 or 0,  based on the  functional covariate $\mbox{\large $\boldsymbol{\chi}$}$, where   
\begin{eqnarray*} \label{Normal1}
\mbox{Class 1 ($Y$\,=\,1):}\hspace{-3mm}&&\mbox{\large $\boldsymbol{\chi}$}(t) = (t - 0.5)^2 A + B,~ \mbox{where~} t\in[0,1],~ A\sim N(5,\, 2^2),\, \mbox{and~} B\sim N(1,\, 0.5)
\\[2pt]
\mbox{Class 0 ($Y$\,=\,0):}\hspace{-3mm}&&\mbox{\large $\boldsymbol{\chi}$}(t) = (t - 0.5)^2 C + D,~ \mbox{where~} t\in[0,1],~ C\sim \mbox{Unif}(0,\, 4),\, \mbox{and~} D\sim \mbox{Unif}(0,\, 2.1)
\end{eqnarray*}
For the simulations, each  discretized curve was generated from 500 equi-spaced  points $t$\,$\in$\,$[0,1]$.  
A sample of 20 of these curves is provided in Figure \ref{Figure-1}. 
\begin{figure}[h]
\centering
\includegraphics[width=11.0cm, height=5.5cm]{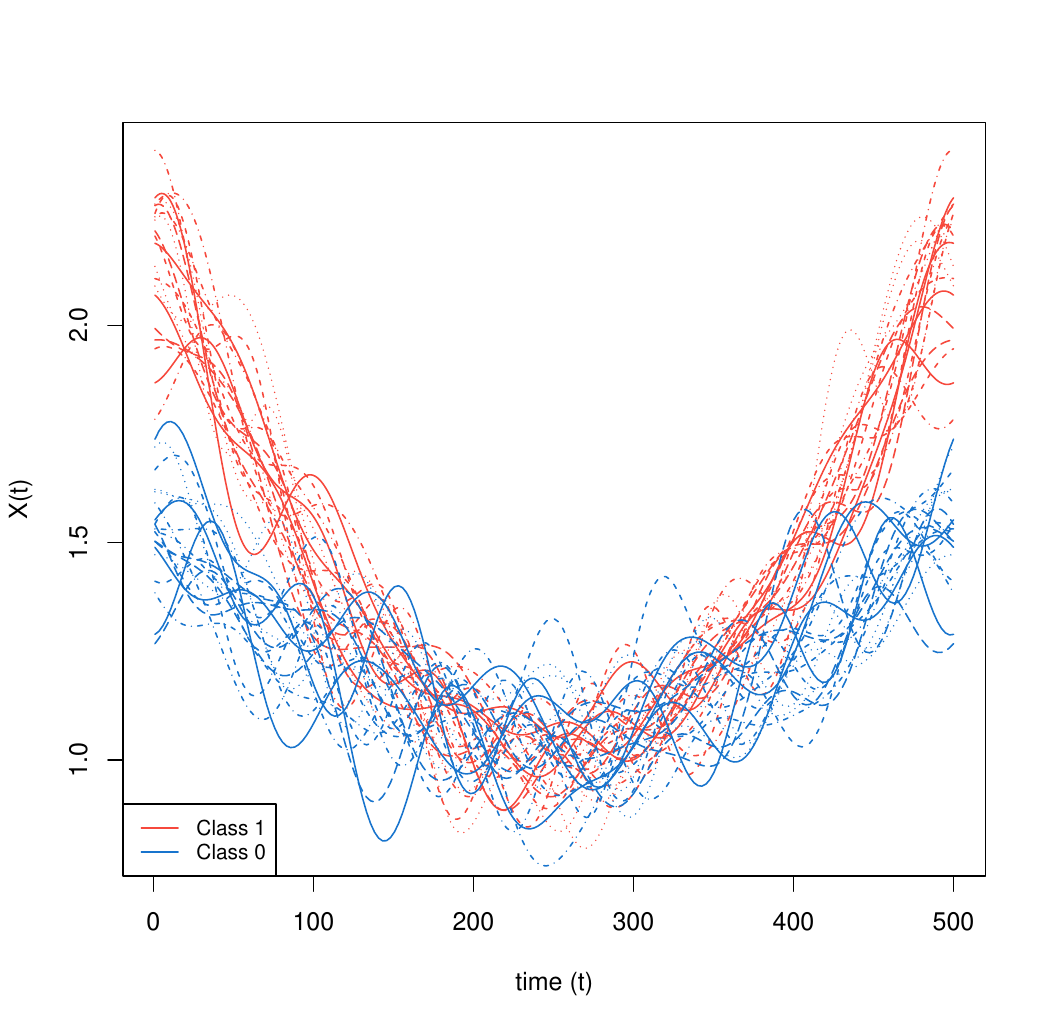}
\caption{\fontsize{10}{6}\selectfont A sample of 20 covariate curves from each of the two populations.}
\label{Figure-1}
\end{figure}
The unconditional class probabilities are $P\{Y=1\}=P\{Y=0\}=1/2$. Here we consider three sample sizes $n$\,= 50, 100, and 200.  
Regarding the choice of the functions $g$ and $\varphi^*$  in the selection probability  (\ref{NonIgnore}), we consider
\begin{equation}\label{g.zeta}
g(\mbox{\small $\boldsymbol{\chi}$}) = \gamma_0 + \gamma_1 \log\Big(\int_{0}^1 \mbox{\small $\boldsymbol{\chi}$}^2(t)\, dt \Big),
\end{equation}
along with two different choices of $\varphi^*$: 
\begin{equation}\label{phiz}
\mbox{\bf Model (i):}~~ \varphi_1^*(y) = \exp\{\gamma y\}~~~~~\mbox{and}~~~~~ \mbox{\bf Model (ii):}~~\varphi_2^*(y) = \exp\{\exp\{\big(y\sqrt{\gamma}\big)\}\},
\end{equation}
where the parameter values $(\gamma_0, \gamma_1, \gamma)$ used in (\ref{g.zeta}) and (\ref{phiz}) are $(0.03, 0.02, -0.8)$ for model (i) and $(-0.21, -1.2, 0.2)$ for model (ii). These values are chosen to produce approximately 40\% missing values under each model. 
For each sample size $n$, we constructed the proposed classifier in (\ref{gFinal2}), 
which will be denoted by $\widehat{\mathcal{T}}_{\hspace{-0.2mm}\mbox{\tiny $\widehat{\varphi},\widehat{h}$}}$ here.  Additionally, we considered two imputation-based kernel classifiers that work by first imputing for the missing values. The first one uses {\it regression imputation} which is denoted by $\widehat{\mathcal{T}}^{\mbox{\tiny Imp1}}_{\hspace{-0.2mm}\mbox{\tiny $\widehat{h}$}}$ here, and the second one is based on {\it mean imputation} which is denoted by $\widehat{\mathcal{T}}^{\mbox{\tiny Imp2}}_{\hspace{-0.2mm}\mbox{\tiny $\widehat{h}$}}$. We also included the complete-case classifier, denoted by $\widehat{\mbox{\Large $\tau$}}^{cc}_{\hspace{-0.8mm}\mbox{\tiny $\widehat{h}$}}$, where $\widehat{\mbox{\Large $\tau$}}^{cc}_{\hspace{-0.8mm}\mbox{\tiny $\widehat{h}$}}({\bf x})$\,=\,1 if\, $\mathcal{R}^{\mbox{\tiny cc}}_n({\bf x}; \widehat{h}) > 1/2$ (and $\widehat{\mbox{\Large $\tau$}}^{cc}_{\hspace{-0.8mm}\mbox{\tiny $\widehat{h}$}}({\bf x})$\,=\,0 otherwise), where $\mathcal{R}^{\mbox{\tiny cc}}_n({\bf x}; h)$ is as in (\ref{mcc1}). As a point of reference, we also 
considered the classifier based on all of the data (i.e., when there are no missing values)  to see how different the results would have been with no missing data; this classifier is denoted by 
$\widehat{\mathcal{T}}^{\mbox{\tiny all}}_{\hspace{-0.2mm}\mbox{\tiny $\widehat{h}$}}$ here. To carry out the estimations, we employed  the R package  ``fda.usc'' developed by Febrero-Bande and Oviedo de la Fuente (2012), where we used the Epanechnikov-type kernel $\mathcal{K}(s)= \frac{3}{2} (1-s^2) \mathds{1}_{\{0\leq s \leq 1\}}$, however, as in general nonparametric estimation, the shape of the kernel is of little importance here. In all these classifiers, the bandwidth estimator $\widehat{h}$ is constructed using the method discussed in section (\ref{bandw-1}). Similarly, the estimation of $\varphi_1^*$ and $\varphi_2^*$ in (\ref{phiz}) was done using the data-splitting approach discussed in section (\ref{Main-1-1}). As for the value of $p_n$ in (\ref{PL-prob}), it is taken to be 
 $\big(\log(n) \cdot n^{0.35} \big)^{-1}$ which results in small average follow-up subsample sizes (see Table  \ref{Table-1}). To study the performance of the proposed classifier numerically, the misclassification error of each classifier is estimated using an independent validation sample of 1000 observations generated in
 the same way as the original data (with 500 from each class). The entire above process is repeated a total of 400 times  and the average misclassification errors are  summarized in Table \ref{Table-1}. The numbers in parentheses are the standard errors and the boldfaced values in square brackets are the average follow-up subsample sizes drawn from the set of non-respondents in $\mathbb{D}_{\ell}$. 
{
	\renewcommand{\arraystretch}{0.99} 
\begin{table}[H]
\caption{\fontsize{10}{8}\selectfont  This table presents average misclassification errors (averaged over 400 simulation runs) at three different sample sizes for the proposed classifier in 	(\ref{gFinal2}), $\widehat{\mbox{\Large $\tau$}}_{\hspace{-0.8mm}\mbox{\tiny $\widehat{\varphi},\widehat{h}$}}$, as well as those of the complete-case  classifier, $\widehat{\mbox{\Large $\tau$}}^{cc}_{\hspace{-0.8mm}\mbox{\tiny $\widehat{h}$}}$, the regression-imputation-based classifier, $\widehat{\mbox{\Large $\tau$}}^{\mbox{\tiny Imp1}}_{\hspace{-0.8mm}\mbox{\tiny $\widehat{h}$}}$,  mean-imputation-based classifier, $\widehat{\mbox{\Large $\tau$}}^{\mbox{\tiny Imp2}}_{\hspace{-0.8mm}\mbox{\tiny $\widehat{h}$}}$, and the  classifier based on no missing data, $\widehat{\mbox{\Large $\tau$}}^{\mbox{\tiny All}}_{\hspace{-0.8mm}\mbox{\tiny $\widehat{h}$}}$.  Here, the missing rate is 40\%. The numbers in parentheses are the standard errors and those in square brackets are the average follow-up subsample sizes drawn from the set of non-respondents in $\mathbb{D}_{\ell}$.} 
\begin{center}
\begin{tabular}{|cc|ccccc|}\label{Table-1}
& Classifier & $\widehat{\mbox{\Large $\tau$}}_{\hspace{-0.8mm}\mbox{\tiny $\widehat{\varphi},\widehat{h}$}}$ 
& $\widehat{\mbox{\Large $\tau$}}^{cc}_{\hspace{-0.8mm}\mbox{\tiny $\widehat{h}$}}$   
& $\widehat{\mbox{\Large $\tau$}}^{\mbox{\tiny Imp1}}_{\hspace{-0.8mm}\mbox{\tiny $\widehat{h}$}}$ 
& $\widehat{\mbox{\Large $\tau$}}^{\mbox{\tiny Imp2}}_{\hspace{-0.8mm}\mbox{\tiny $\widehat{h}$}}$ 
& $\widehat{\mbox{\Large $\tau$}}^{\mbox{\tiny All}}_{\hspace{-0.8mm}\mbox{\tiny $\widehat{h}$}}$   \\ 
\hline\hline
Model (i) & $n=50$  &  0.0532  & 0.0802 & 0.0794 & 0.1302  & 0.0243  \\ 
 & &  (0.0027) & (0.0046) & (0.0051) & (0.0071) & (0.0016) \\
  & &{\bf [0.41]} &  &  &  &  \\
\hline
 & $n=100$ & 0.0244 & 0.0521 & 0.0421 & 0.1532 & 0.0197 \\
  & & (0.0012) & (0.0035) & (0.0033) & (0.0073) & (0.0012) \\
 & & {\bf [0.52]} &  & & & \\
\hline
 & $n=200$    & 0.0178 & 0.0241 & 0.0225 & 0.1493 & 0.0100  \\
  & & (0.0014) & (0.0013) & (0.0025) & (0.0080) & (0.0007) \\
  & & {\bf [0.87]} & & & & \\
\hline\hline
Model (ii) & $n=50$  & 0.0494 & 0.0690 & 0.1835 & 0.2378 & 0.0245 \\
 & & (0.0018) & (0.0033) & (0.0071) & (0.0070) & (0.0015) \\
 & & {\bf [0.33]} & & & & \\
\hline
 & $n=100$ & 0.0286 & 0.0324 & 0.1568 & 0.2776 & 0.0148 \\
 & & (0.0012) & (0.0017) & (0.0074) & (0.0073) & (0.0009) \\
  & & {\bf [0.45]} & & & & \\
\hline
 & $n=200$ & 0.0164 & 0.0188 & 0.1105& 0.2273 & 0.0058 \\
& & (0.0008) & (0.0011) & (0.0070) & (0.0078) & (0.0004) \\    
 & & {\bf [0.83]}  & & & & \\ \hline
\end{tabular}
\end{center}
\end{table}
}

\begin{figure}[H]
\centering
\includegraphics[width=16cm, height=6.2cm]{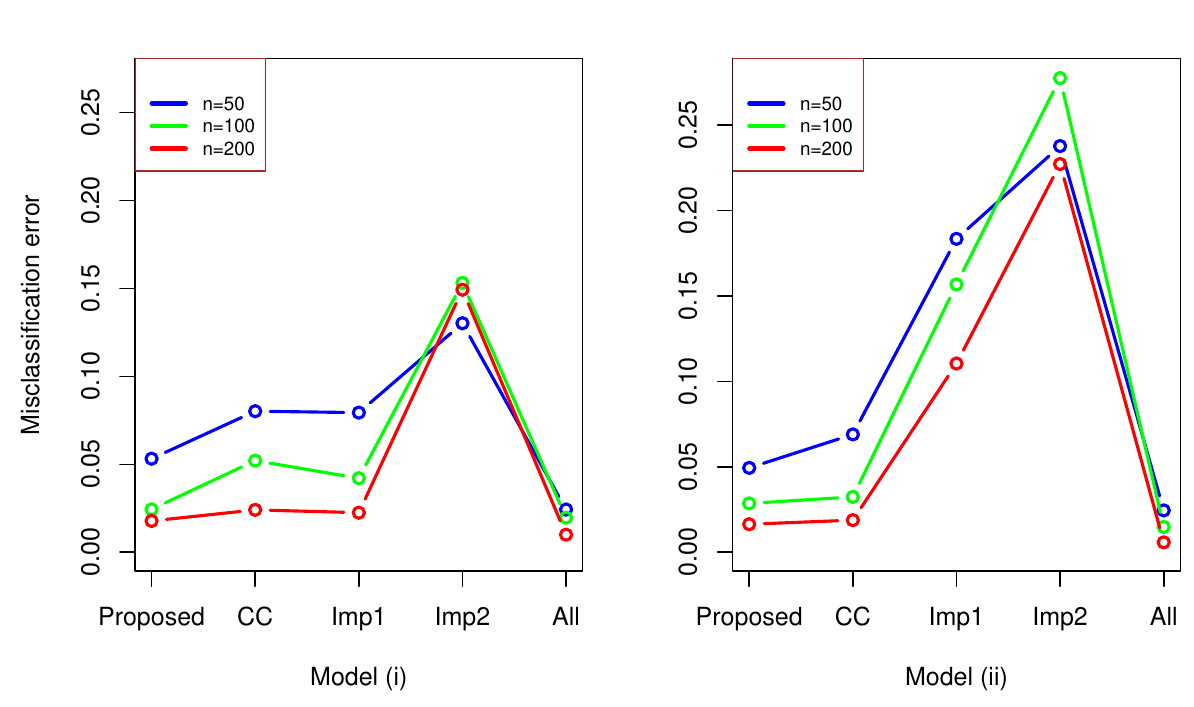}
\caption{\fontsize{10}{6}\selectfont Side-by-side boxplots of misclassification errors of the five classifiers for different values of $n$ under both models (i) and (ii).}
\label{Figure-2}
\end{figure}

\vspace{-4mm}
Table \ref{Table-1} shows that the proposed classifier $\widehat{\mbox{\Large $\tau$}}_{\hspace{-0.8mm}\mbox{\tiny $\widehat{\varphi},\widehat{h}$}}$ in column 1  has the ability to perform well; in fact, it performs better than the classifiers in columns 2, 3, and 4. The table also shows that the average follow-up subsample sizes (boldfaced numbers in the table) are quite small, ranging from an average of only 0.33 to no more than 0.87. This indicates that the seemingly undesirable need for a follow-up subsample can be a non-issue in practice.  Figures \ref{Figure-1} and \ref{Figure-2} provide the side-by-side boxplots and error curves of various classifiers for both models at the three different sample sizes.

\noindent
\begin{figure}[H]
\centering
\includegraphics[width=15cm, height=12.5cm]{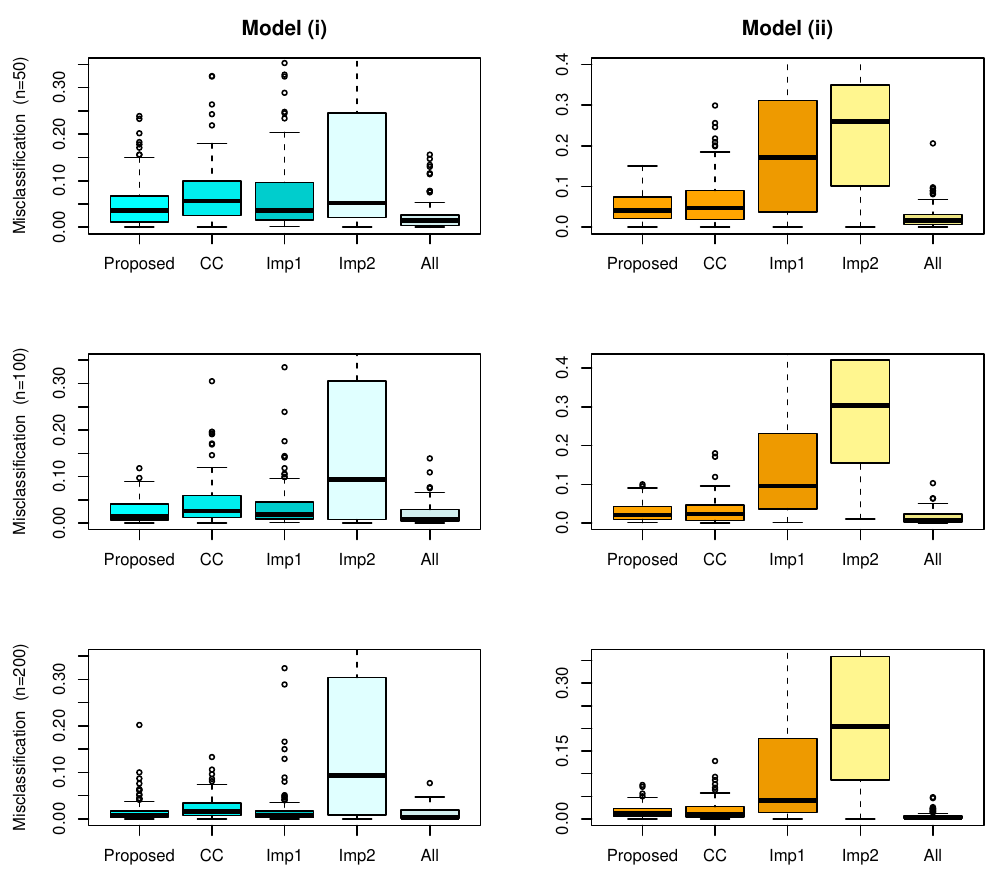}
\caption{\fontsize{10}{6}\selectfont Side-by-side boxplots of misclassification errors of the five classifiers for different values of $n$.}
\label{Figure-1}
\end{figure} 

\vspace{1mm} 
\noindent

\vspace{-2mm}
\section{Proofs of the main results} \label{PRF}
We start by stating a number of lemmas.                    
\begin{lem}\label{LEM-00}
The regression function $\mathcal{R}(\mbox{\scriptsize $\boldsymbol{\chi}$})= E[Y|\mbox{\large $\boldsymbol{\chi}$}= \mbox{\scriptsize $\boldsymbol{\chi}$}]$ can also be represented as
\begin{equation}
\mathcal{R}(\mbox{\scriptsize $\boldsymbol{\chi}$}) \,=\,\mathcal{R}(\mbox{\scriptsize $\boldsymbol{\chi}$};\varphi^*) \,:=~ \eta_1(\mbox{\scriptsize $\boldsymbol{\chi}$}) + \frac{\psi_1(\mbox{\scriptsize $\boldsymbol{\chi}$};\varphi^*)}{\psi_2(\mbox{\scriptsize $\boldsymbol{\chi}$};\varphi^*)}\cdot \left(1- \eta_2(\mbox{\scriptsize $\boldsymbol{\chi}$})\right).
\end{equation}
where   the functions $\psi_k$ and $\eta_k$, $k=1,2,$ are given by (\ref{psieta}) and  $\varphi^*$ is as in (\ref{NonIgnore}). 
\end{lem}

\vspace{3.5mm}\noindent
PROOF OF LEMMA \ref{LEM-00}.

\vspace{0.5mm}\noindent
Let $\pi_{\varphi^*}(\mbox{\scriptsize $\boldsymbol{\chi}$},y)$ be as in (\ref{NonIgnore}) and observe that
\begin{equation}\label{1-PI}
1-\pi_{\varphi^*}(\mbox{\large $\boldsymbol{\chi}$},Y) =\frac{\exp\{g(\mbox{\large $\boldsymbol{\chi}$})\}\,\varphi^*(Y)}{1+\exp\{g(\mbox{\large $\boldsymbol{\chi}$})\}\,\varphi^*(Y)}= \exp\{g(\mbox{\large $\boldsymbol{\chi}$})\}\,\varphi^*(Y)\,\pi_{\varphi^*}(\mbox{\large $\boldsymbol{\chi}$},Y).
\end{equation}
Now, writing $m(\mbox{\scriptsize $\boldsymbol{\chi}$})=E[Y|\mbox{\large $\boldsymbol{\chi}$}=\mbox{\scriptsize $\boldsymbol{\chi}$}]=E[Y\Delta|\mbox{\large $\boldsymbol{\chi}$}=\mbox{\scriptsize $\boldsymbol{\chi}$}] + \frac{E[Y(1-\Delta)|\mbox{\large $\boldsymbol{\chi}$}=\mbox{\scriptsize $\boldsymbol{\chi}$}]}{E[1-\Delta|\mbox{\large $\boldsymbol{\chi}$}=\mbox{\scriptsize $\boldsymbol{\chi}$}]}\cdot E[1-\Delta|\mbox{\large $\boldsymbol{\chi}$}=\mbox{\scriptsize $\boldsymbol{\chi}$}],$ one finds
\begin{eqnarray*}
\frac{E[Y(1-\Delta)|\mbox{\large $\boldsymbol{\chi}$}]}{E[1-\Delta|\mbox{\large $\boldsymbol{\chi}$}]}&=&
\frac{E\big[E\big\{Y(1-\Delta)|\mbox{\large $\boldsymbol{\chi}$},Y\big\}\big|\mbox{\large $\boldsymbol{\chi}$}\big]}{E\big[E\big\{1-\Delta|\mbox{\large $\boldsymbol{\chi}$},Y\big\}\big|\mbox{\large $\boldsymbol{\chi}$}\big]}=\frac{E\big[Y\big\{1-\pi_{\varphi^*}(\mbox{\large $\boldsymbol{\chi}$},Y)\big\}\big|\mbox{\large $\boldsymbol{\chi}$}\big]}{E\big[1-\pi_{\varphi^*}(\mbox{\large $\boldsymbol{\chi}$},Y)\big|\mbox{\large $\boldsymbol{\chi}$}\big]}\\
&\stackrel{\mbox{\tiny by (\ref{1-PI})}}{=}& \frac{E\big[Y\exp\{g(\mbox{\large $\boldsymbol{\chi}$})\}\,\varphi^*(Y)\,\pi_{\varphi^*}(\mbox{\large $\boldsymbol{\chi}$},Y)\big|\mbox{\large $\boldsymbol{\chi}$}\big]}{E\big[\exp\{g(\mbox{\large $\boldsymbol{\chi}$})\}\,\varphi^*(Y)\,\pi_{\varphi^*}(\mbox{\large $\boldsymbol{\chi}$},Y)\big|\mbox{\large $\boldsymbol{\chi}$}\big]}
= \frac{E\big[Y\varphi^*(Y)\,\Delta\big|\mbox{\large $\boldsymbol{\chi}$}\big]}{E\big[\varphi^*(Y)\,\Delta\big|\mbox{\large $\boldsymbol{\chi}$}\big]}=\frac{\psi_1(\mbox{\large $\boldsymbol{\chi}$};\varphi^*)}{\psi_2(\mbox{\large $\boldsymbol{\chi}$};\varphi^*)}\,.
\end{eqnarray*}
The proof of the lemma now follows from this and the definitions of $\psi_k$ and $\eta_k$, $k=1,2$, \,in (\ref{psieta}).

\hfill$\Box$

\begin{lem} \label{BASIC}
For any $\varphi\in \mathcal{F}$, let $\widehat{\mathcal{T}}_{m,\varphi,h}$ be as in (\ref{gFin}). Also, let $\mathcal{T}_{\varphi^*}$ be the optimal classifier given in (\ref{NEW_Bayes}). Then
\begin{eqnarray*}
L_m(\widehat{\mathcal{T}}_{m,\varphi,h}) - L^* &\leq& 2\,E\left[ \left|  
\widehat{\mathcal{R}}_m(\mbox{\large $\boldsymbol{\chi}$}; \varphi,h) - \mathcal{R}(\mbox{\large $\boldsymbol{\chi}$}; \varphi^*)  
\right| \Big| \mathbb{D}_m\right]\,,
\end{eqnarray*}
where $L_m(\widehat{\mathcal{T}}_{m,\varphi,h})=P\big\{  \widehat{\mathcal{T}}_{m,\varphi,h}(\mbox{\large $\boldsymbol{\chi}$}) \neq Y\,\big| \mathbb{D}_m \big\}$ and $L^*= L(\mathcal{T}_{\varphi^*}) = P\{\mathcal{T}_{\varphi^*}(\mbox{\large $\boldsymbol{\chi}$}) \neq Y\}.$
\end{lem} 

\vspace{3mm}\noindent
PROOF OF LEMMA \ref{BASIC}

\vspace{0.5mm}\noindent
The proof is based on standard arguments similar to those of Devroye et al (1996; Theorem 2.3 and Corollary 6.1); also see Abraham et al (2006) and C\'erou and Guyader (2006). First note that 
\begin{eqnarray*}
L_m(\widehat{\mathcal{T}}_{m,\varphi,h}) &=& 1- P\left\{  
\widehat{\mathcal{T}}_{m,\varphi,h}(\mbox{\large $\boldsymbol{\chi}$}) = Y\,\big| \mathbb{D}_m \right\} \\
&=& 1- \left[P\left\{  Y=1, \widehat{\mathcal{T}}_{m,\varphi,h}(\mbox{\large $\boldsymbol{\chi}$}) = 1\,\big| \mathbb{D}_m \right\}  +  P\left\{  Y=0, \widehat{\mathcal{T}}_{m,\varphi,h}(\mbox{\large $\boldsymbol{\chi}$}) = 0\,\big| \mathbb{D}_m \right\}\right]\\
&=&      1- \bigg[ E\left( E\left\{\mbox{\Large $\mathds{1}$}\hspace{-0.5mm}\{Y=1\}\cdot \mbox{\Large $\mathds{1}$}\hspace{-0.5mm}\big\{ \widehat{\mathcal{T}}_{m,\varphi,h}(\mbox{\large $\boldsymbol{\chi}$})=1   \big\}\, \Big| \mathbb{D}_m , \mbox{\large $\boldsymbol{\chi}$} \right\}\,\Big| \mathbb{D}_m \right)\\ 
&&~~~~~~~~~~~ + E\left( E\left\{\mbox{\Large $\mathds{1}$}\hspace{-0.5mm}\{Y=0\}\cdot \mbox{\Large $\mathds{1}$}\hspace{-0.5mm}\big\{ \widehat{\mathcal{T}}_{m,\varphi,h}(\mbox{\large $\boldsymbol{\chi}$})=0   \big\}\, \Big| \mathbb{D}_m , \mbox{\large $\boldsymbol{\chi}$} \right\}\,\Big| \mathbb{D}_m \right) \bigg]\\
&=&      1- E\bigg[ \mbox{\Large $\mathds{1}$}\hspace{-0.5mm}\big\{ \widehat{\mathcal{T}}_{m,\varphi,h}(\mbox{\large $\boldsymbol{\chi}$})=1   \big\}\cdot \mathcal{R}(\mbox{\large $\boldsymbol{\chi}$}; \varphi^*)
+ \mbox{\Large $\mathds{1}$}\hspace{-0.5mm}\big\{ \widehat{\mathcal{T}}_{m,\varphi,h}(\mbox{\large $\boldsymbol{\chi}$})=0   \big\}\cdot \big(1- \mathcal{R}(\mbox{\large $\boldsymbol{\chi}$}; \varphi^*)\big)
\, \Big|\, \mathbb{D}_m \bigg],
\end{eqnarray*}
where the last line follows since $E\big[\mbox{\Large $\mathds{1}$}\hspace{-0.5mm}\{Y=1\}\big| \mathbb{D}_m , \mbox{\large $\boldsymbol{\chi}$}\big]=E[Y|\mbox{\large $\boldsymbol{\chi}$}] = \mathcal{R}(\mbox{\large $\boldsymbol{\chi}$}; \varphi^*)$, by Lemma \ref{LEM-00}. 
It is straightforward to see that we also have
\begin{eqnarray*}
L^* &=& 1- \big[
P\left\{ \mathcal{T}_{\varphi^*}(\mbox{\large $\boldsymbol{\chi}$}) =1, Y=1  \right\}  +  P\left\{ \mathcal{T}_{\varphi^*}(\mbox{\large $\boldsymbol{\chi}$}) =0, Y=0   \right\} \big]\\
&=& 
1-E\big[ \mbox{\Large $\mathds{1}$}\hspace{-0.5mm}\big\{\mathcal{T}_{\varphi^*}(\mbox{\large $\boldsymbol{\chi}$})=1   \big\}\cdot \mathcal{R}(\mbox{\large $\boldsymbol{\chi}$}; \varphi^*)
+ \mbox{\Large $\mathds{1}$}\hspace{-0.5mm}\big\{ \mathcal{T}_{\varphi^*}(\mbox{\large $\boldsymbol{\chi}$})=0   \big\}\cdot \big(1- \mathcal{R}(\mbox{\large $\boldsymbol{\chi}$}; \varphi^*)\big)\big].
\end{eqnarray*}
Now observe that
\begin{eqnarray}
L_m(\widehat{\mathcal{T}}_{m,\varphi,h}) - L^* &=& E\left[\Big(\mbox{\Large $\mathds{1}$}\hspace{-0.5mm}\big\{\mathcal{T}_{\varphi^*}(\mbox{\large $\boldsymbol{\chi}$})=1   \big\} - \mbox{\Large $\mathds{1}$}\hspace{-0.5mm}\big\{\widehat{\mathcal{T}}_{m,\varphi,h}(\mbox{\large $\boldsymbol{\chi}$})=1   \big\}\Big)\big(2\, \mathcal{R}(\mbox{\large $\boldsymbol{\chi}$}; \varphi^*) -1\big)\, \Big| \mathbb{D}_m\right]\nonumber\\
&=& 2\,E\left[
\mbox{\Large $\mathds{1}$}\hspace{-0.5mm}\big\{\mathcal{T}_{\varphi^*}(\mbox{\large $\boldsymbol{\chi}$}) \neq 
\widehat{\mathcal{T}}_{m,\varphi,h}(\mbox{\large $\boldsymbol{\chi}$})\big\}\cdot \Big| \mathcal{R}(\mbox{\large $\boldsymbol{\chi}$}; \varphi^*) -\frac{1}{2}   \Big|\,\Big| \mathbb{D}_m \right]\nonumber\\
&\leq& 2\,E\left[ \left| \widehat{\mathcal{R}}_m(\mbox{\large $\boldsymbol{\chi}$}; \varphi,h) - \mathcal{R}(\mbox{\large $\boldsymbol{\chi}$}; \varphi^*)  \right|\,\Big| \mathbb{D}_m  \right],    \label{Lem7B}
\end{eqnarray}
where (\ref{Lem7B}) follows because on the set $\big\{\mathcal{T}_{\varphi^*}(\mbox{\large $\boldsymbol{\chi}$}) \neq 
\widehat{\mathcal{T}}_{m,\varphi,h}(\mbox{\large $\boldsymbol{\chi}$})\big\}$, one has $\big| \mathcal{R}(\mbox{\large $\boldsymbol{\chi}$}; \varphi^*) -1/2   \big| \leq \big| \widehat{\mathcal{R}}_m(\mbox{\large $\boldsymbol{\chi}$}; \varphi,h) - \mathcal{R}(\mbox{\large $\boldsymbol{\chi}$}; \varphi^*) \big|.$ 

\hfill $\Box$

\begin{lem} \label{MAINN-1}
Suppose that assumptions (A0)--(A8) hold and put $\tau_m=\log m /m$.  Let $\psi_k$ and $\eta_k$  be as in (\ref{psieta}). Also, let $\widehat{\psi}_k$ and $\widehat{\eta}_k$ be as in (\ref{PSI12.hat}) and (\ref{ETA12m.hat}). Then, for $k=1,2$,

\vspace{-5mm}
\begin{eqnarray}
\sup_{\varphi \in \mathcal{F}_{\varepsilon_n}} \sup_{\mbox{\tiny $\boldsymbol{\chi}$}\in\, \mathcal{S}_{\mbox{\tiny $\mathbb{X}$}}}
\Big|\widehat{\psi}_{m,k}(\mbox{\scriptsize $\boldsymbol{\chi}$}; \varphi,h)-\psi_k(\mbox{\scriptsize $\boldsymbol{\chi}$}; \varphi)\Big| 
&=&
\mathcal{O}\big(h^{\beta_k}\big) 
+ \mathcal{O}_{a.co.}\left(
\sqrt{\frac{\log\big[\mathcal{N}_{\varepsilon_n}(\mathcal{F}) \vee \mathcal{N}_{\tau_m}(\mathcal{S}_{\mbox{\tiny $\mathbb{X}$}})\big]}{m\cdot \phi(h)}} 
\right)\label{EQ95A}\\[-3pt]
\sup_{\mbox{\tiny $\boldsymbol{\chi}$}\,\in\, \mathcal{S}_{\mbox{\tiny $\mathbb{X}$}}}
\Big|\widehat{\eta}_{m,k}(\mbox{\scriptsize $\boldsymbol{\chi}$};h)-\eta_k(\mbox{\scriptsize $\boldsymbol{\chi}$})\Big| 
&=& \mathcal{O}\big(h^{\beta_k}\big) + \mathcal{O}_{a.co.}\left(
\sqrt{\frac{\log\big[\mathcal{N}_{\tau_m}(\mathcal{S}_{\mbox{\tiny $\mathbb{X}$}})\big]}{m\cdot \phi(h)}} 
\right),\label{EQ95B}
\end{eqnarray}

\vspace{-1mm}\noindent
where $\beta_1$ and $\beta_2$  are  the positive constants in assumption (A2) and $\mathcal{N}_{\tau_m}$ is as in assumption (A5). 
\end{lem}

\vspace{2mm}\noindent
PROOF OF LEMMA \ref{MAINN-1}

\vspace{1mm}\noindent
We start with the proof of (\ref{EQ95A}). Here, we employ some of the arguments used in Ferraty et al (2010). Let $\psi_k$ and $\widehat{\psi}_k$ be as in (\ref{psieta}) and (\ref{PSI12.hat}), respectively, and observe that 
\begin{eqnarray}
\widehat{\psi}_{m,1}(\mbox{\scriptsize $\boldsymbol{\chi}$}; \varphi, h)-\psi_1(\mbox{\scriptsize $\boldsymbol{\chi}$}; \varphi)
&=& \frac{1}{\widehat{f}_m(\mbox{\scriptsize $\boldsymbol{\chi}$})}\bigg\{\Big[\widehat{g}_m(\mbox{\scriptsize $\boldsymbol{\chi}$}; \varphi)- \mathds{E}\big(\widehat{g}_m(\mbox{\scriptsize $\boldsymbol{\chi}$}; \varphi) \big)\Big]
+ \Big[\mathds{E}\big(\widehat{g}_m(\mbox{\scriptsize $\boldsymbol{\chi}$}; \varphi) \big) - \psi_1(\mbox{\scriptsize $\boldsymbol{\chi}$}; \varphi) \Big]\nonumber\\
&& ~~~~~~~~~~ +\,\Big[1-\widehat{f}_m(\mbox{\scriptsize $\boldsymbol{\chi}$})\Big]\cdot  \psi_1(\mbox{\scriptsize $\boldsymbol{\chi}$}; \varphi)\bigg\}, \label{EQQ11}
\end{eqnarray}
where 
\begin{equation*}
\widehat{f}_m(\mbox{\scriptsize $\boldsymbol{\chi}$})\,=\, 
\frac{\sum_{i\in\boldsymbol{{\cal I}}_m}\mathcal{K}
	\big(h^{-1}d(\mbox{$\mbox{\scriptsize $\boldsymbol{\chi}$}$} ,\, \mbox{\large$\boldsymbol{\chi}$}_i)\big)
}{m\mathds{E}\big[\mathcal{K}
\big(h^{-1}d(\mbox{$\mbox{\scriptsize $\boldsymbol{\chi}$}$} ,\, \mbox{\large$\boldsymbol{\chi}$}_1)\big)\big]}~~~~\mbox{and}~~~~
\widehat{g}_m(\mbox{\scriptsize $\boldsymbol{\chi}$}; \varphi) \,=\, 
\frac{\sum_{i\in\boldsymbol{{\cal I}}_m}\Delta_i Y_i \varphi(Y_i)\,
	\mathcal{K}
	\big(h^{-1}d(\mbox{$\mbox{\scriptsize $\boldsymbol{\chi}$}$} ,\, \mbox{\large$\boldsymbol{\chi}$}_i)\big)
}{m\mathds{E}\big[\mathcal{K}
\big(h^{-1}d(\mbox{$\mbox{\scriptsize $\boldsymbol{\chi}$}$} ,\, \mbox{\large$\boldsymbol{\chi}$}_1)\big)\big]}.
\end{equation*}
Let $\widetilde{\mbox{\scriptsize $\boldsymbol{\chi}$}}_{_j},$ $j=1 \cdots,  \mathcal{N}_{\tau_m}(\mathcal{S}_{\mbox{\tiny $\mathbb{X}$}})$ be a $\tau_m$-cover for $\mathcal{S}_{\mbox{\tiny $\mathbb{X}$}}$, i.e., 
\[
\mathcal{S}_{\mbox{\tiny $\mathbb{X}$}} \subset 
\bigcup_{j=1}^{\mathcal{N}_{\tau}(\mathcal{S}_{\mbox{\tiny $\mathbb{X}$}})}B\left(\widetilde{\mbox{\scriptsize $\boldsymbol{\chi}$}}_{j}, \tau_m \right),
\]
where $\tau_m=\log m/m$ as before, and consider  the basic decomposition
\begin{eqnarray}
\sup_{\varphi \in \mathcal{F}_{\varepsilon_n}} \sup_{\mbox{\tiny $\boldsymbol{\chi}$}\in\, \mathcal{S}_{\mbox{\tiny $\mathbb{X}$}}}\Big|
\widehat{g}_m(\mbox{\scriptsize $\boldsymbol{\chi}$}; \varphi)- \mathds{E}\big(\widehat{g}_m(\mbox{\scriptsize $\boldsymbol{\chi}$}; \varphi) \big)\Big| &\leq& 
\sup_{\varphi \in \mathcal{F}_{\varepsilon_n}} \max_{1\leq j \leq \mathcal{N}_{\tau_m}(\mathcal{S}_{\mbox{\tiny $\mathbb{X}$}})}\,
\sup_{
	\mbox{\tiny $\boldsymbol{\chi}$}\in B(\mbox{\tiny $\widetilde{\boldsymbol{\chi}}$}_{j}, \tau_m)}
\Big|
\widehat{g}_m(\mbox{\scriptsize $\boldsymbol{\chi}$}; \varphi)- \widehat{g}_m(\mbox{\scriptsize $\widetilde{\boldsymbol{\chi}}$}_{j}; \varphi)\Big|\nonumber\\
&& + \sup_{\varphi \in \mathcal{F}_{\varepsilon_n}} \max_{1\leq j \leq \mathcal{N}_{\tau_m}(\mathcal{S}_{\mbox{\tiny $\mathbb{X}$}})}\,
\sup_{
	\mbox{\tiny $\boldsymbol{\chi}$}\in B(\mbox{\tiny $\widetilde{\boldsymbol{\chi}}$}_{j}, \tau_m)}
\Big|
\mathds{E}\left[\widehat{g}_m(\mbox{\scriptsize $\boldsymbol{\chi}$}; \varphi)\right]- \mathds{E}\left[\widehat{g}_m(\mbox{\scriptsize $\widetilde{\boldsymbol{\chi}}$}_{j}; \varphi)\right]\Big|\nonumber\\
&& + \sup_{\varphi \in \mathcal{F}_{\varepsilon_n}} \max_{1\leq j \leq \mathcal{N}_{\tau_m}(\mathcal{S}_{\mbox{\tiny $\mathbb{X}$}})}
\Big|
\widehat{g}_m(\mbox{\scriptsize $\widetilde{\boldsymbol{\chi}}$}_{j}; \varphi) - \mathds{E}\left[\widehat{g}_m(\mbox{\scriptsize $\widetilde{\boldsymbol{\chi}}$}_{j}; \varphi)\right]\Big|\nonumber\\
&:=& I_m+I\!\!I_m+I\!\!I\!\!I_m. \label{EQ4ZZ}
\end{eqnarray}
It can be shown (see Ferraty and Vieu (2006; Lemma 4.4)) that in view of assumptions (A1) and (A4) there are constants $0<C'<C''<\infty$ such that 
\[
\forall ~ \mbox{\scriptsize $\boldsymbol{\chi}$}\in\, \mathcal{S}_{\mbox{\tiny $\mathbb{X}$}}, ~~C'\phi(h) < \mathds{E}\big[\mathcal{K}
\big(h^{-1}d(\mbox{$\mbox{\scriptsize $\boldsymbol{\chi}$}$} ,\, \mbox{\large$\boldsymbol{\chi}$}_1)\big)\big] < C''\phi(h).
\]
Now, this observation together with the fact that $|\Delta_i Y_i \varphi(Y_i)|\leq B$ for all $i=1,\cdots, n$, implies that
\begin{eqnarray}
&& \sup_{
	\mbox{\tiny $\boldsymbol{\chi}$}\in B(\mbox{\tiny $\widetilde{\boldsymbol{\chi}}$}_{j},\, \tau_m)}
\Big|
\widehat{g}_m(\mbox{\scriptsize $\boldsymbol{\chi}$}; \varphi)- \widehat{g}_m(\mbox{\scriptsize $\widetilde{\boldsymbol{\chi}}$}_{j}; \varphi)\Big|\nonumber\\
&&~=\, \sup_{
	\mbox{\tiny $\boldsymbol{\chi}$}\in B(\mbox{\tiny $\widetilde{\boldsymbol{\chi}}$}_{j},\, \tau_m)}\frac{1}{m} \left|
\frac{\sum_{i\in\boldsymbol{{\cal I}}_m}\Delta_i Y_i \varphi(Y_i)\,
	\mathcal{K}
	\big(h^{-1}d(\mbox{$\mbox{\scriptsize $\boldsymbol{\chi}$}$} ,\, \mbox{\large$\boldsymbol{\chi}$}_i)\big)
}{\mathds{E}\big[\mathcal{K}
\big(h^{-1}d(\mbox{$\mbox{\scriptsize $\boldsymbol{\chi}$}$} ,\, \mbox{\large$\boldsymbol{\chi}$}_1)\big)\big]} - 
\frac{\sum_{i\in\boldsymbol{{\cal I}}_m}\Delta_i Y_i \varphi(Y_i)\,
	\mathcal{K}
	\big(h^{-1}d(\mbox{$\mbox{\scriptsize $\widetilde{\boldsymbol{\chi}}$}$}_j ,\, \mbox{\large$\boldsymbol{\chi}$}_i)\big)
}{\mathds{E}\big[\mathcal{K}
\big(h^{-1}d(\mbox{$\mbox{\scriptsize $\widetilde{\boldsymbol{\chi}}$}$}_j ,\, \mbox{\large$\boldsymbol{\chi}$}_1)\big)\big]}
\right|\nonumber\\
&&~\leq\, \sup_{
	\mbox{\tiny $\boldsymbol{\chi}$}\in B(\mbox{\tiny $\widetilde{\boldsymbol{\chi}}$}_{j},\, \tau_m)}\frac{CB}{m\, \phi(h)}\sum_{i\in\boldsymbol{{\cal I}}_m}\Big|\mathcal{K}
\big(h^{-1}d(\mbox{$\mbox{\scriptsize $\boldsymbol{\chi}$}$} ,\, \mbox{\large$\boldsymbol{\chi}$}_i)\big) - \mathcal{K}
\big(h^{-1}d(\mbox{$\mbox{\scriptsize $\widetilde{\boldsymbol{\chi}}$}$}_j ,\, \mbox{\large$\boldsymbol{\chi}$}_i)\big)\Big|\cdot 
\mbox{\Large $\mathds{1}$}\big\{\mbox{\large$\boldsymbol{\chi}_i$} \in \big[B(\mbox{\scriptsize $\boldsymbol{\chi}$},\, h)\, \mbox{$\cup$} B(\mbox{\scriptsize $\widetilde{\boldsymbol{\chi}}$}_{j},\, h)\big] \big\}\nonumber\\
&&~\leq\, \sup_{
	\mbox{\tiny $\boldsymbol{\chi}$}\in B(\mbox{\tiny $\widetilde{\boldsymbol{\chi}}$}_{j},\, \tau_m)}\frac{CB}{m\,\phi(h)}\cdot\frac{\tau_m}{h}
\sum_{i\in\boldsymbol{{\cal I}}_m} 
\mbox{\Large $\mathds{1}$}\big\{\mbox{\large$\boldsymbol{\chi}_i$} \in \big[B(\mbox{\scriptsize $\boldsymbol{\chi}$},\, h)\, \mbox{$\cup$} B(\mbox{\scriptsize $\widetilde{\boldsymbol{\chi}}$}_{j},\, h)\big] \big\}, \label{EQEQ16}
\end{eqnarray}
where the last line follows because $\mathcal{K}$ is Lipschitz on $[0,1]$ which implies that 
\[
\big|\mathcal{K}
\big(h^{-1}d(\mbox{$\mbox{\scriptsize $\boldsymbol{\chi}$}$} ,\, \mbox{\large$\boldsymbol{\chi}$}_i)\big) - \mathcal{K}
\big(h^{-1}d(\mbox{$\mbox{\scriptsize $\widetilde{\boldsymbol{\chi}}$}$}_j ,\, \mbox{\large$\boldsymbol{\chi}$}_i)\big)\big|~\leq~\frac{1}{h}\,
d(\mbox{\scriptsize$\boldsymbol{\chi}$} ,\,\mbox{$\mbox{\scriptsize $\widetilde{\boldsymbol{\chi}}$}$}_j) ~\leq~ \frac{\tau_m}{h},~~~~\forall\,~
\mbox{\scriptsize$\boldsymbol{\chi}$}\in B(\mbox{\scriptsize $\widetilde{\boldsymbol{\chi}}$}_{j},\, \tau_m).
\]
However, if $\mbox{\scriptsize$\boldsymbol{\chi}$}\in B(\mbox{\scriptsize $\widetilde{\boldsymbol{\chi}}$}_{j},\, \tau_m)$, where $\tau_m:=\log m/m \leq h$, then one finds
\[
B(\mbox{\scriptsize $\boldsymbol{\chi}$},\, h)\, \mbox{$\cup$}\, B(\mbox{\scriptsize $\widetilde{\boldsymbol{\chi}}$}_{j},\, h) \subset  B(\mbox{\scriptsize $\widetilde{\boldsymbol{\chi}}$}_{j},\, 2h).
\] 
Consequently
\begin{eqnarray}
\mbox{(right side of }(\ref{EQEQ16})) &\leq&  \frac{CB}{\phi(h)}\cdot\frac{1}{m}\cdot\frac{\tau_m}{h}
\sum_{i\in\boldsymbol{{\cal I}}_m}\, \underbrace{\hspace{-1mm}
	\mbox{\Large $\mathds{1}$}\big\{\mbox{\large$\boldsymbol{\chi}_i$} \in B(\mbox{\scriptsize $\widetilde{\boldsymbol{\chi}}$}_{j},\, 2h) \big\}}_{\mbox{\small free of $\mbox{\scriptsize $\boldsymbol{\chi}$}$}}~:=~\frac{C_1}{m}\sum_{i\in\boldsymbol{{\cal I}}_m} Z_{ij}, \label{EQEQ17}
\end{eqnarray}
where 
\begin{equation}\label{ADD-1}
Z_{ij} \,=\, \frac{\tau_m}{h \cdot \phi(h)}~ \mbox{\Large $\mathds{1}$}\big\{\mbox{\large$\boldsymbol{\chi}_i$} \in B(\mbox{\scriptsize $\widetilde{\boldsymbol{\chi}}$}_{j},\, 2h) \big\}.
\end{equation}
Furthermore, using assumption (A1), one immediately finds
$$
\mathds{E}(Z_{ij}) = C_2\tau_m \phi(2h)/[h \phi(h)]~~~~\mbox{and}~~~~ \mathds{E}(Z^2_{ij}) = C_2\tau^2_m \phi(2h)/[h^2 \phi^2_1(h)],
$$ 
where $C_2$ is a positive constant not depending on $n$. Also, one finds Var$(Z_{ij})=\mathds{E}(Z^2_{ij})- [\mathds{E}(Z_{ij})]^2= C_2\tau^2_m\phi(2h)\cdot[1-C_2\phi(2h)]/ [h^2 \phi^2_1(h)]$. Therefore, in view of (\ref{EQEQ16}) and (\ref{EQEQ17}) (and upon replacing $Z_{ij}$ by $Z_{ij}-\mathds{E}(Z_{ij})+\mathds{E}(Z_{ij})$ in (\ref{EQEQ17})), one arrives at
\begin{eqnarray}
\sup_{
	\mbox{\tiny $\boldsymbol{\chi}$}\in B(\mbox{\tiny $\widetilde{\boldsymbol{\chi}}$}_{j},\, \tau_m)}
\Big|
\widehat{g}_m(\mbox{\scriptsize $\boldsymbol{\chi}$}; \varphi)- \widehat{g}_m(\mbox{\scriptsize $\widetilde{\boldsymbol{\chi}}$}_{j}; \varphi)\Big| &\leq& \frac{1}{m} \left|\sum_{i\in\boldsymbol{{\cal I}}_m} Z'_{ij}\right| + \mathcal{O}\left(\frac{\tau_m \phi(2h)}{h\phi(h)}\right), \label{EQ5}
\end{eqnarray}
where $Z'_{ij}=C_1[Z_{ij}-\mathds{E}(Z_{ij})]$ and where the big-O term does not depend on $\mbox{\scriptsize $\boldsymbol{\chi}$}$, $\mbox{\scriptsize $\widetilde{\boldsymbol{\chi}}$}_{j}$, or $\varphi$. 
Furthermore,  it is not hard to show that for $m$ large enough, 
\begin{equation*}
\mathds{E}\big|Z'_{ij}\big|^2 \,\leq\, C\bigg(\frac{\tau_m \sqrt{\phi(2h)}}{h\, \phi(h)}\bigg)^{2}.
\end{equation*}
Therefore, by Corollary A.8 of Ferraty and Vieu (2006), for any $t>0$ 
\begin{eqnarray}
\mathds{P}\left\{\sup_{\varphi \in \mathcal{F}_{\varepsilon_n}} \max_{1\leq j \leq \mathcal{N}_{\tau_m}(\mathcal{S}_{\mbox{\tiny $\mathbb{X}$}})}\frac{1}{m}\, \bigg|\sum_{i\in\boldsymbol{{\cal I}}_m} Z'_{ij}\bigg| \,\geq\, t
\right\} 
&\leq&
\mathcal{N}_{\tau_m}(\mathcal{S}_{\mbox{\tiny $\mathbb{X}$}}) \max_{1\leq j \leq \mathcal{N}_{\tau_m}(\mathcal{S}_{\mbox{\tiny $\mathbb{X}$}})}\mathds{P}\left\{\frac{1}{m}\, \bigg|\sum_{i\in\boldsymbol{{\cal I}}_m} Z'_{ij}\bigg| \,>\,t
\right\}~~~~ \nonumber\\
&&~~~~~~\mbox{(because $Z'_{ij}$ does not depend on $\varphi$)}\nonumber\\
&\leq& 2\,\mathcal{N}_{\tau_m}(\mathcal{S}_{\mbox{\tiny $\mathbb{X}$}}) \exp\left\{\frac{- m h^2 \phi^2(h)\,t^2}{2(1+t)\tau_m^2 \phi(2h)}\right\} \label{EQEQ18-A}
\end{eqnarray}
Now, for any constant $t_0>0$,  take 
$
t=t_0 \sqrt{\tau_m^2 \phi(2h) \log[\mathcal{N}_{\tau_m}(\mathcal{S}_{\mbox{\tiny $\mathbb{X}$}})]\big/\big(m h^2 \phi^2(h)\big)}
$ 
and observe that in view of (\ref{EQEQ18-A}),
\begin{eqnarray}
&& P(m) ~:=~~\mathds{P}\left\{\sup_{\varphi \in \mathcal{F}_{\varepsilon_n}} \max_{1\leq j \leq \mathcal{N}_{\tau_m}(\mathcal{S}_{\mbox{\tiny $\mathbb{X}$}})}\frac{1}{m}\, \bigg|\sum_{i\in\boldsymbol{{\cal I}}_m} Z'_{ij}\bigg| \,\geq\, t_0\,\sqrt{\frac{\tau_m^2 \phi(2h) \log[\mathcal{N}_{\tau_m}(\mathcal{S}_{\mbox{\tiny $\mathbb{X}$}})]}{m h^2 \phi^2(h)}}\right\} \label{ADD-2}\\ [1pt]
&&~~~~~~~~~\leq~ 2\,\mathcal{N}_{\tau_m}(\mathcal{S}_{\mbox{\tiny $\mathbb{X}$}})\cdot \exp\left\{
\frac{-t_0^2 \log[\mathcal{N}_{\tau_m}(\mathcal{S}_{\mbox{\tiny $\mathbb{X}$}})]}{2
	\left(1+t_0\,\sqrt{\tau_m^2 \phi(2h) \log[\mathcal{N}_{\tau_m}(\mathcal{S}_{\mbox{\tiny $\mathbb{X}$}})]\big/\big(m h^2 \phi^2(h)\big)}\,\right)} \right\} \nonumber\\ [2pt]
&&~~~~~~~~~\leq~ 2\,\mathcal{N}_{\tau_m}(\mathcal{S}_{\mbox{\tiny $\mathbb{X}$}})\cdot\big[ \mathcal{N}_{\tau_m}(\mathcal{S}_{\mbox{\tiny $\mathbb{X}$}})\big]^{-c t_0}\,=\, 2\big[ \mathcal{N}_{\tau_m}(\mathcal{S}_{\mbox{\tiny $\mathbb{X}$}})\big]^{1-c t_0}, 
\nonumber
\end{eqnarray}
for $m$ large enough, where  $c$ is a positive constant not depending on $n$ (or $m$). Consequently, choosing $t_0$ suitably, one finds
\begin{eqnarray*}
	\sum_{m=1}^\infty P(m)
	&\leq& 2\sum_{m=1}^\infty [\mathcal{N}_{\tau_m}(\mathcal{S}_{\mbox{\tiny $\mathbb{X}$}})]^{1-C t_0^2}~ < \infty,
\end{eqnarray*}
which holds due to assumption (A5)(ii). Therefore, by (\ref{EQ5}), assumption (A5)(ii), and the fact that $\phi(2h)/\phi(h)=\mathcal{O}(1)$, one finds
\begin{equation}
I_m \,=\, \mathcal{O}_{a.co.}\left(\sqrt{\frac{\tau^2_m \phi(2h) \log[\mathcal{N}_{\tau_m}(\mathcal{S}_{\mbox{\tiny $\mathbb{X}$}})]}{mh^2\phi^2(h)}}\right) + \mathcal{O}\left(\frac{\tau_m \phi(2h)}{h\phi(h)}\right)
\label{QQ1}
\end{equation}
As for the term $I\!\!I_m$ in (\ref{EQ4ZZ}), first observe that by (\ref{EQEQ16}), (\ref{EQEQ17}), and (\ref{EQ5})
\begin{eqnarray}
I\!\!I_m &\leq& \mathds{E}\left[\sup_{\varphi \in \mathcal{F}_{\varepsilon_n}} \max_{1\leq j \leq \mathcal{N}_{\tau_m}(\mathcal{S}_{\mbox{\tiny $\mathbb{X}$}})}\,
\sup_{
	\mbox{\tiny $\boldsymbol{\chi}$}\in B(\mbox{\tiny $\widetilde{\boldsymbol{\chi}}$}_{j}, \tau_m)}
\Big|
\widehat{g}_m(\mbox{\scriptsize $\boldsymbol{\chi}$}; \varphi)- \widehat{g}_m(\mbox{\scriptsize $\widetilde{\boldsymbol{\chi}}$}_{j}; \varphi)\Big|\right]\nonumber\\
&\leq& \mathds{E}\left[\max_{1\leq j \leq \mathcal{N}_{\tau_m}(\mathcal{S}_{\mbox{\tiny $\mathbb{X}$}})} 
\frac{1}{m} \left|\sum_{i\in\boldsymbol{{\cal I}}_m} Z'_{ij}\right|\right]  + \mathcal{O}\left(\frac{\tau_m \phi(2h)}{h\phi(h)}\right), \label{EQEQ19}
\end{eqnarray}
where, as before, $Z'_{ij}=C_1[Z_{ij}-\mathds{E}(Z_{ij})]$ with $Z_{ij}$ as in (\ref{ADD-1}). On the other hand, since
\begin{equation}
\big|Z'_{ij}\big| 
~\leq~ C_1[1+C_2\phi(2h)]\,\tau_m/
\big(h \phi(h)\big) \,=:\, A(m), \label{EQEQ20}
\end{equation}
one can proceed as follows
\begin{eqnarray}
&& \mathds{E}\left[\max_{1\leq j \leq \mathcal{N}_{\tau_m}(\mathcal{S}_{\mbox{\tiny $\mathbb{X}$}})} 
\frac{1}{m} \left|\sum_{i\in\boldsymbol{{\cal I}}_m} Z'_{ij}\right|\right]
~=~\int_0^{\infty}\mathds{P}\bigg\{\max_{1\leq j \leq \mathcal{N}_{\tau_m}(\mathcal{S}_{\mbox{\tiny $\mathbb{X}$}})}\,
\frac{1}{m}\, \bigg|\sum_{i\in\boldsymbol{{\cal I}}_m} Z'_{ij}\bigg| \,>\, t\bigg\}\, dt \nonumber\\[2pt]
&&~~~\leq~ \int_0^u dt \,+ \int_u^{A(m)}\mathds{P}\bigg\{\max_{1\leq j \leq \mathcal{N}_{\tau_m}(\mathcal{S}_{\mbox{\tiny $\mathbb{X}$}})}\,
\frac{1}{m}\, \bigg|\sum_{i\in\boldsymbol{{\cal I}}_m} Z'_{ij}\bigg| \,>\, t\bigg\}\, dt \nonumber\\[3pt]
&&~~~~~~~~~~\mbox{(becuase $|Z'_{ij}| \leq A(m)$, where $A(m)$ is as in (\ref{EQEQ20}))}\nonumber\\
&& ~~~\leq~ u + 2\,\mathcal{N}_{\tau_m}(\mathcal{S}_{\mbox{\tiny $\mathbb{X}$}}) \int_u^{A(m)}\exp\left\{-m h^2 \phi^2_1(h)\, t^2\big/\big(2(1+A(m)) \tau^2_m\phi(2h)\big)\right\} dt, \nonumber\\[2pt] 
&&~~~~~~~~~~\mbox{(via the exponential bound in (\ref{EQEQ18-A}) and the fact that $m^{-1}|\sum_{i\in\boldsymbol{{\cal I}}_m}Z'_{ij}|\leq A(m)$)}\nonumber\\[3pt]
&&~~~\leq~ u + \frac{2\,\mathcal{N}_{\tau_m}(\mathcal{S}_{\mbox{\tiny $\mathbb{X}$}})}{\sqrt{m h^2 \phi^2_1(h)\big/\big((1+A(m)) \tau^2_m\phi(2h)\big)}}
\int^{\infty}_{u\,\sqrt{m h^2 \phi^2_1(h)/((1+A(m)) \tau^2_m\phi(2h))}} e^{-v^2/2}\, dv \nonumber\\[3pt]
&&~~~~~~~~~~\mbox{\big(by the change of variable, $v = t\, \sqrt{m h^2 \phi^2_1(h)\big/\big(\big(1+A(m)) \tau^2_m\phi(2h)\big)}$~\big)} \nonumber\\[3pt]
&&~~~\leq~ u + \frac{2\,\mathcal{N}_{\tau_m}(\mathcal{S}_{\mbox{\tiny $\mathbb{X}$}})\cdot \exp\Big\{\mbox{$-$}\big[m h^2 \phi^2_1(h)\big/\big((1+A(m)) \tau^2_m\phi(2h)\big)\big] u^2/2 \Big\}}{\big[m h^2 \phi^2_1(h)\big/\big((1+A(m)) \tau^2_m\phi(2h)\big)\big]\cdot u}
\nonumber\\[2pt]
&&~~~~~~~~~~\mbox{\big(via the upper bound in Mill's ratio (see Mitrinovic (1970; p.177))\big)}\nonumber\\[3pt]
&&~~~=:~ u + \frac{2\,\mathcal{N}_{\tau_m}(\mathcal{S}_{\mbox{\tiny $\mathbb{X}$}})}{4N u}\,e^{-2N u^2}, ~~~\mbox{where~ $N =\, m h^2 \phi^2_1(h)\big/\big[4(1+A(m)) \tau^2_m\phi(2h)\big].$} \label{EQ7}
\end{eqnarray}
But the expression\,  $u + \big[2\,\mathcal{N}_{\tau_m}(\mathcal{S}_{\mbox{\tiny $\mathbb{X}$}})/(4N u)\big]\,e^{-2N u^2}$ in (\ref{EQ7}) is approximately minimized by taking $u=\sqrt{\log(2\,\mathcal{N}_{\tau_m}(\mathcal{S}_{\mbox{\tiny $\mathbb{X}$}}))/(2N)}$, and the corresponding value of the right side of (\ref{EQ7}) becomes
\begin{eqnarray*}
	&&\sqrt{\frac{\log(2\,\mathcal{N}_{\tau_m}(\mathcal{S}_{\mbox{\tiny $\mathbb{X}$}}))}{2N}} + \sqrt{\frac{1}{8 N \log(2\,\mathcal{N}_{\tau_m}(\mathcal{S}_{\mbox{\tiny $\mathbb{X}$}}))}}\\
	&&~~=~\sqrt{\frac{\log(2\,\mathcal{N}_{\tau_m}(\mathcal{S}_{\mbox{\tiny $\mathbb{X}$}}))[2(1+A(m)) \tau^2_m\phi(2h)]}{m h^2 \phi^2_1(h)}} + \sqrt{\frac{(1+A(m)) \tau^2_m\phi(2h)}{2m h^2 \phi^2_1(h)\cdot \log(2\,\mathcal{N}_{\tau_m}(\mathcal{S}_{\mbox{\tiny $\mathbb{X}$}}))}} \\
	&&~~=~\mathcal{O}\left(\sqrt{\frac{\tau_m^2 \log(\mathcal{N}_{\tau_m}(\mathcal{S}_{\mbox{\tiny $\mathbb{X}$}}))}{m h^2 \phi(h)}}\right).
\end{eqnarray*}
This last bound together with (\ref{EQ7}) and (\ref{EQEQ19})  implies that the term $I\!\!I_m$ defined via (\ref{EQ4ZZ}) satisfies
\begin{eqnarray}
I\!\!I_m &=& \mathcal{O}\left(\sqrt{\frac{\tau_m^2 \log(\mathcal{N}_{\tau_m}(\mathcal{S}_{\mbox{\tiny $\mathbb{X}$}}))}{m h^2 \phi(h)}}\right) + \mathcal{O}\left(\frac{\tau_m \phi(2h)}{h\phi(h)}\right). \label{QQ2}
\end{eqnarray}
To deal with $I\!\!I\!\!I_m$, i.e., the last term in (\ref{EQ4ZZ}), define the quantity
\[
\mathds{U}_{ij}(\varphi) = \frac{\Delta_i Y_i \varphi(Y_i)\,
	\mathcal{K}
	\big(h^{-1}d(\mbox{$\mbox{\scriptsize $\widetilde{\boldsymbol{\chi}}$}$}_j ,\, \mbox{\large$\boldsymbol{\chi}$}_i)\big)
	- \mathds{E}\Big[\Delta_i Y_i \varphi(Y_i)\,
	\mathcal{K}
	\big(h^{-1}d(\mbox{$\mbox{\scriptsize $\widetilde{\boldsymbol{\chi}}$}$}_j ,\, \mbox{\large$\boldsymbol{\chi}$}_i)\big)\Big]
}{\mathds{E}\big[\mathcal{K}
\big(h^{-1}d(\mbox{$\mbox{\scriptsize $\widetilde{\boldsymbol{\chi}}$}$}_j ,\, \mbox{\large$\boldsymbol{\chi}$}_1)\big)\big]}
\]
and observe that for every $t>0$
\begin{eqnarray}
\mathds{P}\left\{I\!\!I\!\!I_m \geq t\right\} &>&
\mathcal{N}_{\varepsilon_n}(\mathcal{F})\,
\mathcal{N}_{\tau_m}(\mathcal{S}_{\mbox{\tiny $\mathbb{X}$}}) \cdot
\sup_{\varphi \in \mathcal{F}_{\varepsilon_n}} \max_{1\leq j \leq \mathcal{N}_{\tau_m}(\mathcal{S}_{\mbox{\tiny $\mathbb{X}$}})}
\mathds{P}\left\{\frac{1}{m}\, \left|\sum_{i\in\boldsymbol{{\cal I}}_m} \mathds{U}_{ij}(\varphi)\right| \,>\, t\right\} \label{EQ8}.
\end{eqnarray}
But, for each fixed $\varphi$, it can be shown that  $\mathds{E}\big|\mathds{U}_{ij}(\varphi)\big|^k = \mathcal{O}\big((\phi(h))^{-k+1}\big),$ for all $k\geq2$; see Ferraty et al (2010; p.347) as well as Ferraty and Vieu (2006; p.66). Therefore, by Corollary A.8 of Ferraty and Vieu (2006), for any arbitrary $t_0>0$,
\begin{eqnarray*}
	\mathds{P}\left\{\frac{1}{m}\, \left|\sum_{i\in\boldsymbol{{\cal I}}_m} \mathds{U}_{ij}(\varphi)\right| \,>\, t_0 \sqrt{\frac{\log\big[\mathcal{N}_{\varepsilon_n}(\mathcal{F}) \vee \mathcal{N}_{\tau_m}(\mathcal{S}_{\mbox{\tiny $\mathbb{X}$}})\big]}{m\phi(h)}}\right\} &\leq& 2\big[\mathcal{N}_{\varepsilon_n}(\mathcal{F}) \vee \mathcal{N}_{\tau_m}(\mathcal{S}_{\mbox{\tiny $\mathbb{X}$}})\big]^{-c\,t_0^2},~~c>0.
\end{eqnarray*}
Therefore, in view of (\ref{EQ8}), 
\begin{eqnarray}
\mathds{P}\left\{I\!\!I\!\!I_m \,>\, t_0 \sqrt{\frac{\log\big[\mathcal{N}_{\varepsilon_n}(\mathcal{F}) \vee \mathcal{N}_{\tau_m}(\mathcal{S}_{\mbox{\tiny $\mathbb{X}$}})\big]}{m\phi(h)}}\right\} &\leq&
2 \big[\mathcal{N}_{\varepsilon_n}(\mathcal{F}) \vee \mathcal{N}_{\tau_m}(\mathcal{S}_{\mbox{\tiny $\mathbb{X}$}})\big]^{2-c\,t_0^2}\nonumber
\end{eqnarray}
Choosing $t_0$ suitably so that $2-c\,t_0^2 \leq 1-\beta$, where $\beta>1$ is as in assumption (A5)(ii), one finds
\[
\sum_{m=1}^\infty \mathds{P}\left\{I\!\!I\!\!I_m \,>\, t_0 \sqrt{\frac{\log\big[\mathcal{N}_{\varepsilon_n}(\mathcal{F}) \vee \mathcal{N}_{\tau_m}(\mathcal{S}_{\mbox{\tiny $\mathbb{X}$}})\big]}{m\phi(h)}}\right\} \,\leq\,
2 \sum_{m=1}^\infty  \big[\mathcal{N}_{\varepsilon_n}(\mathcal{F}) \vee \mathcal{N}_{\tau_m}(\mathcal{S}_{\mbox{\tiny $\mathbb{X}$}})\big]^{1-\beta} \,<\, \infty,
\]
which then yields
\begin{eqnarray}
I\!\!I\!\!I_m &=& \mathcal{O}_{a.co.}\left(\sqrt{\frac{\log\big[\mathcal{N}_{\varepsilon_n}(\mathcal{F}) \vee \mathcal{N}_{\tau_m}(\mathcal{S}_{\mbox{\tiny $\mathbb{X}$}})\big]}{m\phi(h)}}
\right). \label{QQ3}
\end{eqnarray}
Putting together (\ref{EQ4ZZ}), (\ref{QQ1}), (\ref{QQ2}), and (\ref{QQ3}), one finds
\begin{eqnarray}
\sup_{\varphi \in \mathcal{F}_{\varepsilon_n}} \sup_{\mbox{\tiny $\boldsymbol{\chi}$}\in\, \mathcal{S}_{\mbox{\tiny $\mathbb{X}$}}}\Big|
\widehat{g}_m(\mbox{\scriptsize $\boldsymbol{\chi}$}; \varphi)- \mathds{E}\big(\widehat{g}_m(\mbox{\scriptsize $\boldsymbol{\chi}$}; \varphi) \big)\Big| 
&=&
\mathcal{O}_{a.co.}\left(\sqrt{\frac{\log\big[\mathcal{N}_{\varepsilon_n}(\mathcal{F}) \vee \mathcal{N}_{\tau_m}(\mathcal{S}_{\mbox{\tiny $\mathbb{X}$}})\big]}{m\phi(h)}}
\right). \label{EQ9}
\end{eqnarray}
Regarding the term $\big[\mathds{E}\big(\widehat{g}_m(\mbox{\scriptsize $\boldsymbol{\chi}$}; \varphi) \big) - \psi_1(\mbox{\scriptsize $\boldsymbol{\chi}$}; \varphi) \big]$ in (\ref{EQQ11}), one can argue as in the proof of Lemma 10 of Ferraty et al (2010) and  Lemma 4.4 of Ferraty and Vieu (2006) that, under assumptions (A1), (A2), (A3), and (A4),
\begin{eqnarray}
\Big|\mathds{E}\big(\widehat{g}_m(\mbox{\scriptsize $\boldsymbol{\chi}$}; \varphi) \big) - \psi_1(\mbox{\scriptsize $\boldsymbol{\chi}$}; \varphi) \Big| &\leq& 
\frac{1}{\mathds{E}\big[\mathcal{K}
	\big(h^{-1}d(\mbox{$\mbox{\scriptsize $\boldsymbol{\chi}$}$} ,\, \mbox{\large$\boldsymbol{\chi}$}_1)\big)\big]}\hspace{0.1mm}
\mathds{E}\Big[
\mathcal{K}\big(h^{-1}d(\mbox{$\mbox{\scriptsize $\boldsymbol{\chi}$}$} ,\,\mbox{\large$\boldsymbol{\chi}$}_1)\big)
\cdot\underbrace{\big|\psi_1(\mbox{\large$\boldsymbol{\chi}$}_1; \varphi) - \psi_1(\mbox{\scriptsize $\boldsymbol{\chi}$}; \varphi)\big|}_{
	\mbox{\scriptsize \,$\leq\, C_1 d^{^{\beta_1}}(\mbox{\scriptsize $\boldsymbol{\chi}$},\, \mbox{\large$\boldsymbol{\chi}$}_1)$}} \Big]\nonumber\\
&\leq& \frac{C_1}{\mathds{E}\big[\mathcal{K}
	\big(h^{-1}d(\mbox{$\mbox{\scriptsize $\boldsymbol{\chi}$}$} ,\, \mbox{\large$\boldsymbol{\chi}$}_1)\big)\big]}
\mathds{E}\Big[
\mathcal{K}\big(h^{-1}d(\mbox{$\mbox{\scriptsize $\boldsymbol{\chi}$}$} ,\,\mbox{\large$\boldsymbol{\chi}$}_1)\big)\,
\mbox{\Large $\mathds{1}$}\big\{\mbox{\large$\boldsymbol{\chi}_1$} 
	\in B(\mbox{\scriptsize $\boldsymbol{\chi}$},\, h)\big\}
\cdot d^{^{\beta_1}}(\mbox{\scriptsize $\boldsymbol{\chi}$},\,\mbox{\large$\boldsymbol{\chi}$}_1) \Big]\nonumber\\[4pt]
&\leq& C_1 h^{\beta_1},\nonumber
\end{eqnarray}
where  $\beta_1$ and $C_1$ are the positive constants in assumption (A2). Since $C_1$ does not depend on $\mbox{\scriptsize $\boldsymbol{\chi}$}$ or $\varphi$, we find
\begin{eqnarray}
\sup_{\varphi \in \mathcal{F}_{\varepsilon_n}} \sup_{\mbox{\tiny $\boldsymbol{\chi}$}\in\, \mathcal{S}_{\mbox{\tiny $\mathbb{X}$}}}\Big|\mathds{E}\big(\widehat{g}_m(\mbox{\scriptsize $\boldsymbol{\chi}$}; \varphi) \big) - \psi_1(\mbox{\scriptsize $\boldsymbol{\chi}$}; \varphi)\Big| &=& \mathcal{O}\big(h^{\beta_1}\big). \label{EQ10}
\end{eqnarray}
Furthermore, Lemma 8 and Corollary 9 of Ferraty et al (2010) imply that under assumptions (A1) and (A3)\,--\,(A5), one has
\begin{equation}
\sup_{\mbox{\tiny $\boldsymbol{\chi}$}\in\, \mathcal{S}_{\mbox{\tiny $\mathbb{X}$}}}\Big|1-\widehat{f}_m(\mbox{\scriptsize $\boldsymbol{\chi}$})\Big| =\,\mathcal{O}_{a.co.}
\left(
\sqrt{\frac{\log[
		\mathcal{N}_{\tau_m}(\mathcal{S}_{\mbox{\tiny $\mathbb{X}$}})]}{ m\phi(h)}}\right)~~~~\mbox{and} ~~~~
\sum_{m=1}^{\infty}\mathds{P}\left\{\inf_{\mbox{\tiny $\boldsymbol{\chi}$}\in\, \mathcal{S}_{\mbox{\tiny $\mathbb{X}$}}} \widehat{f}_m(\mbox{\scriptsize $\boldsymbol{\chi}$}) \,<\,\frac{1}{2}\right\} \,<\, \infty. \label{EQ11}
\end{equation}
Putting together (\ref{EQQ11}), (\ref{EQ9}), (\ref{EQ10}), and (\ref{EQ11}), one finds 
\begin{eqnarray}
\sup_{\varphi \in \mathcal{F}_{\varepsilon_n}} \sup_{\mbox{\tiny $\boldsymbol{\chi}$}\in\, \mathcal{S}_{\mbox{\tiny $\mathbb{X}$}}}\Big|\widehat{\psi}_{m,1}(\mbox{\scriptsize $\boldsymbol{\chi}$}; \varphi, h)-\psi_1(\mbox{\scriptsize $\boldsymbol{\chi}$}; \varphi)
\Big| &=& \mathcal{O}\big(h^{\beta_1}\big) + \mathcal{O}_{a.co.}\left(\sqrt{\frac{\log\big[\mathcal{N}_{\varepsilon_n}(\mathcal{F}) \vee \mathcal{N}_{\tau_m}(\mathcal{S}_{\mbox{\tiny $\mathbb{X}$}})\big]}{m\phi(h)}}
\right). \label{EQ12}
\end{eqnarray}
This completes the proof of (\ref{EQ95A}) of Lemma \ref{MAINN-1} for the case of $k$\,=\,1. The case of $k$\,=\,2 is easier since it amounts to using (\ref{PSI12.hat}) with $k$\,=\,2.  The proof of (\ref{EQ95B}) of Lemma \ref{MAINN-1} is similar (and in fact easier) and will not be given.

\hfill $\Box$

\begin{lem} \label{LEM-1}  
Let $\widehat{\mathcal{T}}_{m,\varphi,h}$ be the classifier in (\ref{gFin}). Also, let  $L^*$ be the misclassification error of the best classifier as given by (\ref{LnL*}). Then,  under assumptions (A0) -- (A8), 
\begin{eqnarray} \label{Bound5}
\inf_{\varphi\in \mathcal{F}_{\varepsilon_n}} L_m\big(\widehat{\mathcal{T}}_{m,\varphi,h}\big) -L^* &=& \mathcal{O}(h^{\alpha}) 	+ \mathcal{O}(\varepsilon_n)
		+ \mathcal{O}_{a.\,co.}\left( \sqrt{\frac{\log[\mathcal{N}_{\tau_m}(\mathcal{S}_{\mbox{\tiny $\mathbb{X}$}})]}{m\cdot \phi(h)}} \right).
\end{eqnarray}
where  $\alpha$ is a positive constant not depending on $m$ or $n$.
\end{lem}

\vspace{2mm}\noindent
PROOF OF LEMMA \ref{LEM-1}.    

\vspace{1mm}
\noindent
Let $\varphi'\in \mathcal{F}_{\varepsilon_n}$ 
be such that $\varphi^* \in B\big( \varphi' , \varepsilon_n\big)$, where $\varphi^*$ is as in (\ref{NonIgnore}) and $B\big( \varphi' , \varepsilon_n\big)$ is the ball of functions centered at $\varphi'$ with radius $\varepsilon_n$. 
Clearly such a $\varphi'$ exists because $\varphi^*\in \mathcal{F}$ and $\mathcal{F}_{\varepsilon_n}$ is an $\varepsilon_n$-cover of $\mathcal{F}$. Now let $\widehat{\mathcal{R}}_m(\mbox{\large $\boldsymbol{\chi}$}; \varphi,h)$ and $\mathcal{R}(\mbox{\scriptsize $\boldsymbol{\chi}$}; \varphi^*)$ be as in (\ref{mhat3}) and (\ref{repr2}) and observe that
\begin{eqnarray}
&& \inf_{\varphi\in \mathcal{F}_{\varepsilon_n}} L_m\big(\widehat{\mathcal{T}}_{m,\varphi,h}\big) -L^* ~\leq~ L_m\big(\widehat{\mathcal{T}}_{m,\varphi',h}\big) -L^* \nonumber\\
&& ~~\leq\, 2E\left[ \Big|  \widehat{\mathcal{R}}_m(\mbox{\large $\boldsymbol{\chi}$}; \varphi',h) - \mathcal{R}(\mbox{\large $\boldsymbol{\chi}$}; \varphi^*)  \Big|\Big|\mathbb{D}_m\right],~~~\mbox{(by Lemma \ref{BASIC})} \nonumber\\
&& ~~\leq\,
2E\left[ \Big|  \widehat{\mathcal{R}}_m(\mbox{\large $\boldsymbol{\chi}$}; \varphi',h) - \widehat{\mathcal{R}}_m(\mbox{\large $\boldsymbol{\chi}$}; \varphi^*,h) \Big|\Big|\mathbb{D}_m\right]
+   2E\left[ \Big|  \widehat{\mathcal{R}}_m(\mbox{\large $\boldsymbol{\chi}$}; \varphi^*,h) - \mathcal{R}(\mbox{\large $\boldsymbol{\chi}$}; \varphi^*)  \Big|\Big|\mathbb{D}_m\right] \nonumber \\
&& ~~=:\, {\bf I}_m + {\bf I\!I}_m . \label{ImIIm}
\end{eqnarray}
Also, observe that
\begin{eqnarray}
&& \left|  \widehat{\mathcal{R}}_m(\mbox{\scriptsize $\boldsymbol{\chi}$}; \varphi',h) - \widehat{\mathcal{R}}_m(\mbox{\scriptsize $\boldsymbol{\chi}$}; \varphi^*,h)  \right|
~\leq~ \left|
\frac{\widehat{\psi}_{m,1}(\mbox{\scriptsize $\boldsymbol{\chi}$}; \varphi',h)}{\widehat{\psi}_{m,2}(\mbox{\scriptsize $\boldsymbol{\chi}$}; \varphi',h)} - 
\frac{\widehat{\psi}_{m,1}(\mbox{\scriptsize $\boldsymbol{\chi}$}; \varphi^*,h)}{\widehat{\psi}_{m,2}(\mbox{\scriptsize $\boldsymbol{\chi}$}; \varphi^*,h)}
\right|\nonumber\\
&&~~\leq~ 
\left|
\frac{\widehat{\psi}_{m,1}(\mbox{\scriptsize $\boldsymbol{\chi}$}; \varphi',h)}{\widehat{\psi}_{m,2}(\mbox{\scriptsize $\boldsymbol{\chi}$}; \varphi',h)} - 
\frac{\psi_{1}(\mbox{\scriptsize $\boldsymbol{\chi}$}; \varphi')}{\psi_{2}(\mbox{\scriptsize $\boldsymbol{\chi}$}; \varphi')}
\right| + 
\left|
\frac{\widehat{\psi}_{m,1}(\mbox{\scriptsize $\boldsymbol{\chi}$}; \varphi^*,h)}{\widehat{\psi}_{m,2}(\mbox{\scriptsize $\boldsymbol{\chi}$};\varphi^*,h)} - 
\frac{\psi_{1}(\mbox{\scriptsize $\boldsymbol{\chi}$}; \varphi^*)}{\psi_{2}(\mbox{\scriptsize $\boldsymbol{\chi}$}; \varphi^*)}
\right| +
\left|
\frac{ \psi_{1}(\mbox{\scriptsize $\boldsymbol{\chi}$}; \varphi')}{\psi_{2}(\mbox{\scriptsize $\boldsymbol{\chi}$}; \varphi')}  - 
\frac{\psi_{1}(\mbox{\scriptsize $\boldsymbol{\chi}$}; \varphi^*)}{\psi_{2}(\mbox{\scriptsize $\boldsymbol{\chi}$}; \varphi^*)}
\right|  ~~~ \nonumber \\[3pt]
&&~~=:~ {\bf I}_{m,1}(\mbox{\scriptsize $\boldsymbol{\chi}$}) + {\bf I}_{m,2}(\mbox{\scriptsize $\boldsymbol{\chi}$}) + {\bf I}_{3}(\mbox{\scriptsize $\boldsymbol{\chi}$}),  \label{Im123}
\end{eqnarray}
where $\psi_k$, $k=1,2$, is as in (\ref{psieta}).  Therefore, 
\begin{eqnarray}
{\bf I}_{m} &\leq& 2\,\sum_{k=1}^2 E\left[ {\bf I}_{m,k}(\mbox{\large $\boldsymbol{\chi}$}) \big| \mathbb{D}_m\right] + E\left[{\bf I}_{3}(\mbox{\large $\boldsymbol{\chi}$}) \right]. \label{IntIm} 
\end{eqnarray}
However, by the definition of $\psi_{2}(\mbox{\scriptsize $\boldsymbol{\chi}$}; \varphi^*)$ in (\ref{psieta}) and $\varrho_0$  in assumption (A7), one has
\begin{eqnarray}              
{\bf I}_{3}(\mbox{\scriptsize $\boldsymbol{\chi}$}) &=& \left|
- \frac{ \psi_{1}(\mbox{\scriptsize $\boldsymbol{\chi}$}; \varphi')}{\psi_{2}(\mbox{\scriptsize $\boldsymbol{\chi}$}; \varphi')}\cdot \frac{ \psi_{2}(\mbox{\scriptsize $\boldsymbol{\chi}$}; \varphi') - \psi_{2}(\mbox{\scriptsize $\boldsymbol{\chi}$}; \varphi^*)}{\psi_{2}(\mbox{\scriptsize $\boldsymbol{\chi}$}; \varphi^*)} +
\frac{ \psi_{1}(\mbox{\scriptsize $\boldsymbol{\chi}$}; \varphi') - \psi_{1}(\mbox{\scriptsize $\boldsymbol{\chi}$}; \varphi^*)}{\psi_{2}(\mbox{\scriptsize $\boldsymbol{\chi}$}; \varphi^*)}\right|\nonumber\\
&\leq& \frac{1}{\varrho_0} \Big|
\psi_{2}(\mbox{\scriptsize $\boldsymbol{\chi}$}; \varphi') - \psi_{2}(\mbox{\scriptsize $\boldsymbol{\chi}$}; \varphi^*)\Big|     +
\frac{1}{\varrho_0} \Big|\psi_{1}(\mbox{\scriptsize $\boldsymbol{\chi}$}; \varphi') - \psi_{1}(\mbox{\scriptsize $\boldsymbol{\chi}$}; \varphi^*)   \Big|. \label{Im3}
\end{eqnarray}
Furthermore, 
$|\psi_1(\mbox{\scriptsize $\boldsymbol{\chi}$}; \varphi') - \psi_1(\mbox{\scriptsize $\boldsymbol{\chi}$}; \varphi^*)| \leq E\left[|\Delta Y| \big|\varphi'(Y)-\varphi^*(Y)\big| \big|\mbox{\large $\boldsymbol{\chi}$}=\mbox{\scriptsize $\boldsymbol{\chi}$} \right] \leq \sup_{0\leq y \,\leq 1}\big|\varphi'(y)-\varphi^*(y)\big|$ and, similarly, $|\psi_2(\mbox{\scriptsize $\boldsymbol{\chi}$}; \varphi')-\psi_2(\mbox{\scriptsize $\boldsymbol{\chi}$}; \varphi^*)|\,\leq\, \sup_{0\leq y \,\leq 1}\big|\varphi'(y)-\varphi^*(y)\big|$\,. Consequently
\begin{eqnarray}
E\left[ {\bf I}_{3}(\mbox{\large $\boldsymbol{\chi}$})\right] &\leq&  \frac{1}{\varrho_0} E\Big|
\psi_{2}(\mbox{\large $\boldsymbol{\chi}$}; \varphi') - \psi_{2}(\mbox{\large $\boldsymbol{\chi}$}; \varphi^*)\Big|     +
\frac{1}{\varrho_0} E\Big|\psi_{1}(\mbox{\large $\boldsymbol{\chi}$}; \varphi') - \psi_{1}(\mbox{\large $\boldsymbol{\chi}$}; \varphi^*)   \Big| \nonumber\\
&\leq& (2/\varrho_0) \sup_{0\leq y \,\leq 1}\big|\varphi'(y)-\varphi^*(y)\big|~\leq~  C_{10} \cdot\varepsilon_n, ~~~~\mbox{(because $\varphi^* \in B\big( \varphi' , \varepsilon_n\big)$).}~~~~~ \label{IntIm3}
\end{eqnarray}
Next, to deal with the terms $E\left[ {\bf I}_{m,k}(\mbox{\large $\boldsymbol{\chi}$}) \big| \mathbb{D}_m\right]$, $k=1,2$, in (\ref{IntIm}), we first note that
\begin{eqnarray*}
{\bf I}_{m,1}(\mbox{\scriptsize $\boldsymbol{\chi}$}) &=& 
\left|
-\, \frac{ \widehat{\psi}_{m,1}(\mbox{\scriptsize $\boldsymbol{\chi}$}; \varphi',h)}{\widehat{\psi}_{m,2}(\mbox{\scriptsize $\boldsymbol{\chi}$}; \varphi',h)}\cdot \frac{ \widehat{\psi}_{m,2}(\mbox{\scriptsize $\boldsymbol{\chi}$}; \varphi',h) - \psi_{2}(\mbox{\scriptsize $\boldsymbol{\chi}$}; \varphi')}{\psi_{2}(\mbox{\scriptsize $\boldsymbol{\chi}$}; \varphi')} +
\frac{ \widehat{\psi}_{m,1}(\mbox{\scriptsize $\boldsymbol{\chi}$}; \varphi',h) - \psi_{1}(\mbox{\scriptsize $\boldsymbol{\chi}$}; \varphi')}{\psi_{2}(\mbox{\scriptsize $\boldsymbol{\chi}$}; \varphi')}\right|,\\[2pt]
&\leq& \frac{1}{\varrho_0}\, \sum_{k=1}^2\Big|
\widehat{\psi}_{m,k}(\mbox{\scriptsize $\boldsymbol{\chi}$}; \varphi',h) - \psi_{k}(\mbox{\scriptsize $\boldsymbol{\chi}$}; \varphi')\Big|,
\end{eqnarray*}
where the last line follows because 
$|\widehat{\psi}_{m,1}(\mbox{\scriptsize $\boldsymbol{\chi}$}; \varphi',h) / \widehat{\psi}_{m,2}(\mbox{\scriptsize $\boldsymbol{\chi}$}; \varphi',h)| \leq 1$. 
Similarly,
\begin{eqnarray*}
{\bf I}_{m,2}(\mbox{\scriptsize $\boldsymbol{\chi}$})
&\leq& \frac{1}{\varrho_0}\, \sum_{k=1}^2\Big|
\widehat{\psi}_{m,k}(\mbox{\scriptsize $\boldsymbol{\chi}$}; \varphi^*,h) - \psi_{k}(\mbox{\scriptsize $\boldsymbol{\chi}$}; \varphi^*)\Big|.
\end{eqnarray*}
Therefore, by (\ref{IntIm}),
\begin{eqnarray}
{\bf I}_{m} &\leq& 2\,C_{10}\, \varepsilon_n + \frac{2}{\varrho_0} \sum_{k=1}^2 E\left[\big| \widehat{\psi}_{m,k}(\mbox{\large $\boldsymbol{\chi}$};\varphi',h)-\psi_k(\mbox{\large $\boldsymbol{\chi}$};\varphi') \big|\Big| \mathbb{D}_m \right]\nonumber\\
&&~~~~~~~~~~~~~~~~~~ + \frac{2}{\varrho_0} \sum_{k=1}^2 E\left[\big| \widehat{\psi}_{m,k}(\mbox{\large $\boldsymbol{\chi}$};\varphi^*,h)-\psi_k(\mbox{\large $\boldsymbol{\chi}$};\varphi^*) \big|\Big| \mathbb{D}_m \right].
\label{69-N2} 
\end{eqnarray}
Next, to deal with the term ${\bf I\!I}_m$ in (\ref{ImIIm}), first note that by the definitions of $\mathcal{R}(\mbox{\scriptsize $\boldsymbol{\chi}$}; \varphi^*)$ and $\widehat{\mathcal{R}}_{m}(\mbox{\scriptsize $\boldsymbol{\chi}$};\varphi,h)$ in (\ref{repr2}) and (\ref{mhat3}) we have
\begin{eqnarray}
\left| \widehat{\mathcal{R}}_{m}(\mbox{\scriptsize $\boldsymbol{\chi}$};\varphi^*,h) - \mathcal{R}(\mbox{\scriptsize $\boldsymbol{\chi}$}; \varphi^*)\right|
&\leq& \left| \widehat{\eta}_{m,1}(\mbox{\scriptsize $\boldsymbol{\chi}$};h) - \eta_1(\mbox{\scriptsize $\boldsymbol{\chi}$})\right| +
\Bigg|
\frac{\widehat{\psi}_{m,1}(\mbox{\scriptsize $\boldsymbol{\chi}$}; \varphi^*,h)}{\widehat{\psi}_{m,2}(\mbox{\scriptsize $\boldsymbol{\chi}$}; \varphi^*,h)}\left(1-\widehat{\eta}_{m,2}(\mbox{\scriptsize $\boldsymbol{\chi}$};h)\right) \nonumber\\
&& ~~~~~~~~ - \frac{\psi_1(\mbox{\scriptsize $\boldsymbol{\chi}$}; \varphi^*)}{\psi_2(\mbox{\scriptsize $\boldsymbol{\chi}$}; \varphi^*)} (1-\eta_2 (\mbox{\scriptsize $\boldsymbol{\chi}$}))\Bigg|~=:~
{\bf I\!I}_{m,1}(\mbox{\scriptsize $\boldsymbol{\chi}$}) + {\bf I\!I}_{m,2}(\mbox{\scriptsize $\boldsymbol{\chi}$}), \label{IIm12}
\end{eqnarray}
where $\psi_k$ and $\eta_k$, $k=1,2$, are as in (\ref{psieta}) and $\widehat{\eta}_{m,k}$ and $\widehat{\psi}_{m,k}$ are given by (\ref{PSI12.hat}) and (\ref{ETA12m.hat}). 
Furthermore observe that 
\begin{eqnarray}  \label{72-N}
{\bf I\!I}_{m,2}(\mbox{\scriptsize $\boldsymbol{\chi}$}) &\leq& 
\left|
\frac{\widehat{\psi}_{m,1}(\mbox{\scriptsize $\boldsymbol{\chi}$}; \varphi*,h)}{\widehat{\psi}_{m,2}(\mbox{\scriptsize $\boldsymbol{\chi}$}; \varphi*,h)} - 
\frac{\psi_{1}(\mbox{\scriptsize $\boldsymbol{\chi}$}; \varphi*)}{\psi_{2}(\mbox{\scriptsize $\boldsymbol{\chi}$}; \varphi*)}
\right| + \left| \widehat{\eta}_{m,2}(\mbox{\scriptsize $\boldsymbol{\chi}$};h) - \eta_2(\mbox{\scriptsize $\boldsymbol{\chi}$}) \right|  \nonumber\\
&\leq&  \frac{1}{\varrho_0}\, \sum_{k=1}^2\Big|
\widehat{\psi}_{m,k}(\mbox{\scriptsize $\boldsymbol{\chi}$}; \varphi^*,h) - \psi_{k}(\mbox{\scriptsize $\boldsymbol{\chi}$}; \varphi^*)\Big| + \left| \widehat{\eta}_{m,2}(\mbox{\scriptsize $\boldsymbol{\chi}$};h) - \eta_2(\mbox{\scriptsize $\boldsymbol{\chi}$}) \right|.
\end{eqnarray} 
Therefore, by (\ref{ImIIm}), (\ref{IIm12}), and (\ref{72-N}),
\begin{eqnarray} \label{72-N2}
{\bf I\!I}_m &\leq & \frac{2}{\varrho_0}\, \sum_{k=1}^2E\left[\Big|
\widehat{\psi}_{m,k}(\mbox{\large $\boldsymbol{\chi}$}; \varphi^*,h) - \psi_{k}(\mbox{\large $\boldsymbol{\chi}$}; \varphi^*)\Big|\Big|\mathbb{D}_m\right] +
2\, \sum_{k=1}^2 E\left[
 \left| \widehat{\eta}_{m,k}(\mbox{\large $\boldsymbol{\chi}$};h) - \eta_k(\mbox{\large $\boldsymbol{\chi}$}) \right|\Big|\mathbb{D}_m\right].~~~
\end{eqnarray}
Furthermore, 
since $\varphi'\in \mathcal{F}_{\varepsilon_n}$ 
is such that $\varphi^* \in B\big( \varphi' , \varepsilon_n\big)$,
it is straightforward to see that for $k=1,2$,
\begin{eqnarray*}
\Big|\widehat{\psi}_{m,k}(\mbox{\scriptsize $\boldsymbol{\chi}$}; \varphi^*,h) - \psi_{k}(\mbox{\scriptsize $\boldsymbol{\chi}$}; \varphi^*)\Big|
&\leq& \Big|\widehat{\psi}_{m,k}(\mbox{\scriptsize $\boldsymbol{\chi}$}; \varphi',h) - \psi_{k}(\mbox{\scriptsize $\boldsymbol{\chi}$}; \varphi')\Big| +
\Big|\widehat{\psi}_{m,k}(\mbox{\scriptsize $\boldsymbol{\chi}$};\, \mbox{$\varphi^*$$-$$\varphi'$},h)\Big| + \Big| \psi_{k}(\mbox{\scriptsize $\boldsymbol{\chi}$};\, \mbox{$\varphi^*$$-$$\varphi'$})\Big|\\
&\leq& \Big|\widehat{\psi}_{m,k}(\mbox{\scriptsize $\boldsymbol{\chi}$}; \varphi',h) - \psi_{k}(\mbox{\scriptsize $\boldsymbol{\chi}$}; \varphi')\Big| + 2 \varepsilon_n.
\end{eqnarray*}
Putting together the above result together with (\ref{ImIIm}), (\ref{69-N2}), and (\ref{72-N2}), we find
\begin{eqnarray*} 
\inf_{\varphi\in \mathcal{F}_{\varepsilon_n}} L_m\big(\widehat{\mathcal{T}}_{m,\varphi,h}\big) -L^* &\leq & C_{11}\cdot\varepsilon_n + C_{12}\, \sum_{k=1}^2E\left[\Big|
\widehat{\psi}_{m,k}(\mbox{\large $\boldsymbol{\chi}$}; \varphi',h) - \psi_{k}(\mbox{\large $\boldsymbol{\chi}$}; \varphi')\Big|\Big|\mathbb{D}_m\right] \\
&&~~~~~~~~~~~~~~~~~~+
C_{13}\, \sum_{k=1}^2 E\left[
 \left| \widehat{\eta}_{m,k}(\mbox{\large $\boldsymbol{\chi}$};h) - \eta_k(\mbox{\large $\boldsymbol{\chi}$}) \right|\Big|\mathbb{D}_m\right]\\
&=&  \mathcal{O}(\varepsilon_n) + \mathcal{O}(h^{\beta_1 \wedge \beta_2})
		+ \mathcal{O}_{a.\,co.}\left( \sqrt{\frac{\log[\mathcal{N}_{\tau_m}(\mathcal{S}_{\mbox{\tiny $\mathbb{X}$}})]}{m\cdot \phi(h)}} \right),
\end{eqnarray*}
where the last line follows from Lemma \ref{MAINN-1}.
This completes the proof of Lemma \ref{LEM-1}.

\hfill  $\Box$

\begin{lem}\label{LEM-VAP}
For each $\varphi\in \mathcal{F}$ let $\mathcal{C}_m(\varphi)$ be the family of classifiers in (\ref{CMPHI}). Also let $\mathcal{A}_{_{\mathcal{C}_m(\varphi)}}$ be the collection 
of all sets of the form (\ref{AMPHI}). 
For each $A_{m, \varphi, h}\in \mathcal{A}_{_{\mathcal{C}_m(\varphi)}}$ define
\begin{eqnarray}
\mbox{\Large $\nu$}\big(A_{m, \varphi, h}\big|\,\mathbb{D}_m\big) &=& P\big\{   (\mbox{\large $\boldsymbol{\chi}$}, Y) \in A_{m, \varphi, h} \big|\, \mathbb{D}_m\big\} \label{NEW-A1} \\
\mbox{\Large $\nu$}_{\ell} \big(A_{m, \varphi, h}\big) &=& \frac{1}{\ell} 
\sum_{i\in \boldsymbol{{\cal I}}_\ell}
\big[\Delta_i +(1-\Delta_i) \delta_i/p_n\big]\cdot \mbox{\Large $\mathds{1}$}\big\{(\mbox{\large $\boldsymbol{\chi}$}_i, Y_i) \in A_{m, \varphi, h}\big\}. \label{NEW-A2}
\end{eqnarray}
Then for every $t>0$, 
\begin{equation}
P\left\{ \sup_{A_{m, \varphi, h} \,\in\, \mathcal{A}_{_{\mathcal{C}_m(\varphi)}}}\Big| \mbox{\Large $\nu$}_{\ell} \big(A_{m, \varphi, h})  - \mbox{\Large $\nu$}\big(A_{m, \varphi, h}\big|\,\mathbb{D}_m\big)\Big| > t \,\bigg|\, \mathbb{D}_m   \right\} ~\leq~ 8\,\mathcal{S}( 
\mathcal{A}_{_{\mathcal{C}_m(\varphi)}},\, \ell) \cdot e^{-\ell\, p_n^2 t^2/2},
\end{equation}
where $\mathcal{S}(\mathcal{A}_{_{\mathcal{C}_m(\varphi)}},\, \ell)$ is the $\ell^{\mbox{\tiny th}}$  shatter coefficient of $\mathcal{A}_{_{\mathcal{C}_m(\varphi)}}$.
\end{lem}

\vspace{2mm}\noindent
PROOF OF LEMMA \ref{LEM-VAP}    

\vspace{1mm}
\noindent
Our proof below is based on standard symmetrization arguments to deal with the supremum of empirical processes; see, for example, Dudley (1978; p.\,925), Pollard (1984; Sec.\,II.3), and van der Vaart and Wellner (1996; Sec.\,2.3). 

\vspace{2mm}\noindent
{\it (i) The first symmetrization (w.r.t. a hypothetical sample).} \\
Let $\mathbb{D}'_\ell = \{(\mbox{\large $\boldsymbol{\chi}$}'_1, Y'_1, \Delta'_1),\cdots,(\mbox{\large $\boldsymbol{\chi}$}'_\ell, Y'_\ell, \Delta'_\ell)\}$ be a ghost sample, i.e., $(\mbox{\large $\boldsymbol{\chi}$}'_i, Y'_i, \Delta'_i)\stackrel{\mbox{\tiny iid}}{=} (\mbox{\large $\boldsymbol{\chi}$}_1, Y_1, \Delta_1)$, independent of  $(\delta_1,\cdots,\delta_\ell)$. Also, let $(\delta'_1,\cdots,\delta'_\ell)$ be a ghost sample independent of $\mathbb{D}'_\ell$, $\mathbb{D}_n$, and $(\delta_1,\cdots,\delta_\ell)$, and define 
\[
\mbox{\Large $\nu$}'_{\ell} \big(A_{m, \varphi, h}\big) \,=\, \frac{1}{\ell} 
\sum_{i=1}^{\ell}
\big[\Delta'_i +(1-\Delta'_i) \delta'_i\,/p_n\big]\cdot \mbox{\Large $\mathds{1}$}\big\{(\mbox{\large $\boldsymbol{\chi}$}'_i, Y'_i) \in A_{m, \varphi, h}\big\}. 
\]
We note that $\mbox{\Large $\nu$}'_{\ell} \big(A_{m, \varphi, h}\big)$ is not exactly the ghost  counterpart of (\ref{NEW-A2}) as they both use the set $A_{m, \varphi, h}$ which depends on $\mathbb{D}_m$. Next, fix the data $\mathbb{D}_n$ and observe that for any $\varphi\in\mathcal{F}$, if\, $\sup_{A_{m, \varphi, h} \in \mathcal{A}_{_{\mathcal{C}_m(\varphi)}}}\big| \mbox{\Large $\nu$}_{\ell} \big(A_{m, \varphi, h}\big)  - \mbox{\Large $\nu$}\big(A_{m, \varphi, h}\big|\,\mathbb{D}_m\big)\big| > t$ then there is at least one set $A^{^{(t)}}_{m,\varphi,h}$ in 
$\mathcal{A}_{_{\mathcal{C}_m(\varphi)}}$ that depends on $\mathbb{D}_n$ (but not $\mathbb{D}'_{\ell}$) such that $\big| \mbox{\Large $\nu$}_{\ell} \big(A^{^{(t)}}_{m,\varphi,h}\big)  - \mbox{\Large $\nu$}\big(A^{^{(t)}}_{m,\varphi,h}\big|\,\mathbb{D}_n\big)\big| > t$, where 
\[
\mbox{\Large $\nu$}\big(A^{^{(t)}}_{m,\varphi,h}\big|\,\mathbb{D}_n\big) = P\big\{   (\mbox{\large $\boldsymbol{\chi}$}, Y) \in A^{^{(t)}}_{m,\varphi,h} \big|\, \mathbb{D}_n\big\}.
\]
Also, observe that 
\begin{eqnarray*}
&& P\left\{ \Big| \mbox{\Large $\nu$}'_{\ell} \big(A^{^{(t)}}_{m,\varphi,h}\big)  - \mbox{\Large $\nu$}\big(A^{^{(t)}}_{m,\varphi,h}\big|\,\mathbb{D}_n\big)\Big| \,<\, \frac{t}{2} \,\Big|\,\mathbb{D}_n  \right\} \\
&& ~\geq~ 1-\sup_{A_{m, \varphi, h} \,\in\, \mathcal{A}_{_{\mathcal{C}_m(\varphi)}}} P\left\{ \Big| \mbox{\Large $\nu$}'_{\ell} \big(A_{m, \varphi, h}\big)  - \mbox{\Large $\nu$}\big(A_{m, \varphi, h}\big|\,\mathbb{D}_m\big)\Big| \,\geq\, \frac{t}{2} \,\Big|\,\mathbb{D}_m  \right\} \\
&& ~\geq~ 1-\frac{4}{\ell t^2} \sup_{A_{m, \varphi, h} \,\in\, \mathcal{A}_{_{\mathcal{C}_m(\varphi)}}} 
\mbox{Var}\Big( \big[\Delta'_1 +(1-\Delta'_1) \delta'_1/p_n\big]\cdot 
\mbox{\Large $\mathds{1}$}\big\{(\mbox{\large $\boldsymbol{\chi}$}'_1, Y'_1) \in A_{m, \varphi, h}\big\}   \Big|\,\mathbb{D}_m\Big)\\
&& ~\geq~ 1-\frac{4}{\ell p_n^2 t^2}\cdot\frac{1}{4}~~~ (\mbox{because the variance of the indicator function is bounded by $\frac{1}{4}$})\\
&& ~\geq~ \frac{1}{2}, ~~~\mbox{whenever $\ell p_n^2 t^2 \geq 2$.}
\end{eqnarray*}
Therefore, for $\ell p_n^2 t^2 \geq 2$ (the theorem follows trivially when $\ell p_n^2 t^2 < 2$),
\begin{eqnarray}
\frac{1}{2} 
&\leq& P\left\{ \Big| \mbox{\Large $\nu$}'_{\ell} \big(A^{^{(t)}}_{m,\varphi,h}\big)  - \mbox{\Large $\nu$}\big(A^{^{(t)}}_{m,\varphi,h}\big|\,\mathbb{D}_n\big)\Big| \,<\, \frac{t}{2} \,\Big|\,\mathbb{D}_n  \right\} \nonumber\\
&\leq&  P\Bigg\{ -\Big| \mbox{\Large $\nu$}'_{\ell} \big(A^{^{(t)}}_{m,\varphi,h}\big)  -
\mbox{\Large $\nu$}_{\ell} \big(A^{^{(t)}}_{m,\varphi,h}\big)\Big| + 
\underbrace{\Big|\mbox{\Large $\nu$}_{\ell} \big(A^{^{(t)}}_{m,\varphi,h}\big) - \mbox{\Large $\nu$}\big(A^{^{(t)}}_{m,\varphi,h}\big|\,\mathbb{D}_n\big)\Big|}_{>\,t} \,<\, \frac{t}{2} \,\bigg|\,\mathbb{D}_n  \Bigg\} \nonumber\\
&\leq& P\Bigg\{\sup_{A_{m, \varphi, h} \in \mathcal{A}_{_{\mathcal{C}_m(\varphi)}}}\Big| \mbox{\Large $\nu$}'_{\ell} \big(A_{m, \varphi, h}\big)  -
\mbox{\Large $\nu$}_{\ell} \big(A_{m, \varphi, h}\big)\Big|  \,>\, \frac{t}{2} \,\bigg|\,
\mathbb{D}_n  \Bigg\}. \label{VAP-1}
\end{eqnarray}
But the far right and far left sides of (\ref{VAP-1}) do not depend on any particular $A_{m, \varphi, h}$  and the chain of inequalities between them remains valid on the set\,$\big\{\sup_{A_{m, \varphi, h} \in \mathcal{A}_{_{\mathcal{C}_m(\varphi)}}}
\big| \mbox{\Large $\nu$}_{\ell} (A_{m, \varphi, h})  - \mbox{\Large $\nu$}\big(A_{m, \varphi, h}|\,\mathbb{D}_m\big)\big| > t\big\}$. Therefore, integrating the two sides with respect to the distribution of $\mathbb{D}_\ell$ over this set yields
\begin{eqnarray}\label{VAP-2} 
&& P\left\{ \sup_{A_{m, \varphi, h} \in \mathcal{A}_{_{\mathcal{C}_m(\varphi)}}}\Big| \mbox{\Large $\nu$}_{\ell} \big(A_{m, \varphi, h})  - \mbox{\Large $\nu$}\big(A_{m, \varphi, h}\big|\,\mathbb{D}_m\big)\Big| > t \,\bigg|\, \mathbb{D}_m   \right\} \nonumber\\
&&~~~~~~~~~~~~~~~~~~~~~~~~~~~~~\leq~  \, 2\,P\Bigg\{\sup_{A_{m, \varphi, h} \in \mathcal{A}_{_{\mathcal{C}_m(\varphi)}}}\Big| \mbox{\Large $\nu$}'_{\ell} \big(A_{m, \varphi, h}\big)  - \mbox{\Large $\nu$}_{\ell} \big(A_{m, \varphi, h}\big)\Big|  \,>\, \frac{t}{2} \,\bigg|\,
\mathbb{D}_m  \Bigg\}. ~~~~~~~~~~~~
\end{eqnarray}
{\it (ii) The second symmetrization (w.r.t. an iid Rademacher sequence).}\\
In what follows, without loss  of generality we may assume $\mathbb{D}_{\ell} = \{(\mbox{\large $\boldsymbol{\chi}$}_1, Y_1, \Delta_1),\cdots,(\mbox{\large $\boldsymbol{\chi}$}_\ell, Y_\ell, \Delta_\ell)\}$, which can be achieved by a re-indexing of the observations in $\mathbb{D}_{\ell}$ (because the data are iid). Now, since $\mathbb{D}'_{\ell}$, $\mathbb{D}_{\ell}$, and $\mathbb{D}_{m}$ are all independent of each other, the joint distributions of the $\ell$-dimensional vectors
\[
\Big(
\big[\Delta_1 +(1-\Delta_1) \delta_1/p_n\big]\mbox{\Large $\mathds{1}$}\big\{(\mbox{\large $\boldsymbol{\chi}$}_1, Y_1) \in A_{m, \varphi, h}\big\},\cdots\cdot,\big[\Delta_\ell +(1-\Delta_\ell) \delta_\ell/p_n\big] \mbox{\Large $\mathds{1}$}\big\{(\mbox{\large $\boldsymbol{\chi}$}_\ell, Y_\ell) \in A_{m, \varphi, h}\big\}
\Big)
\]
and 
\[
\Big(
\big[\Delta'_1 +(1-\Delta'_1) \delta'_1/p_n\big]\mbox{\Large $\mathds{1}$}\big\{(\mbox{\large $\boldsymbol{\chi}$}'_1, Y'_1) \in A_{m, \varphi, h}\big\},\cdots\cdot,\big[\Delta'_\ell +(1-\Delta'_\ell) \delta'_\ell/p_n\big] \mbox{\Large $\mathds{1}$}\big\{(\mbox{\large $\boldsymbol{\chi}$}'_\ell, Y'_\ell) \in A_{m, \varphi, h}\big\}
\Big)
\]
are the same and this joint distribution is not affected if one randomly interchanges the corresponding components of these two vectors. Therefore, if we let $\sigma_1,\cdots, \sigma_{\ell}$ be an iid symmetric Rademacher sequence, i.e., $P\{\sigma_i = -1\} = 0.5 = P\{\sigma_i = +1\}$, then 
\begin{eqnarray}
(\ref{VAP-2}) \hspace{-1mm}&=& \hspace{-1mm}2 P\Bigg\{
\sup_{A_{m, \varphi, h} \in \mathcal{A}_{_{\mathcal{C}_m(\varphi)}}} \frac{1}{\ell}\bigg|
\sum_{i=1}^{\ell}\bigg[
\Big(\Delta'_i +\frac{(1-\Delta'_i) \delta'_i}{p_n}\Big)\mbox{\Large $\mathds{1}$}\big\{(\mbox{\large $\boldsymbol{\chi}$}'_i, Y'_i) \in A_{m, \varphi, h}\big\} \nonumber\\
&& ~~~~~~~~~~~~~~~~~~~~~~~~~~~~ - \Big(\Delta_i +\frac{(1-\Delta_i) \delta_i}{p_n}\Big)\mbox{\Large $\mathds{1}$}\big\{(\mbox{\large $\boldsymbol{\chi}$}_i, Y_i) \in A_{m, \varphi, h}\big\}\bigg]\bigg| > \frac{t}{2} \,\Bigg|\, \mathbb{D}_m
\Bigg\}\nonumber\\
\hspace{-1mm} &=& \hspace{-1mm} 2 P\Bigg\{
\sup_{A_{m, \varphi, h} \in \mathcal{A}_{_{\mathcal{C}_m(\varphi)}}} \frac{1}{\ell}\bigg|
\sum_{i=1}^{\ell}\sigma_i\cdot\bigg[
\Big(\Delta'_i +\frac{(1-\Delta'_i) \delta'_i}{p_n}\Big)\mbox{\Large $\mathds{1}$}\big\{(\mbox{\large $\boldsymbol{\chi}$}'_i, Y'_i) \in A_{m, \varphi, h}\big\} \nonumber\\
&& ~~~~~~~~~~~~~~~~~~~~~~~~~~~~ - \Big(\Delta_i +\frac{(1-\Delta_i) \delta_i}{p_n}\Big)\mbox{\Large $\mathds{1}$}\big\{(\mbox{\large $\boldsymbol{\chi}$}_i, Y_i) \in A_{m, \varphi, h}\big\}\bigg]\bigg| > \frac{t}{2} \,\Bigg|\, \mathbb{D}_m
\Bigg\}\nonumber\\
\hspace{-1mm}
&\leq&\hspace{-1mm} 4 E\Bigg[P\Bigg\{
\sup_{A_{m, \varphi, h} \in \mathcal{A}_{_{\mathcal{C}_m(\varphi)}}} \frac{1}{\ell}\bigg|
\sum_{i=1}^{\ell}\sigma_i\cdot
\Big(\Delta_i +\frac{(1-\Delta_i) \delta_i}{p_n}\Big)\mbox{\Large $\mathds{1}$}\big\{(\mbox{\large $\boldsymbol{\chi}$}_i, Y_i) \in A_{m, \varphi, h}\big\} \bigg| > \frac{t}{4} \,\Bigg|\, \mathbb{D}_n\Bigg\}\Bigg|\mathbb{D}_m
\Bigg] \nonumber\\\label{VAP-3}
\end{eqnarray}
However for fixed $(\mbox{\scriptsize $\boldsymbol{\chi}$}_1, y_1),\cdots, (\mbox{\scriptsize $\boldsymbol{\chi}$}_{\ell},y_{\ell})$ and fixed $\mathbb{D}_m$, the number of different vectors 
\[
\Big(
\mbox{\Large $\mathds{1}$}\big\{(\mbox{\scriptsize $\boldsymbol{\chi}$}_1, y_1) \in A_{m, \varphi, h}\big\}, \cdots\cdot, \mbox{\Large $\mathds{1}$}\big\{(\mbox{\scriptsize $\boldsymbol{\chi}$}_\ell, y_\ell) \in A_{m, \varphi, h}\big\}
\Big)
\]
obtained as $A_{m, \varphi, h}$ ranges over all possible sets in $\mathcal{A}_{_{\mathcal{C}_m(\varphi)}}$ is just the number of different sets in 
\[
\left\{ \big\{
(\mbox{\scriptsize $\boldsymbol{\chi}$}_1, y_1),\cdots, (\mbox{\scriptsize $\boldsymbol{\chi}$}_{\ell},y_{\ell})\big\}\cap A_{m, \varphi, h}\,\Big|
A_{m, \varphi, h} \in \mathcal{A}_{_{\mathcal{C}_m(\varphi)}}
\right\}
\]
which is bounded by $\mathcal{S}\big(  \mathcal{A}_{_{\mathcal{C}_m(\varphi)}}, \ell  \big)$,  the $\ell^{\text{th}}$ shatter coefficient of $\mathcal{A}_{_{\mathcal{C}_m(\varphi)}}$. Therefore
\begin{eqnarray}  \label{VAP-4}
&& P\Bigg\{
\sup_{A_{m, \varphi, h} \in \mathcal{A}_{_{\mathcal{C}_m(\varphi)}}} \frac{1}{\ell}\bigg|
\sum_{i=1}^{\ell}\sigma_i\cdot
\Big(\Delta_i +\frac{(1-\Delta_i) \delta_i}{p_n}\Big)\mbox{\Large $\mathds{1}$}\big\{(\mbox{\large $\boldsymbol{\chi}$}_i, Y_i) \in A_{m, \varphi, h}\big\} \bigg| > \frac{t}{4} \,\Bigg|\, \mathbb{D}_n\Bigg\}\nonumber\\
&&\,\leq\,\mathcal{S}\big(  \mathcal{A}_{_{\mathcal{C}_m(\varphi)}}, \ell  \big)
\sup_{A_{m, \varphi, h} \in \mathcal{A}_{_{\mathcal{C}_m(\varphi)}}} \hspace{-1mm}
P\Bigg\{
 \frac{1}{\ell}\bigg|
\sum_{i=1}^{\ell}\sigma_i\cdot
\Big(\Delta_i +\frac{(1-\Delta_i) \delta_i}{p_n}\Big)\mbox{\Large $\mathds{1}$}\big\{(\mbox{\large $\boldsymbol{\chi}$}_i, Y_i) \in A_{m, \varphi, h}\big\} \bigg| > \frac{t}{4} \,\Bigg|\, \mathbb{D}_n\Bigg\} \nonumber\\
\end{eqnarray}
But, conditional on $\mathbb{D}_n$, the terms $\sigma_i\cdot
\big(\Delta_i +(1-\Delta_i) \delta_i/p_n\big)\mbox{\Large $\mathds{1}$}\big\{(\mbox{\large $\boldsymbol{\chi}$}_i, Y_i) \in A_{m, \varphi, h}\big\}$, $i=1,\cdots,\ell$, are independent zero-mean random variables bounded by $-1/p_n$ and $+1/p_n$. Therefore, using Hoeffding's inequality, we have
\begin{eqnarray*}  
P\Bigg\{
 \frac{1}{\ell}\,\bigg|
\sum_{i=1}^{\ell}\sigma_i\cdot
\Big(\Delta_i +\frac{(1-\Delta_i) \delta_i}{p_n}\Big)\mbox{\Large $\mathds{1}$}\big\{(\mbox{\large $\boldsymbol{\chi}$}_i, Y_i) \in A_{m, \varphi, h}\big\} \bigg| > \frac{t}{4} \,\Bigg|\, \mathbb{D}_n\Bigg\} 
&\leq& 2\,e^{-\ell \,p_n^2 t^2/2}\,.
\end{eqnarray*}
This result together with (\ref{VAP-4}), (\ref{VAP-3}), and (\ref{VAP-2}) yield, for every $t>0$,
\begin{equation*}
P\left\{ \sup_{A_{m, \varphi, h} \in \mathcal{A}_{_{\mathcal{C}_m(\varphi)}}}\Big| \mbox{\Large $\nu$}_{\ell} \big(A_{m, \varphi, h})  - \mbox{\Large $\nu$}\big(A_{m, \varphi, h}\big|\,\mathbb{D}_m\big)\Big| > t \,\bigg|\, \mathbb{D}_m   \right\} ~\leq~ 8\,\mathcal{S}( 
\mathcal{A}_{_{\mathcal{C}_m(\varphi)}},\, \ell) \cdot e^{-\ell\, p_n^2 t^2/2},
\end{equation*}

\hfill
$\Box$

\vspace{3mm}\noindent
PROOF OF THEOREM \ref{THM-BBC}

\vspace{1mm}\noindent
To prove the theorem, consider the basic decomposition
\begin{eqnarray}
&& L_n(\widehat{\mathcal{T}}_{n, \widehat{\varphi},h}) - L^* \nonumber\\
&&=
\left[L_n(\widehat{\mathcal{T}}_{n, \widehat{\varphi},h}) - \widehat{L}_{m,\ell}(\widehat{\mathcal{T}}_{n, \widehat{\varphi},h})\right] + \left[\widehat{L}_{m,\ell}(\widehat{\mathcal{T}}_{n, \widehat{\varphi},h}) - \inf_{\varphi\in \mathcal{F}_{\varepsilon_n}} L_m\big(\widehat{\mathcal{T}}_{m,\varphi,h}\big)
\right] + \left[  \inf_{\varphi\in \mathcal{F}_{\varepsilon_n}} L_m\big(\widehat{\mathcal{T}}_{m,\varphi,h}\big) -L^*    \right] \nonumber\\
&&=:\, R_n(1) + R_n(2) + R_m(3), \label{Rn123}
\end{eqnarray}
where $L_m(\widehat{\mathcal{T}}_{m,\varphi,h})$ is given in Lemma \ref{BASIC} and $\widehat{L}_{m,\ell}$ is as in (\ref{NEW-Lhat}). But, first note that 
\begin{equation*}
R_n(1)\, \leq \,\sup_{\varphi\in \mathcal{F}_{\varepsilon_n}}\left|L_m(\widehat{\mathcal{T}}_{m, \varphi,h}) - \widehat{L}_{m,\ell}(\widehat{\mathcal{T}}_{m, \varphi,h})\right|.
\end{equation*}
Next, let $\breve{\varphi} = \argmin_{\varphi \in \mathcal{F}_{\varepsilon_n}}L_m(\widehat{\mathcal{T}}_{m,\varphi,h})$; here,  $\breve{\varphi}$ depends on $\mathbb{D}_m$ (because $\widehat{\mathcal{T}}_{m,\varphi,h}$ does). Hence, by the definition  $\widehat{L}_{m,\ell}$ in (\ref{NEW-Lhat}) one can write
\begin{eqnarray*}
R_n(2) &=& \widehat{L}_{m,\ell}(\widehat{\mathcal{T}}_{n, \widehat{\varphi},h}) -  L_m\big(\widehat{\mathcal{T}}_{m,\breve{\varphi},h}\big) \,\leq \,
\widehat{L}_{m,\ell}(\widehat{\mathcal{T}}_{n, \breve{\varphi},h}) -  L_m\big(\widehat{\mathcal{T}}_{m,\breve{\varphi},h}\big) \\
&\leq& \sup_{\varphi\in \mathcal{F}_{\varepsilon_n}}\left|
\widehat{L}_{m,\ell}(\widehat{\mathcal{T}}_{m,\varphi,h}) -  L_m\big(\widehat{\mathcal{T}}_{m,\varphi,h}\big)\right|.
\end{eqnarray*}
Thus, 
\begin{eqnarray}
\big| R_n(1) + R_n(2)\big| &\leq & 2 \sup_{\varphi\in \mathcal{F}_{\varepsilon_n}}\left|
\widehat{L}_{m,\ell}(\widehat{\mathcal{T}}_{m, \varphi,h}) -  L_m\big(\widehat{\mathcal{T}}_{m,\varphi,h}\big)\right|.
\end{eqnarray}
Now observe that for any $t>0$, 
\begin{eqnarray}
&& P\left\{ \big| R_n(1) + R_n(2) \big| > t  \right\} \nonumber\\
&& \leq
P\left\{  \sup_{\varphi\in \mathcal{F}_{\varepsilon_n}}\left|
\widehat{L}_{m,\ell}(\widehat{\mathcal{T}}_{m, \varphi,h}) -  L_m\big(\widehat{\mathcal{T}}_{m,\varphi,h}\big)
\right| > \frac{t}{2}   \right\} \label{Bound86}\\
&& = P\left\{  \sup_{\varphi\in \mathcal{F}_{\varepsilon_n}}\left| \frac{1}{\ell}
\sum_{i\in \boldsymbol{{\cal I}}_\ell}\left[
\left(\Delta_i +\frac{(1-\Delta_i) \delta_i}{p_n}\right)
\mbox{\Large $\mathds{1}$}\hspace{-0.5mm}\big\{\widehat{\mathcal{T}}_{m,\varphi,h}(\mbox{\large $\boldsymbol{\chi}$}_i)\neq Y_i\big\}\right] - L_m\big(\widehat{\mathcal{T}}_{m,\varphi,h}\big) \right| >\frac{t}{2}
 \right\}~~~~ \label{Bound6-A}
\end{eqnarray}
On the other hand, for each $(\mbox{\large $\boldsymbol{\chi}$}_i, Y_i,\Delta_i) \in\mathbb{D}_{\ell}$ we have
\begin{eqnarray*}
&& E\left[ \left(\Delta_i +\frac{(1-\Delta_i) \delta_i}{p_n}\right)
\mbox{\Large $\mathds{1}$}\hspace{-0.5mm}\big\{\widehat{\mathcal{T}}_{m,\varphi,h}(\mbox{\large $\boldsymbol{\chi}$}_i)\neq Y_i\big\} \,\Big| \mathbb{D}_m\right]\\
&& = E\left[ E\left\{ 
\left(\Delta_i +\frac{(1-\Delta_i) \delta_i}{p_n}\right)
\mbox{\Large $\mathds{1}$}\hspace{-0.5mm}\big\{\widehat{\mathcal{T}}_{m,\varphi,h}(\mbox{\large $\boldsymbol{\chi}$}_i)\neq Y_i\big\}\,\bigg| \mathbb{D}_m, \mbox{\large $\boldsymbol{\chi}$}_i, Y_i, \delta_i
\right\}\,\bigg| \mathbb{D}_m\right]\\
&& = E\left[ \mbox{\Large $\mathds{1}$}\hspace{-0.5mm}\big\{\widehat{\mathcal{T}}_{m,\varphi,h}(\mbox{\large $\boldsymbol{\chi}$}_i)\neq Y_i\big\}\, \pi_{\varphi^*}(\mbox{\large $\boldsymbol{\chi}$}_i, Y_i)\,\Big|\mathbb{D}_m\right]
 + \frac{E(\delta_i)}{p_n} \left[\mbox{\Large $\mathds{1}$}\hspace{-0.5mm}\big\{\widehat{\mathcal{T}}_{m,\varphi,h}(\mbox{\large $\boldsymbol{\chi}$}_i)\neq Y_i\big\}\, \Big( 1-  \pi_{\varphi^*}(\mbox{\large $\boldsymbol{\chi}$}_i, Y_i) \Big)\,\Big| \mathbb{D}_m   \right]\\
&& = E\left[ \mbox{\Large $\mathds{1}$}\hspace{-0.5mm}\big\{\widehat{\mathcal{T}}_{m,\varphi,h}(\mbox{\large $\boldsymbol{\chi}$}_i)\neq Y_i\big\}\,\big| \mathbb{D}_m   \right] \,=\, L_m(\widehat{\mathcal{T}}_{m,\varphi,h})
\end{eqnarray*}
because $\delta_i$ is independent of the data with $E(\delta_i)= p_n$, and the fact that $\Delta_i\in \mathbb{D}_{\ell}$ is independent of $\mathbb{D}_m$. Furthermore, conditional on $\mathbb{D}_m$, the terms $\big(\Delta_i +(1-\Delta_i) \delta_i/p_n\big)\cdot
\mbox{\Large $\mathds{1}$}\hspace{-0.5mm}\big\{\widehat{\mathcal{T}}_{m,\varphi,h}(\mbox{\large $\boldsymbol{\chi}$}_i)\neq Y_i\big\}$, \,$i\in \mathbb{D}_{\ell}$, are independent nonnegative random variables bounded by $1/p_n$. Therefore,
\begin{eqnarray}
(\mbox{r.h.s. of (\ref{Bound6-A})}) &\leq& \mathcal{N}_{\varepsilon_n}(\mathcal{F})\cdot\sup_{\varphi\in \mathcal{F}_{\varepsilon_n}}  E\left[ P\left\{ \left|
\frac{1}{\ell}
\sum_{i\in \boldsymbol{{\cal I}}_\ell}\left[
\left(\Delta_i +\frac{(1-\Delta_i) \delta_i}{p_n}\right)
\mbox{\Large $\mathds{1}$}\hspace{-0.5mm}\big\{\widehat{\mathcal{T}}_{m,\varphi,h}(\mbox{\large $\boldsymbol{\chi}$}_i)\neq Y_i\big\}\right]\right. \right. \right. \nonumber\\
&& ~~~~~~~~~~~~~~~~~~~~~~~~~~~~~~~~~~~~~~~~~~~~~~~~~~~~~~ \left.\left.   -\, L_m\big(\widehat{\mathcal{T}}_{m,\varphi,h}\big)
\Bigg| >\frac{t}{2}\,\Bigg| \mathbb{D}_m\right\}\right]\nonumber\\
&\leq& 2\,\mathcal{N}_{\varepsilon_n}(\mathcal{F}) \cdot\exp\big\{ -2\,p_n^2 \ell t^2/4  \big\}, ~~\mbox{(via Hoeffding's inequality).} \label{Bound7}
\end{eqnarray}
Since the above bound holds for all $t$\,$>$\,0, taking $t=t_0\cdot\sqrt{\log(\mathcal{N}_{\varepsilon_n}(\mathcal{F}))/\ell p_n^2\,}$, for any $t_0$\,$>$\,0, yields 
$
(\ref{Bound6-A}) \leq 2\big(\mathcal{N}_{\varepsilon_n}(\mathcal{F})\big)^{1-ct^2_0},
$
where $c>0$ is a constant not depending on $n$. Choosing $t_0$ large enough, we find (in view of (\ref{Bound86}) and (\ref{Bound6-A}))
$
\sum_{n=1}^{\infty} P\left\{ \big| R_n(1) + R_n(2) \big| > t  \right\} \,\leq\, 2 \sum_{n=1}^{\infty} \big(\mathcal{N}_{\varepsilon_n}(\mathcal{F})\big)^{1-ct^2_0} < \infty.
$
Therefore
\begin{equation}
\big| R_n(1) + R_n(2) \big| \,=\, \mathcal{O}_{a.co.}\left(\sqrt{\frac{\log(\mathcal{N}_{\varepsilon_n}(\mathcal{F}))}{\ell p_n^2}}\right). \label{EQ92}
\end{equation}

\noindent
Furthermore, in view of Lemma \ref{LEM-1}, 
\begin{eqnarray*}
|R_m(3)| &=& \mathcal{O}(h^{\beta_1 \wedge \beta_2}) 	+ \mathcal{O}(\varepsilon_n)
		+ \mathcal{O}_{a.\,co.}\left( \sqrt{\frac{\log[\mathcal{N}_{\tau_m}(\mathcal{S}_{\mbox{\tiny $\mathbb{X}$}})]}{m\cdot \phi(h)}} \right).
\end{eqnarray*}
This together with (\ref{EQ92}) completes the proof of Theorem  \ref{THM-BBC}.

\hfill $\Box$

\vspace{3mm}\noindent
PROOF OF THEOREM \ref{THM-H1}

\vspace{1mm}\noindent
Let $L_m(\widehat{\mathcal{T}}_{m,\varphi,h})$ be the misclassification error probability in (\ref{Lm-Ln}) and put 
\[
(\breve{\varphi}, \breve{h}) = \argmin_{\varphi\in \mathcal{F}_{\varepsilon_n} , h\in H} L_m(\widehat{\mathcal{T}}_{m,\varphi,h}).
\]
Also, observe the fundamental decomposition
\begin{eqnarray*}
L_n(\widehat{\mathcal{T}}_{n, \widehat{\varphi},\widehat{h}}) - \inf_{\varphi\in\mathcal{F}_{\varepsilon}} \inf_{h\in H} 
L_m(\widehat{\mathcal{T}}_{m,\varphi,h})  = 
\left\{ L_n(\widehat{\mathcal{T}}_{n, \widehat{\varphi},\widehat{h}}) - \widehat{L}_{m,\ell}(\widehat{\mathcal{T}}_{n, \widehat{\varphi},\widehat{h}})  \right\}
+ \left\{  \widehat{L}_{m,\ell}(\widehat{\mathcal{T}}_{n, \widehat{\varphi},\widehat{h}})  -  L_m(\widehat{\mathcal{T}}_{m,\breve{\varphi}, \breve{h}}) \right\}.
\end{eqnarray*}
But the first bracketed term above is bounded by $\sup_{\varphi\in \mathcal{F}_{\varepsilon}} \sup_{h\in H} \big|    \widehat{L}_{m,\ell}(\widehat{\mathcal{T}}_{n, \varphi,h})  -  L_m(\widehat{\mathcal{T}}_{m,\varphi, h}) \big|$. Furthermore, by the definition of 
$(\widehat{\varphi}, \widehat{h})$ in (\ref{phi.h.hat}),
$\widehat{L}_{m,\ell}(\widehat{\mathcal{T}}_{n, \widehat{\varphi},\widehat{h}})  \leq \widehat{L}_{m,\ell}(\widehat{\mathcal{T}}_{n, \breve{\varphi},\breve{h}})$ 
and thus 
$\big\{  \widehat{L}_{m,\ell}(\widehat{\mathcal{T}}_{n, \widehat{\varphi},\widehat{h}})  -  L_m(\widehat{\mathcal{T}}_{m,\breve{\varphi}, \breve{h}}) \big\} \leq 
\big\{\widehat{L}_{m,\ell}(\widehat{\mathcal{T}}_{n, \breve{\varphi},\breve{h}})   -  L_m(\widehat{\mathcal{T}}_{m,\breve{\varphi}, \breve{h}})  \big\} \leq \sup_{\varphi\in \mathcal{F}_{\varepsilon}} \sup_{h\in H} \big|    \widehat{L}_{m,\ell}(\widehat{\mathcal{T}}_{n, \varphi,h})  -  L_m(\widehat{\mathcal{T}}_{m,\varphi, h}) \big|$. Therefore, for every $t>0$,
\begin{eqnarray*}
	P\left\{  L_n(\widehat{\mathcal{T}}_{n, \widehat{\varphi},\widehat{h}}) - \inf_{\varphi\in\mathcal{F}_{\varepsilon}} \inf_{h\in H} 
L_m(\widehat{\mathcal{T}}_{m,\varphi,h}) 	>  t \right\}
	&\leq& 
P\left\{  \sup_{\varphi\in \mathcal{F}_{\varepsilon_n}} \sup_{h\in H} \Big|    \widehat{L}_{m,\ell}(\widehat{\mathcal{T}}_{n, \varphi,h})  -  L_m(\widehat{\mathcal{T}}_{m,\varphi, h}) \Big| >\frac{t}{2}
\right\}	
\end{eqnarray*}
However, we can write 
\begin{eqnarray}
&& P\left\{  \sup_{\varphi\in \mathcal{F}_{\varepsilon_n}} \sup_{h\in H} \Big|    \widehat{L}_{m,\ell}(\widehat{\mathcal{T}}_{n, \varphi,h})  -  L_m(\widehat{\mathcal{T}}_{m,\varphi, h}) \Big| >\frac{t}{2}
\right\}	\nonumber\\
&&~=\, 
P\left\{  \sup_{\varphi\in \mathcal{F}_{\varepsilon_n}}\sup_{h\in H}\left| \frac{1}{\ell}
\sum_{i\in \boldsymbol{{\cal I}}_\ell}\left[
\left(\Delta_i +\frac{(1-\Delta_i) \delta_i}{p_n}\right)
\mbox{\Large $\mathds{1}$}\hspace{-0.5mm}\big\{\widehat{\mathcal{T}}_{m,\varphi,h}(\mbox{\large $\boldsymbol{\chi}$}_i)\neq Y_i\big\}\right] - L_m\big(\widehat{\mathcal{T}}_{m,\varphi,h}\big) \right| >\frac{t}{2}
 \right\}~~~~~ \label{Bound6}
 \end{eqnarray}
On the other hand, for each $(\mbox{\large $\boldsymbol{\chi}$}_i, Y_i,\Delta_i) \in\mathbb{D}_{\ell}$ we have
\begin{eqnarray*}
&& E\left[ \left(\Delta_i +\frac{(1-\Delta_i) \delta_i}{p_n}\right)
\mbox{\Large $\mathds{1}$}\hspace{-0.5mm}\big\{\widehat{\mathcal{T}}_{m,\varphi,h}(\mbox{\large $\boldsymbol{\chi}$}_i)\neq Y_i\big\} \,\Big| \mathbb{D}_m\right]\\
&& = E\left[ E\left\{ 
\left(\Delta_i +\frac{(1-\Delta_i) \delta_i}{p_n}\right)
\mbox{\Large $\mathds{1}$}\hspace{-0.5mm}\big\{\widehat{\mathcal{T}}_{m,\varphi,h}(\mbox{\large $\boldsymbol{\chi}$}_i)\neq Y_i\big\}\,\bigg| \mathbb{D}_m, \mbox{\large $\boldsymbol{\chi}$}_i, Y_i, \delta_i
\right\}\,\bigg| \mathbb{D}_m\right]\\
&& = E\left[ \mbox{\Large $\mathds{1}$}\hspace{-0.5mm}\big\{\widehat{\mathcal{T}}_{m,\varphi,h}(\mbox{\large $\boldsymbol{\chi}$}_i)\neq Y_i\big\}\, \pi_{\varphi^*}(\mbox{\large $\boldsymbol{\chi}$}_i, Y_i)\,\Big|\mathbb{D}_m\right]
 + \frac{E(\delta_i)}{p_n} \left[\mbox{\Large $\mathds{1}$}\hspace{-0.5mm}\big\{\widehat{\mathcal{T}}_{m,\varphi,h}(\mbox{\large $\boldsymbol{\chi}$}_i)\neq Y_i\big\}\, \Big( 1-  \pi_{\varphi^*}(\mbox{\large $\boldsymbol{\chi}$}_i, Y_i) \Big)\,\Big| \mathbb{D}_m   \right]\\
&& = E\left[ \mbox{\Large $\mathds{1}$}\hspace{-0.5mm}\big\{\widehat{\mathcal{T}}_{m,\varphi,h}(\mbox{\large $\boldsymbol{\chi}$}_i)\neq Y_i\big\}\,\big| \mathbb{D}_m   \right] \,=\, L_m(\widehat{\mathcal{T}}_{m,\varphi,h})
\end{eqnarray*}
because $\delta_i$ is independent of the data with $E(\delta_i)= p_n$, and the fact that $\Delta_i\in \mathbb{D}_{\ell}$ is independent of $\mathbb{D}_m$. Furthermore, conditional on $\mathbb{D}_m$, the terms $\big(\Delta_i +(1-\Delta_i) \delta_i/p_n\big)\cdot
\mbox{\Large $\mathds{1}$}\hspace{-0.5mm}\big\{\widehat{\mathcal{T}}_{m,\varphi,h}(\mbox{\large $\boldsymbol{\chi}$}_i)\neq Y_i\big\}$, \,$i\in \mathbb{D}_{\ell}$, are independent nonnegative random variables bounded by $1/p_n$. Therefore,
\begin{eqnarray}
\mbox{(r.h.s. of (\ref{Bound6}))} &\leq&  
N\cdot \mathcal{N}_{\varepsilon_n}(\mathcal{F})\,\sup_{\varphi\in \mathcal{F}_{\varepsilon_n}} \max_{h\in H} \,
E\Bigg[ P\Bigg\{ \Bigg| \frac{1}{\ell}  \sum_{i\in \boldsymbol{{\cal I}}_\ell}\left[
\left(\Delta_i +\frac{(1-\Delta_i) \delta_i}{p_n}\right)
\mbox{\Large $\mathds{1}$}\big\{\widehat{\mathcal{T}}_{m,\varphi,h}(\mbox{\large $\boldsymbol{\chi}$}_i)\neq Y_i\big\}\right] \nonumber\\
&&~~~~~~~~~~~~~~~~~~~~~~~~~~~~~~~~~~~~~~~~~~ - L_m\big(\widehat{\mathcal{T}}_{m,\varphi,h}\big)
\Bigg| >\frac{t}{2}\,\Bigg|\, \mathbb{D}_m\Bigg\}\Bigg]\nonumber\\
&\leq&
 2\,N\cdot \mathcal{N}_{\varepsilon_n}(\mathcal{F}) \cdot\exp\big\{ -\ell p_n^2  t^2/2  \big\},
\end{eqnarray}
via Hoeffding's inequality. This completes the proof of Theorem \ref{THM-H1}.

\hfill $\Box$

\vspace{3.5mm}\noindent
PROOF OF THEOREM \ref{THM-BBC-2}

\vspace{0.5mm}\noindent
To prove this theorem, let $L_m(\widehat{\mathcal{T}}_{m,\varphi,h})\,:=P\big\{  \widehat{\mathcal{T}}_{m,\varphi,h}(\mbox{\large $\boldsymbol{\chi}$}) \neq Y\,\big| \mathbb{D}_m \big\}$ and observe that 
\begin{eqnarray}
L_n(\widehat{\mathcal{T}}_{n, \widehat{\varphi},\widehat{h}}) - L^* &=& \left\{   L_n(\widehat{\mathcal{T}}_{n, \widehat{\varphi},\widehat{h}}) - \inf_{\varphi\in\mathcal{F}_{\varepsilon_n}} \inf_{h\in H} 
L_m(\widehat{\mathcal{T}}_{m,\varphi,h})  \right\}
+ \left\{\inf_{\varphi\in\mathcal{F}_{\varepsilon_n}} \inf_{h\in H} L_m(\widehat{\mathcal{T}}_{m,\varphi,h})- L^*   \right\}\nonumber\\[3pt]
&=& T_n + T_m. \label{TnTm}
\end{eqnarray}
We deal with $T_m$ first. Let $\tilde{h}\equiv \tilde{h}_n$ be a sequence in $H$ satisfying $\tilde{h}\to 0$ and $m \tilde{h}^d\to\infty$, as $m\to\infty$ (take, for example, $\tilde{h}=m^{-1/(d+c)},$ $c>0$ arbitrary). Then $T_m \leq \inf_{\varphi\in\mathcal{F}_{\varepsilon_n}}  L_m(\widehat{\mathcal{T}}_{m,\varphi,\tilde{h}})- L^*$. Now by Lemma \ref{LEM-1}, 
\begin{equation}\label{Tm.bound}
T_m \,=\, \mathcal{O}(h^{\alpha}) 	+ \mathcal{O}(\varepsilon_n)
		+ \mathcal{O}_{a.\,co.}\left( \sqrt{\frac{\log[\mathcal{N}_{\tau_m}(\mathcal{S}_{\mbox{\tiny $\mathbb{X}$}})]}{m\cdot \phi(h)}} \right),
\end{equation}
for some positive constant $\alpha$.
 Next, to deal with the term $T_n$ in (\ref{TnTm}), put 
\[
(\breve{\varphi}, \breve{h}) = \argmin_{\varphi\in \mathcal{F}_{\varepsilon_n} ,\, h\in H} L_m(\widehat{\mathcal{T}}_{m,\varphi,h})
\]
and note that with $\widehat{L}_{m,\ell}$ as in (\ref{NEW-Lhat}),
\begin{eqnarray*}
T_n = L_n(\widehat{\mathcal{T}}_{n, \widehat{\varphi},\widehat{h}}) -  
L_m(\widehat{\mathcal{T}}_{m,\breve{\varphi}, \breve{h}})  = 
\left\{ L_n(\widehat{\mathcal{T}}_{n, \widehat{\varphi},\widehat{h}}) - \widehat{L}_{m,\ell}(\widehat{\mathcal{T}}_{n, \widehat{\varphi},\widehat{h}})  \right\}
+ \left\{  \widehat{L}_{m,\ell}(\widehat{\mathcal{T}}_{n, \widehat{\varphi},\widehat{h}})  -  L_m(\widehat{\mathcal{T}}_{m,\breve{\varphi}, \breve{h}}) \right\}
\end{eqnarray*}
But, by the definition of 
$(\widehat{\varphi}, \widehat{h})$ in (\ref{phi.h.hat}), one has
$\widehat{L}_{m,\ell}(\widehat{\mathcal{T}}_{n, \widehat{\varphi},\widehat{h}})  \leq \widehat{L}_{m,\ell}(\widehat{\mathcal{T}}_{n, \tilde{\varphi},\tilde{h}})$. Therefore, we have
$T_n\leq 
\big\{ L_n(\widehat{\mathcal{T}}_{n, \widehat{\varphi},\widehat{h}}) - \widehat{L}_{m,\ell}(\widehat{\mathcal{T}}_{n, \widehat{\varphi},\widehat{h}})  \big\}
+ \big\{  \widehat{L}_{m,\ell}(\widehat{\mathcal{T}}_{n, \breve{\varphi},\breve{h}})  -  L_m(\widehat{\mathcal{T}}_{m,\breve{\varphi}, \breve{h}}) \big\}$, which yields
\begin{eqnarray*}
T_n &\leq& 2 \sup_{\varphi\in \mathcal{F}_{\varepsilon_n}} \sup_{h\in H} \Big|    \widehat{L}_{m,\ell}(\widehat{\mathcal{T}}_{n, \varphi,h})  -  L_m(\widehat{\mathcal{T}}_{m,\varphi, h}) \Big|.
\end{eqnarray*}
Now let  $\mathcal{A}_{_{\mathcal{C}_m(\varphi)}}$ be the collection of all sets of the form  $A_{m, \varphi, h}$ in (\ref{AMPHI}). 
Also, let $\mbox{\Large $\nu$}\big(A_{m, \varphi, h}\big|\,\mathbb{D}_m\big)$ and
$\mbox{\Large $\nu$}_{\ell} \big(A_{m, \varphi, h}\big)$ be as in (\ref{NEW-A1}) and (\ref{NEW-A2}), respectively. Then, in view of (\ref{PROB2}), 
\begin{eqnarray}
L_m(\widehat{\mathcal{T}}_{m,\varphi,h})\,
&=& P\Big\{   (\mbox{\large $\boldsymbol{\chi}$}, Y) \in  A_{m, \varphi, h} \,\Big| \mathbb{D}_m \Big\} \,=\, \mbox{\Large $\nu$}\big(A_{m, \varphi, h}\big|\,\mathbb{D}_m\big) \label{L.equiv}\\
\widehat{L}_{m,\ell}(\widehat{\mathcal{T}}_{n, \varphi,h}) &=& 
\frac{1}{\ell} 
\sum_{i\in \boldsymbol{{\cal I}}_\ell}
\big[\Delta_i +(1-\Delta_i) \delta_i/p_n\big]\cdot \mbox{\Large $\mathds{1}$}\big\{(\mbox{\large $\boldsymbol{\chi}$}_i, Y_i) \in A_{m, \varphi, h}\big\} \,=\,
\mbox{\Large $\nu$}_{\ell} \big(A_{m, \varphi, h}\big),  \label{Lhat.equiv}
\end{eqnarray}
where the second expression above follows because for each $i\in \boldsymbol{{\cal I}}_\ell$, one has $\mbox{\Large $\mathds{1}$}\big\{ \widehat{\mathcal{T}}_{m,\varphi,h}(\mbox{\large $\boldsymbol{\chi}$}_i) \neq Y_i  \big\} = \mbox{\Large $\mathds{1}$}\big\{(\mbox{\large $\boldsymbol{\chi}$}_i, Y_i) \in A_{m, \varphi, h}\big\}$. Since for each $\varphi\in \mathcal{F}_{\varepsilon_n}$ the set $\mathcal{C}_m(\varphi)$ in  (\ref{CMPHI}) is the family of all classifiers indexed by $h\in H$, we find that for every $t>0$,
\begin{eqnarray}
P\{ T_n >t\} &\leq& P\left\{  \sup_{\varphi\in \mathcal{F}_{\varepsilon_n}}\,
\sup_{\widehat{\mathcal{T}}_{n, \varphi,h} \in\, \mathcal{C}_m(\varphi)} 
\left|    \widehat{L}_{m,\ell}(\widehat{\mathcal{T}}_{n, \varphi,h})  -  L_m(\widehat{\mathcal{T}}_{m,\varphi, h}) \right| > \frac{t}{2}  \right\}\nonumber\\
&=& P\left\{  \sup_{\varphi\in \mathcal{F}_{\varepsilon_n}}\,
\sup_{ A_{m, \varphi, h} \in \, \mathcal{A}_{_{\mathcal{C}_m(\varphi)}}} 
\Big| \mbox{\Large $\nu$}_{\ell} \big(A_{m, \varphi, h}\big) -
\mbox{\Large $\nu$}\big(A_{m, \varphi, h}\big|\,\mathbb{D}_m \big) \Big| > \frac{t}{2}  \right\},~\,\mbox{(by (\ref{L.equiv}) and (\ref{Lhat.equiv}))}\nonumber\\
&\leq& \mathcal{N}_{\varepsilon_n}(\mathcal{F})\cdot \sup_{\varphi\in \mathcal{F}_{\varepsilon_n}}  P\left\{  
\sup_{ A_{m, \varphi, h} \in \, \mathcal{A}_{_{\mathcal{C}_m(\varphi)}}} 
\Big| \mbox{\Large $\nu$}_{\ell} \big(A_{m, \varphi, h}\big) -
\mbox{\Large $\nu$}\big(A_{m, \varphi, h}\big|\,\mathbb{D}_m \big) \Big| > \frac{t}{2}  \right\}\nonumber\\
&=&  \mathcal{N}_{\varepsilon_n}(\mathcal{F}) \cdot \sup_{\varphi\in \mathcal{F}_{\varepsilon_n}}  E\left[ P\left\{  
\sup_{ A_{m, \varphi, h} \in \, \mathcal{A}_{_{\mathcal{C}_m(\varphi)}}} 
\Big| \mbox{\Large $\nu$}_{\ell} \big(A_{m, \varphi, h}\big) -
\mbox{\Large $\nu$}\big(A_{m, \varphi, h}\big|\,\mathbb{D}_m \big) \Big| > \frac{t}{2} \,\bigg| \mathbb{D}_m \right\}\right]\nonumber\\
&\leq&  8\,  \mathcal{N}_{\varepsilon_n}(\mathcal{F})\cdot \sup_{\varphi\in \mathcal{F}_{\varepsilon_n}} E\left[
\mathcal{S}\big( 
\mathcal{C}_m(\varphi),\, \ell\big)\right] \cdot e^{-\ell\, p_n^2 t^2/2},~~~ \mbox{(by (\ref{SALCm}) and Lemma \ref{LEM-VAP})}. \label{PTnt}
\end{eqnarray}
Since the above bound holds for all $t$\,$>$\,0, taking $t=t_0\cdot\sqrt{\log\big
\{\mathcal{N}_{\varepsilon_n}(\mathcal{F}) \cdot \sup_{\varphi\in \mathcal{F}_{\varepsilon_n}} E\left[
\mathcal{S}\big( \mathcal{C}_m(\varphi),\, \ell\big)\right]
\big\}/\ell p_n^2\,}$, for any $t_0$\,$>$\,0, yields 
$
P\{T_n>t\} \leq 8\big(\mathcal{N}_{\varepsilon_n}(\mathcal{F})\cdot\sup_{\varphi\in \mathcal{F}_{\varepsilon_n}} E\left[
\mathcal{S}\big( 
\mathcal{C}_m(\varphi),\, \ell\big)\right]\big)^{1-ct^2_0},
$
where $c>0$ is a constant not depending on $n$. Choosing $t_0$ large enough, we find 
\begin{equation}\label{Tnt8}
\sum_{n=1}^{\infty} P\left\{ T_n > t  \right\} \,\leq\, 8 \sum_{n=1}^{\infty} \big(\mathcal{N}_{\varepsilon_n}(\mathcal{F})\cdot\sup_{\varphi\in \mathcal{F}_{\varepsilon_n}} E\left[
\mathcal{S}\big( 
\mathcal{C}_m(\varphi),\, \ell\big)\right]\big)^{1-ct^2_0} < \infty.
\end{equation} 
Therefore
\begin{equation}
T_n \,=\, \mathcal{O}_{a.co.}\left(
\sqrt{\frac{\log(  \mathcal{N}_{\varepsilon_n}(\mathcal{F})   ) + 
\log\left( \sup_{\varphi\in \mathcal{F}_{\varepsilon_n}}E\left[\mathcal{S}\big( 
\mathcal{C}_m(\varphi),\, \ell\big)\right]  \right) }{\ell p_n^2}}\right). \label{Fin_B}
\end{equation}
Now, putting together this bound with the one on $T_m$, we find
\begin{eqnarray*}
L_n(\widehat{\mathcal{T}}_{n, \widehat{\varphi},\widehat{h}}) - L^* &=&
\mathcal{O}(h^{\alpha}) 	+ \mathcal{O}(\varepsilon_n)
		+ \mathcal{O}_{a.\,co.}\left( \sqrt{\frac{\log[\mathcal{N}_{\tau_m}(\mathcal{S}_{\mbox{\tiny $\mathbb{X}$}})]}{m\cdot \phi(h)}} \right) \\
&&~~~		+ \mathcal{O}_{a.co.}\left(
\sqrt{\frac{\log(  \mathcal{N}_{\varepsilon_n}(\mathcal{F})   ) + 
\log\left( \sup_{\varphi\in \mathcal{F}_{\varepsilon_n}}E\left[\mathcal{S}\big( 
\mathcal{C}_m(\varphi),\, \ell\big)\right]  \right) }{\ell p_n^2}}\right).		
\end{eqnarray*}
This completes the proof of Theorem \ref{THM-BBC-2}.

\hfill $\Box$

\vspace{3.5mm}\noindent
PROOF OF THEOREM \ref{THM-RATE2} 

\vspace{0.5mm}\noindent
We start by bounding $\mathcal{S}(\mathcal{C}_m(\varphi),\, \ell)$ in (\ref{BND10-B}). First let
\begin{eqnarray} \label{HmPHI}
\mathcal{H}_{m,\varphi, h}(\mbox{\scriptsize $\boldsymbol{\chi}$}) &:=& 
\frac{\sum_{i\in \boldsymbol{{\cal I}}_m}\Delta_i (2Y_i-1) \,\mathcal{K}\big(h^{-1}d(\mbox{\scriptsize $\boldsymbol{\chi}$}, \mbox{\large $\boldsymbol{\chi}$}_i)\big)}{\sum_{i\in \boldsymbol{{\cal I}}_m} 
\mathcal{K}\big(h^{-1}d(\mbox{\scriptsize $\boldsymbol{\chi}$}, \mbox{\large $\boldsymbol{\chi}$}_i)\big)}
+ 
\frac{\sum_{i\in \boldsymbol{{\cal I}}_m}\Delta_i (2Y_i-1)  \,\varphi(Y_i)\,
\mathcal{K}\big(h^{-1}d(\mbox{\scriptsize $\boldsymbol{\chi}$}, \mbox{\large $\boldsymbol{\chi}$}_i)\big)
}{\sum_{i\in \boldsymbol{{\cal I}}_m} \Delta_i  \,\varphi(Y_i)
\,\mathcal{K}\big(h^{-1}d(\mbox{\scriptsize $\boldsymbol{\chi}$}, \mbox{\large $\boldsymbol{\chi}$}_i)\big)
}\nonumber\\
&&~~~~~~~~~~~~~~~~~~~~~~~~~~~~~~~~~~~~~~~~~ \times \left(1-\frac{\sum_{i\in \boldsymbol{{\cal I}}_m}\Delta_i \,
\mathcal{K}\big(h^{-1}d(\mbox{\scriptsize $\boldsymbol{\chi}$}, \mbox{\large $\boldsymbol{\chi}$}_i)\big)
}{ \sum_{i\in \boldsymbol{{\cal I}}_m} 
\mathcal{K}\big(h^{-1}d(\mbox{\scriptsize $\boldsymbol{\chi}$}, \mbox{\large $\boldsymbol{\chi}$}_i)\big)
}
\right)
\end{eqnarray}
with the convention $0/0 = 0$, and observe that  
\[\widehat{\mathcal{R}}_{m}(\mbox{\scriptsize $\boldsymbol{\chi}$};\varphi,h)>\frac{1}{2}~~~ \mbox{if and only if}~~
~\mathcal{H}_{m,\varphi, h}(\mbox{\scriptsize $\boldsymbol{\chi}$}) >0,
\] 
where $\widehat{\mathcal{R}}_{m}(\mbox{\scriptsize $\boldsymbol{\chi}$};\varphi,h)$ is as in (\ref{mhat3}). Therefore the classifier in (\ref{gFin}) can alternatively be expressed as
\begin{equation}\label{NEW-T.hat}
\widehat{\mathcal{T}}_{m,\varphi,h}(\mbox{\scriptsize $\boldsymbol{\chi}$}) = 1~~\mbox{if~ $\mathcal{H}_{m,\varphi, h}(\mbox{\scriptsize $\boldsymbol{\chi}$}) 
>0$\, (otherwise $\widehat{\mathcal{T}}_{m,\varphi,h}(\mbox{\scriptsize $\boldsymbol{\chi}$})=0)$.}
\end{equation}
Now, for each $\varphi\in \mathcal{F}$ define the   {\it kernel complexity} $\kappa_m(\varphi)$ as follows (see, for example, Sec. 25 of Devroye et al (1996) for more on this):
\begin{equation}\label{KMPHI}
\kappa_m(\varphi) =\hspace{-2mm} \sup_{\mbox{\scriptsize $\boldsymbol{\chi}$},\, (\mbox{\scriptsize $\boldsymbol{\chi}$}_1, y_1),\dots, (\mbox{\scriptsize $\boldsymbol{\chi}$}_m, y_m)} \Big\{
\mbox{number of sign changes of $\mathcal{H}_{m,\varphi, h}(\mbox{\scriptsize $\boldsymbol{\chi}$})$ as $h$ varies from 0 to $\infty$}
\Big\}.
\end{equation}
Then, as $h$ varies from 0 to infinity, the binary $\ell$-dimensional vector 
\[
\left\{\mathcal{H}_{m,\varphi, h}(\mbox{\large $\boldsymbol{\chi}$}_j)\right\}_{j\in \boldsymbol{{\cal I}}_\ell}
\]
changes at most $\ell\cdot\kappa_m(\varphi)$ times. It therefore takes at most $\ell\cdot\kappa_m(\varphi)+1$ different values which implies that for each $\varphi\in \mathcal{F}$
\begin{equation}
\mathcal{S}\big(\mathcal{C}_m(\varphi),\, \ell\big) ~\leq~ \ell\cdot\kappa_m(\varphi)+1 .  \label{BND8}
\end{equation}
Next, we will show that $\kappa_m(\varphi)\leq (2m)^2$ for all $\varphi$. 
 To this end, let 
\begin{eqnarray*}
A_{m,h}(\mbox{\scriptsize $\boldsymbol{\chi}$} &=& 1/\sum_{i\in \boldsymbol{{\cal I}}_m} 
\mathcal{K}\big(h^{-1}d(\mbox{\scriptsize $\boldsymbol{\chi}$}, \mbox{\large $\boldsymbol{\chi}$}_i)\big)\\
B_{m,\varphi,h}(\mbox{\scriptsize $\boldsymbol{\chi}$}) &=& \bigg(1-\frac{\sum_{i\in \boldsymbol{{\cal I}}_m}\Delta_i \,
\mathcal{K}\big(h^{-1}d(\mbox{\scriptsize $\boldsymbol{\chi}$}, \mbox{\large $\boldsymbol{\chi}$}_i)\big)
}{ \sum_{i\in \boldsymbol{{\cal I}}_m} 
\mathcal{K}\big(h^{-1}d(\mbox{\scriptsize $\boldsymbol{\chi}$}, \mbox{\large $\boldsymbol{\chi}$}_i)\big)
}
\bigg) \div \sum_{i\in \boldsymbol{{\cal I}}_m} \Delta_i  \,\varphi(Y_i)
\,\mathcal{K}\big(h^{-1}d(\mbox{\scriptsize $\boldsymbol{\chi}$}, \mbox{\large $\boldsymbol{\chi}$}_i)\big),
\end{eqnarray*}
and observe that the quantity $\mathcal{H}_{m,\varphi, h}(\mbox{\scriptsize $\boldsymbol{\chi}$})$  in (\ref{HmPHI}) can be written as
\[
\mathcal{H}_{m,\varphi, h}(\mbox{\scriptsize $\boldsymbol{\chi}$}) = A_{m,h}(\mbox{\scriptsize $\boldsymbol{\chi}$})\hspace{-1mm} \sum_{i\in \boldsymbol{{\cal I}}_m}\hspace{-1mm} \Delta_i (2Y_i-1) \,\mathcal{K}\big(h^{-1}d(\mbox{\scriptsize $\boldsymbol{\chi}$}, \mbox{\large $\boldsymbol{\chi}$}_i)\big) \,+\, B_{m,\varphi,h}(\mbox{\scriptsize $\boldsymbol{\chi}$})\hspace{-1mm}  \sum_{i\in \boldsymbol{{\cal I}}_m}\Delta_i (2Y_i-1)  \,\varphi(Y_i)\,
\mathcal{K}\big(h^{-1}d(\mbox{\scriptsize $\boldsymbol{\chi}$}, \mbox{\large $\boldsymbol{\chi}$}_i)\big). 
\]
Arguing as in Devroye et al (1996; Ch.\,25), since $0<A_{m,h}(\mbox{\scriptsize $\boldsymbol{\chi}$}) <\infty$ holds for each $\mbox{\scriptsize $\boldsymbol{\chi}$}$ and $h$, the total number of sign changes of the expression $A_{m,h}(\mbox{\scriptsize $\boldsymbol{\chi}$})\sum_{i\in \boldsymbol{{\cal I}}_m}\hspace{-1mm} \Delta_i (2Y_i-1) \,\mathcal{K}\big(h^{-1}d(\mbox{\scriptsize $\boldsymbol{\chi}$}, \mbox{\large $\boldsymbol{\chi}$}_i)\big)$ is the same as that of $\sum_{i\in \boldsymbol{{\cal I}}_m}\hspace{-1mm} \Delta_i (2Y_i-1) \,\mathcal{K}\big(h^{-1}d(\mbox{\scriptsize $\boldsymbol{\chi}$}, \mbox{\large $\boldsymbol{\chi}$}_i)\big)$, as $h$ varies from 0 to $\infty$, which is at most $2m$. Similarly, since $0< B_{m,\varphi,h}(\mbox{\scriptsize $\boldsymbol{\chi}$})<\infty$ holds for all $\mbox{\scriptsize $\boldsymbol{\chi}$}$, $h$, and $\varphi\in \mathcal{F}_{\varepsilon_n}$, the function $B_{m,\varphi,h}(\mbox{\scriptsize $\boldsymbol{\chi}$}) \sum_{i\in \boldsymbol{{\cal I}}_m}\Delta_i (2Y_i-1)  \,\varphi(Y_i)\,
\mathcal{K}\big(h^{-1}d(\mbox{\scriptsize $\boldsymbol{\chi}$}, \mbox{\large $\boldsymbol{\chi}$}_i)\big)$ has at most $2m$ sign changes as $h$ varies from 0 to $\infty$. Therefore the combinatorial quantity $\kappa_m(\varphi)$ in (\ref{KMPHI}) is bounded by $(2m)^2$ for any $\varphi$. This together with  (\ref{BND8}) implies that  the expression in (\ref{Tnt8}) becomes
\begin{eqnarray}
\sum_{n=1}^{\infty} P\left\{ T_n > t  \right\} &\leq& 8 \sum_{n=1}^{\infty} \big(\mathcal{N}_{\varepsilon_n}(\mathcal{F})\cdot 4m^2\ell\big)^{1-ct^2_0} \nonumber\\
&\leq& 32  \Big(\sum_{n=1}^{\infty} \big|\mathcal{N}_{\varepsilon_n}(\mathcal{F})\big|^{2(1-ct^2_0)} \Big)^{1/2} 
\Big(\sum_{n=1}^{\infty} n^{3(1-ct^2_0)} \Big)^{1/2} \label{star9}\\
&<& \infty,~~~\mbox{(for $t_0$ large enough by assumption (A5)(ii)),} \nonumber
\end{eqnarray}
where (\ref{star9}) follows from Holder's inequality and the fact that $m^2\ell < n^3$. Therefore (\ref{Fin_B}) becomes


\begin{equation}
T_n \,=\, \mathcal{O}_{a.co.}\left(
\sqrt{\frac{\log(  \mathcal{N}_{\varepsilon_n}(\mathcal{F})   ) + 
\log(m^2\ell) }{\ell p_n^2}}\right). \label{Fin_B7}
\end{equation}
Theorem \ref{THM-RATE2} now follows from (\ref{Fin_B7}), (\ref{Tm.bound}), and (\ref{TnTm}).

\hfill    $\Box$

\vspace{10mm}\noindent
{\bf Acknowledgements} \\
This work was supported by the National Science Foundation (NSF) under Grant DMS-2310504 of Majid Mojirsheibani

\vspace{12mm}\noindent
{\bf \Large References}

\vspace{2mm}\noindent
Abraham C, Biau G, and Cadre B (2006) On the kernel rule for functional classification.  AISM  58:619--633.  \href{https://doi.org/10.1007/s10463-006-0032-1}{https://doi.org/10.1007/s10463-006-0032-1}

\vspace{2.6mm}\noindent
Azizyan M, Singh A, Wasserman L (2013) Density-sensitive semisupervised inference. Ann. Statist 41(2):751--771.
  \href{https://doi.org/10.1214/13-AOS1092}{https://doi.org/10.1214/13-AOS1092}



\vspace{2.6mm}\noindent
Biau G, Bunea F, and Wegkamp M (2005) Functional classification in Hilbert spaces. IEEE Trans. Inform. Theory 51:2163--2172. \href{https://doi.org/10.1109/TIT.2005.847705}{https://doi.org/10.1109/TIT.2005.847705}

\vspace{2.6mm}\noindent
Biau G, C\'erou F, and Guyader A (2010) Rates of convergence of the functional k-nearest neighbor estimate. IEEE Trans. Inform. Theory 56(4):2034--2040. \href{https://doi.org/10.1109/TIT.2010.2040857}{https://doi.org/10.1109/TIT.2010.2040857}

\vspace{2.6mm}\noindent 
Biau G, Fischer A, Guedj B, and Malley J (2016) Cobra: a combined regression strategy.  J. Multivar. Anal 146:18--28.
\href{https://doi.org/10.1016/j.jmva.2015.04.007}{https://doi.org/10.1016/j.jmva.2015.04.007}

\vspace{2.6mm}\noindent
Bindele H,  Zhao Y (2018) Rank-based estimating equation with non-ignorable missing responses via empirical likelihood. Statistica Sinica 28(4):1787--1820.
\href{https://www.jstor.org/stable/26511189}{https://www.jstor.org/stable/26511189}

\vspace{2.6mm}\noindent
Bouzebda S, Souddi Y, and Madani F (2024) Weak convergence of the conditional set-indexed empirical process for missing at random functional ergodic data.  Mathematics 12(3), 448.\\ \href{https://doi.org/10.3390/math12030448}{https://doi.org/10.3390/math12030448} 

\vspace{2.6mm}\noindent
Carroll RJ, Delaigle A, and Hall P (2013) Unexpected properties of bandwidth choice when smoothing discrete data for constructing a functional data classifier. Ann. Statist. 41(6):2739--2767. \href{https://doi.org/10.1214/13-AOS1158}{https://doi.org/10.1214/13-AOS1158}

\vspace{2.6mm}\noindent
C\'erou F and Guyader A (2006) Nearest neighbor classification in infinite dimensions. ESAIM-Probab. Stat 10:340--355. \href{https://doi.org/10.1051/ps:2006014}{https://doi.org/10.1051/ps:2006014}

\vspace{2.6mm}\noindent
Chen X, Diao G, and Qin J (2020) Pseudo likelihood-based estimation and testing of missingness mechanism function in nonignorable missing data problems. Scand. J. Stat 47(4):1377--1400. \href{https://doi.org/10.1111/sjos.12493}{https://doi.org/10.1111/sjos.12493}

\vspace{2.6mm}\noindent
Cheng PE and Chu CK (1996) Kernel estimation of distribution functions and quantiles with missing data.  Statistica Sinica 6(1):63–78. \href{https://www.jstor.org/stable/24305999}{https://www.jstor.org/stable/24305999}

\vspace{2.6mm}\noindent
Delaigle A and Hall P (2012) Achieving near perfect classification for functional data. J. R. Stat. Soc. Ser. B. Stat. Methodol. 74(2):267--286. \href{https://doi.org/10.1111/j.1467-9868.2011.01003.x}{https://doi.org/10.1111/j.1467-9868.2011.01003.x}

\vspace{2.6mm}\noindent
Devroye L, Gy\"{o}rfi L, and Lugosi G (1996) A probabilistic theory of pattern recognition. Springer-Verlag, New York. 



\vspace{2.6mm}\noindent
Dudley R (1978) Central limit theorems for empirical measures.  Ann. Probab  6(6):899--929.
\href{https://doi.org/10.1214/aop/1176995384}{https://doi.org/10.1214/aop/1176995384}


\vspace{2.6mm}\noindent
Febrero-Bande M and Oviedo de la Fuente M (2012) Statistical Computing in Functional Data Analysis: The R Package fda.usc. J. Stat. Soft 51:1-28. \href{https://doi.org/10.18637/jss.v051.i04}{https://doi.org/10.18637/jss.v051.i04}

\vspace{2.6mm}\noindent
Ferraty F,  Laksaci A, Tadj A, and Vieu P (2010) Rate of uniform consistency for nonparametric estimates with functional variables. J. Statist. Plann. Inf 140:335--352. \href{https://doi.org/10.1016/j.jspi.2009.07.019}{doi.org/10.1016/j.jspi.2009.07.019}

\vspace{2.6mm}\noindent
Ferraty F, Sued M, and Vieu P (2013) Mean estimation with data missing at random for functional covariables.  Statistics  47(4):688--706. \href{https://doi.org/10.1080/02331888.2011.650172}{https://doi.org/10.1080/02331888.2011.650172}

\vspace{2.6mm}\noindent
Ferraty F and Vieu P (2006) Nonparametric Functional Data Analysis: Theory and Practice. Springer, New York.


\vspace{2.6mm}\noindent
Greenlees W, Reese J, and Zieschang K (1982) Imputation of missing values when the probability of response depends on the variable being imputed. J. Am. Statist. Assoc 77(378):251--261. \href{https://doi.org/10.1080/01621459.1982.10477793}{https://doi.org/10.1080/01621459.1982.10477793}

\vspace{2.6mm}\noindent
Guo X, Song Y, and Zhu L (2019) Model checking for general linear regression with nonignorable missing response. Computational Statistics \& Data Analysis 138:1--12. \\
\href{https://doi.org/10.1016/j.csda.2019.03.009}{https://doi.org/10.1016/j.csda.2019.03.009}




\vspace{2.6mm}\noindent
Kim JK,  Yu CL (2011) A semiparametric estimation of mean functionals with nonignorable missing data. J. Am. Statist. Assoc 106(493):157--65. \href{https://doi.org/10.1198/jasa.2011.tm10104}{https://doi.org/10.1198/jasa.2011.tm10104}




\vspace{2.6mm}\noindent
Ling N, Liang L, and Vieu P (2015) Nonparametric regression estimation for functional stationary ergodic data with missing at random. J. Stat. Plan. Inference 162:75--87.  \\
\href{https://doi.org/10.1016/j.jspi.2015.02.001}{https://doi.org/10.1016/j.jspi.2015.02.001}




\vspace{2.6mm}\noindent
Meister A (2016) Optimal classification and nonparametric regression for functional data.  Bernoulli 22(3):1729--1744. \href{http://dx.doi.org/10.3150/15-BEJ709}{http://dx.doi.org/10.3150/15-BEJ709}

\vspace{2.6mm}\noindent
Miao W, Li X, and Sun B (2024) A stableness of resistance model for nonresponse adjustment with callback data. \href{https://doi.org/10.48550/arXiv.2112.02822}{https://doi.org/10.48550/arXiv.2112.02822} 

\vspace{2.6mm}\noindent
Mitrinovic DS (1970) Analytic Inequalities. Springer-Verlag, New York. 

\vspace{2.6mm}\noindent
Mojirsheibani M (2022) On the maximal deviation of kernel regression estimators with MNAR response variables. Statistical Papers 63:1677--1705. \href{https://doi.org/10.1007/s00362-022-01293-0}{https://doi.org/10.1007/s00362-022-01293-0}



\vspace{2.6mm}\noindent
Morikawa K, Kim JK, and  Kano Y (2017)  Semiparametric maximum likelihood estimation with data missing not at random. Canad. J. Statist.  45(4):393--409. \href{https://doi.org/10.1002/cjs.11340}{https://doi.org/10.1002/cjs.11340}


\vspace{2.6mm}\noindent
Niu C, Guo X, Xu W, and Zhu L (2014) Empirical likelihood inference in linear regression with nonignorable missing response. Computational Statistics \& Data Analysis 79():91--112. \\ \href{https://doi.org/10.1016/j.csda.2014.05.005}{https://doi.org/10.1016/j.csda.2014.05.005} 

\vspace{2.6mm}\noindent
Pollard D (1984) Convergence of Stochastic Processes. Springer, New York.





\vspace{2.6mm}\noindent
Shao J and  Wang L (2016) Semiparametric inverse propensity weighting for nonignorable missing data. Biometrika 103(1):175--187. \href{https://doi.org/10.1093/biomet/asv071}{https://doi.org/10.1093/biomet/asv071}

\vspace{2.6mm}\noindent
Tang G, Little RJA, and Raghunathan TE (2003) Analysis of multivariate missing data with nonignorable nonresponse. BiometriKa 90(4):747--764. \href{https://doi.org/10.1093/biomet/90.4.747}{https://doi.org/10.1093/biomet/90.4.747}


\vspace{2.6mm}\noindent
Uehara M, Lee D, and Kim JK (2023) Statistical inference with semiparametric nonignorable nonresponse models. Scan J. Stat 50(4):1795--1817.  \href{https://doi.org/10.1111/sjos.12652}{https://doi.org/10.1111/sjos.12652}

\vspace{2.6mm}\noindent
van der Vaart A, Wellner J (1996) Weak Convergence and Empirical Processes with Applications to Statistics. Springer, New York. 




\vspace{2.6mm}\noindent
Wang J,  Shen X (2007) Large margin semi-supervised  learning. J. Mach. Learn. Res  8:1867--1891. 

\vspace{2.6mm}\noindent
Zhao J,  Ma Y  (2022) A versatile estimation procedure without estimating the nonignorable missingness mechanism. J. Am. Statist. Assoc 117(540):1916--1930. \href{https://doi.org/10.1080/01621459.2021.1893176}{https://doi.org/10.1080/01621459.2021.1893176}



\end{document}